\documentclass[preprintnumbers,amsmath,amssymb]{revtex4}
\usepackage{graphicx,color}
\begin{document}

\vspace*{1.0cm}

\begin{center}
{\Large {\bf Calculation of BR($\overline{B}^0\to
\Lambda_{c}^++\bar p)$ in the PQCD Approach}}
\end{center}
\vspace*{0.3cm}
\begin{center}
{Xiao-Gang He$^{1,2}$, Tong Li$^1$, Xue-Qian Li$^1$, and Yu-Ming Wang$^1$}\\
\vspace*{0.3cm}
$^1$Department of Physics, Nankai University, Tianjin\\

\vspace*{0.1in}
 $^2$Department of Physics and Center for Theoretical Sciences,
 National Taiwan University,
 Taipei\\
\end{center}

 \vspace*{0.5cm}

\begin{center}
\begin{minipage}{12cm}

\noindent Abstract:\\

We calculate the branching ratio of $\overline{B}^0 \to
\Lambda_c^+ \overline{p}$ in the PQCD approach. Most previous
model calculations obtained branching ratios significantly larger
than experimental data. We find that the predicted branching ratio
for BR$(\overline{B}^0 \to \Lambda_c^+ \overline{p})$ in the PQCD
approach can vary over a range of $(2.3\sim 5.1)\times 10^{-5}$
with the largest uncertainty coming from the parameters in the
wave function of $\Lambda_c$. With the favored values  for the
parameters in the $\Lambda_c^+$ wave function,  $\beta = 1$ GeV
and $m_q = 0.3$ GeV, the branching ratio is about $2.3\times
10^{-5}$ which is satisfactorily consistent with the value
measured by experiments.
\end{minipage}

\end{center}

\vspace{2 cm} PACS numbers: 13.30.Ce, 12.38.Bx, 14.20.Mr

\newpage

\section{Introduction}

B-physics stands as an ideal laboratory for studying hadron
structure and fundamental mechanisms which govern the reactions
for a heavy quark and the related hadronization processes. It also
offers a window to search for new physics. The mesonic decays of
B-meson have already been carefully studied by many authors
because such processes are theoretically simpler and have a rich
data accumulation. Nevertheless much data on the baryonic decays
of B-mesons have also been collected. In 1987, $B\rightarrow
p\overline{p}\pi^{\pm}$ and $B\rightarrow p\overline{p}\pi^+\pi^-$
were firstly observed by ARGUS\cite{ARGUS} and since then, various
baryonic B baryonic-decay modes were measured at CLEO, Belle and
BaBar\cite{other, Belle,babar,CLEO,data0}. All the information
offers us an opportunity to more seriously investigate the
processes where baryons are involved. However, the case for
baryonic decays is much more complicated than the mesonic one,
especially in the perturbative QCD (PQCD)
approach\cite{hadronmesonLi} where there is not any real spectator
constituent, namely all the constituents must be connected by
exchanged gluons. One also needs to have detailed information for
the wave functions of baryons which contain three valence quarks.
All these make the problem much more complicated and cause
theoretical uncertainties.

In this work we study the process $\overline{B}^0\to \Lambda_c^+
\overline{p}$. This process has been measured
\cite{Belle,babar,CLEO} with a branching ratio of \cite{data0}
$(2.2\pm 0.8)\times 10^{-5}$. There have been some theoretical
evaluations for this branching ratio using various
phenomenological models, such as the constituent quark model, the
pole model, the QCD sum rule, the diquark model and
others\cite{hewu,Chernyak,Lu,Jarfi,Ball}. All the theoretical
predictions made in these models are substantially larger than the
data. Later Cheng and Yang\cite{Cheng} considered the pole
structure and applied the bag model approach for calculating the
hadronic matrix element. Their result is close to the experimental
data. The related theoretical estimations on the branching ratio
and the experimental data are listed in Table I.

Since the success of the PQCD approach in studying mesonic decays
of B-meson is remarkable, it is natural to extend this approach to
evaluate the decay rates of baryonic decays of B-mesons. In
$\overline{B}^0\to \Lambda_c^+ \overline{p}$, the two final
baryons are much lighter than B-meson, they possess relatively
large linear momentum ${\bf
p}=\lambda^{1/2}(M_B^2,M_{\Lambda_c}^2,M_p^2)/2M_B$ in the rest
frame of B-meson, thus the momentum match of the quarks in the
final baryons requires that the exchanged gluons must be ''hard'',
or say, have large $k^2$. Here $k$ indicates the generic momentum
carried by a gluon. Moreover, unlike in some cases where a pair of
quark-antiquark is created from vacuum, the pair of
quark-antiquark is created by a hard gluon, therefore one expects
that the whole process is calculable in the framework of
PQCD\cite{Cheng1}.

In the PQCD approach, there are unknown parameters in the wave
functions of the B-meson and the two baryons in the final state.
However, since these parameters are universal, they can be fixed
by fitting data of other relevant decays. This treatment of the
unknown parameters reduces the model dependence of the theoretical
calculations. Therefore one has good reasons to believe that the
result obtained in PQCD would be credible. Indeed, our numerical
result is closer to the experimental data than those by other
models.

\begin{table}[h]
\caption{Branching ratio of $\overline{B}^0\to \Lambda_c^+
\overline{p}$ for different models and experiments (last two
numbers), respectively. There is also a preliminary Babar
result\cite{babar} $(2.15\pm 0.36\pm 0.13\pm 0.56)\times
10^{-5}$.}
\begin{center}
\begin{tabular}{|c|c|c|c|c|c||c|c|}
  \hline
  $\overline{B}^0\to \Lambda_c^+
\overline{p}$ & \cite{Chernyak} & \cite{Lu} & \cite{Jarfi} & \cite{Ball} &
\cite{Cheng} &  \cite{Belle} & \cite{CLEO}\\
  \hline
  BR & $4\times 10^{-4}$ & $8.5\times 10^{-4}$ &
$1.1\times 10^{-3}$
  & $(1.7\sim 1.9)\times 10^{-3}$ & $1.1\times 10^{-5}$ &
  $(2.19^{+0.56}_{-0.49}\pm0.32\pm0.57)\times 10^{-5}$ & $<9\times 10^{-5}$\\
  \hline
\end{tabular}
\end{center}\label{BR}
\end{table}

The effective Hamiltonian responsible at the quark level for
$\overline{B}^0\to \Lambda_c^+ \overline{p}$ is given by\cite{H}:
\begin{eqnarray}
H_{eff} &=&
\frac{G_{F}}{\sqrt{2}}V_{cb}V_{ud}^{*}\left[C_1(\overline{c}b)_{V-A}(\overline{d}u)_{V-A}+
C_2(\overline{c}u)_{V-A}(\overline{d}b)_{V-A}\right]+h.c.,
\label{gamma}
\end{eqnarray}
where $(\bar a b)_{V-A}=\bar a\gamma^\mu (1-\gamma_5) b$, and
$C_1^{LO}(m_b)=1.12, C_2^{LO}(m_b)=-0.27$\cite{H}. In our
calculation the $\mu$ dependence of $C_{1,2}$ at the leading order
remains.

At the hadron level, the decay amplitude for $\overline{B}^0\to
\Lambda_c^+ \overline{p}$ is obtained by sandwiching  the
effective Hamiltonian between the initial and final hadron states,
\begin{eqnarray}
M(\overline{B}^0\to \Lambda_c^+ \overline{p}) &=& \langle
\Lambda_c^+ \overline{p}\vert H_{eff}\vert
\overline{B}^0\rangle\nonumber \\
&=&\frac{G_{F}}{\sqrt{2}}V_{cb}V_{ud}^{*}\overline{\Lambda_c^+}\left(
A+B\gamma_5\right)p,\label{formfactor}
\end{eqnarray}
where $A$ and $B$ are two form factors which we will calculate in
the PQCD approach.

The partial decay width is then given by
\begin{eqnarray}
\Gamma(\overline{B}^0\to \Lambda_c^+ \overline{P}) &=&
{G_F^2|V_{cb}|^2|V_{ud}|^2\over 16 m_{B}^3\pi}
\sqrt{(m_B^2-(m_{\Lambda_c}+m_P)^2)(m_B^2-(m_{\Lambda_c}-m_P)^2)}\nonumber
\\
&&\left[|A|^2(m_B^2-(m_{\Lambda_c}+m_P)^2) +
|B|^2(m_B^2-(m_{\Lambda_c}-m_P)^2)\right].
\end{eqnarray}

In the following sections we present details for the calculations.
In Section II, we describe the approach for calculations. In
Section III, we present our numerical results along with all
necessary input parameters. The last section is devoted to our
discussions and conclusions. Detailed calculations for each diagram are
given in the Appendices.\\

\section{PQCD calculation of the hadronic matrix elements}

The effective Hamiltonian at quark level has been studied by many
authors and the PQCD approach has also extensively been discussed
in literature. In this section we discuss how to calculate the
hadronic matrix elements defined in the previous section in terms
of the PQCD method\cite{hadronmesonLi,Lip,Lic,LiJ}. As usual, we
define, in the rest frame of $\overline{B^0}$, $q$, $p$ and $p'$
to be the four-momenta of $\overline{B^0}$, $\Lambda_{c}^+$ and
antiproton, $q$ is the $b$ quark momentum, $q_i(i=1,2)$,
$k_i(i=1,2,3)$ and $k_i'$ to be the momenta of the valance light
quarks (anti-quark) inside $\overline{B^0}$, $\Lambda_{c}^+$, and
$\overline{p}$ respectively. We parameterize the corresponding
light cone momenta with the masses of all light quarks and
antiproton  being neglected as
\begin{eqnarray}
&&q = (q^{+},q^{-},\mathbf{0}_{T}) = \frac
{m_{B}}{\sqrt{2}}(1,1,\mathbf{0}_{T}), \;\;p =
(p^{+},p^-,\mathbf{0}_{T}) = (q^{+},r^2q^-,\mathbf{0}_{T}) = \frac
{m_{B}}{\sqrt{2}}(1,r^2,\mathbf{0}_{T})\nonumber \\
&&p' = (0,p'^-,\mathbf{0}_{T}) = \frac
{m_{B}}{\sqrt{2}}(0,1-r^2,\mathbf{0}_{T})\nonumber \\
&&q_1 = (yq^+,q^-,\mathbf{q}_{T}),\;\;q_2 = ((1-y)q^+,0,-\mathbf{q}_{T})\nonumber \\
&&k_{1} = (x_1p^{+},p^{-},\textbf{k}_{1T}),\;\;k_{2} =
(x_2p^{+},0,\textbf{k}_{2T}),
\;\;k_{3} = (x_3p^{+},0,\textbf{k}_{3T})\nonumber \\
&&k'_{1} = (0,x_{1}'p'^{-},\textbf{k}_{1T}'),\;\;k'_{2} =
(0,x_{2}'p'^{-},\textbf{k}_{2T}'),\;\;k'_{3} =
(0,x_{3}'p'^{-},\textbf{k}_{3T}')
\end{eqnarray}
where $r=m_{\Lambda_c}/m_B$, and $y$, $x_i$, $x'_i$ are the
fractions of the longitudinal momenta of the valence quarks with
$x_1+x_2+x_3 = 1$ and $x'_1+x'_2+x'_3 = 1$. $\mathbf{q}_{T}$,
$\mathbf{k}_{iT}$ and $\mathbf{k}_{iT}'$ are the transverse
momenta of the valence quarks inside $\overline{B^0}$,
$\Lambda_c^+$ and antiproton, respectively.

Note that because $\overline{B^0}$ is in its rest frame, even
though the momenta of the valance quarks inside the final states
($\Lambda_c^+, \overline{p}$) are fixed, the four-momenta of the
valance quarks inside $\overline{B^0}$ ($q_1, q_2$) are not
uniquely determined. One can also choose to have:
\begin{eqnarray}
&&q_1 = (q^+,yq^-,\mathbf{q}_{T}),\;\;q_2 =
(0,(1-y)q^-,-\mathbf{q}_{T}).
\end{eqnarray}
Obviously, the first case ($q^1_i$) corresponds to that the
momenta of $\overline{d}$-quarks in initial state $\overline{B^0}$
and in final state $\overline{p}$ are opposite to each other, and
the second one ($q^2_i$) is in the same direction as the
antiproton momentum (i.e. they are parallel.). One has to
convolute the initial light-cone wave functions with the final
light-cone wave function to make the correct choice.

In the PQCD picture, hadrons are made of valence quarks whose
momenta-distributions are described by appropriate wave functions.
The wave function of $\Lambda_c$ is usually defined through the
correlator\cite{lambdacwave},
\begin{eqnarray}
&&(Y_{\Lambda_c})_{\alpha \beta \gamma}(k_{i},\nu) =
\frac{1}{2\sqrt{2}N_{c}}
 \int \prod_{l=2}^{3}\frac{d w_{l}^{+} d \bf{w}_l }{(2\pi)^{3}} e^{ik_{l}w_{l}}
 \varepsilon^{abc}\langle0|T[c_{\alpha}^{a}(0)u_{\beta}^{b}(w_{2})d_{\gamma}^{c}(w_{3})]|
 \Lambda_c(p)\rangle\nonumber \\
&&=\frac{f_{\Lambda_c}}{8\sqrt{2}N_{c}}[(p\!\!\! \slash
+m_{\Lambda_c})\gamma_{5}C]_{\beta \gamma}
[\Lambda_c(p)]_{\alpha}\Psi(k_{i},\nu), \label{YYb}
\end{eqnarray}
where $f_{\Lambda_c}$ is a normalization constant, $\Lambda_c(p)$
is the $\Lambda_c$ spinor, and $\Psi(k_i, \nu)$ is the wave
function. $\nu$ is the hard sub-amplitude energy scale for the
process which is within the range of $\Lambda_{QCD}<\nu<m_b$ and
will be further discussed below. Similarly, the leading-twist wave
function of proton is defined by\cite{Lip,protonwave}:
\begin{eqnarray}
&&(Y_{P})_{\alpha \beta \gamma}(k_{i}',\nu) =
\frac{1}{2\sqrt{2}N_{c}} \int \prod_{l=1}^{2}\frac{d w_{l}^{-} d
\mathbf{w_{l}} }{(2\pi)^{3}} e^{ik_{l}'w_{l}}
\varepsilon^{abc}\langle0|T[u_{\alpha}^{a}(w_{1})
u_{\beta}^{b}(w_{2})d_{\gamma}^{c}(0)]|
P(p')\rangle \nonumber \\
&&=\frac{f_P(\nu)}{8\sqrt{2}N_{c}}\{(p'\!\!\! \slash C)_{
\beta\gamma} [\gamma_{5}P(p')]_{\alpha}\Phi^{V}(k_{i}',\nu)
+(p'\!\!\! \slash \gamma_{5} C)_{\beta\gamma}[P(p')]_{\alpha}\Phi^{A}(k_{i}',\nu)\nonumber \\
&&-(\sigma_{\mu \nu}p'^{\nu}C)_{\beta\gamma}
[\gamma^{\mu}\gamma_{5}P(p')]_{\alpha}\Phi^{T}(k_{i}',\nu)\},
\label{YY}
\end{eqnarray}
where $f_P$ is the normalization constant, and $P(p')$ is the
proton spinor.
\begin{eqnarray}
&&\Phi^V(k_1',k_2',k_3',\nu)=\frac{1}{2}\left[\Phi(k_2',k_1',k_3',\nu)
+\Phi(k_1',k_2',k_3',\nu)\right]\;,
\nonumber \\
&&\Phi^A(k_1',k_2',k_3',\nu)=\frac{1}{2}\left[\Phi(k_2',k_1',k_3',\nu)
-\Phi(k_1',k_2',k_3',\nu)\right]\;,
\nonumber \\
&&\Phi^T(k_1',k_2',k_3',\nu)=\frac{1}{2}\left[\Phi(k_1',k_3',k_2',\nu)+
\Phi(k_2',k_3',k_1',\nu)\right]\;. \label{u2}
\end{eqnarray}
The $B$ meson wave function is expressed as\cite{B}
\begin{eqnarray}
(Y_B)_{\alpha\beta}(q_i,\nu) &=& \frac{1}{N_c} \int
\frac{d^4w}{(2\pi)^4} e^{iq_2\cdot w} \langle
0|T[\overline{d}_\alpha (w) b_\beta (0)] |\overline{B^0}(q)\rangle
\nonumber \\
&=& \frac{i}{\sqrt{2N_c}}\left[(\rlap /q+m_B)\gamma_5\left({\rlap
/n_+\over \sqrt{2}}\phi_B^+(q_2,\nu)+{\rlap /n_-\over
\sqrt{2}}\phi_B^-(q_2,\nu)\right)\right]_{\beta\alpha}.
\end{eqnarray}
where $\rlap /n_+$ term corresponds to the first choice of momenta
of the valance quark $\overline{d}$ in $\overline{B^0}$ and $\rlap
/n_-$ term corresponds to the second one with the relation:
\begin{eqnarray}
{\rlap /n_{\pm}\over \sqrt{2}} &=& {\rlap /q_2^{1,2}\over
(1-y)m_B}.
\end{eqnarray}
When convoluting the hadron wave functions with the hard
amplitude, one should use corresponding light cone momenta
$q^{1,2}_i$ consistently for the above two terms.

There is also a simple relation between wave functions $\phi_B^+,
\phi_B^-$ and $\phi_B, \overline{\phi}_B$ used in the usual PQCD
calculation:
\begin{eqnarray}
\phi_B^+ &=& \phi_B+\overline{\phi}_B, \;\;\;\;\; \phi_B^- =
\phi_B-\overline{\phi}_B
\end{eqnarray}
It has been argued that the distribution amplitude
$\overline{\phi}_B$ is negligible compared to $\phi_B$\cite{B},
therefore $\phi_B^+=\phi_B^-\simeq\phi_B$. The specific form of
$\phi_B$ will be given later when carrying our numerical
calculations.

Including the Sudakov factor with infrared cut-offs
$\omega(\omega', \omega_q)$, and running the wave function from
$\nu$ down to $\omega(\omega', \omega_q)$, the wave functions are
obtained as \cite{Lip}:
\begin{eqnarray}
&&\Psi(x_{i},b_{i},p,\nu)=\mathrm{exp}\left[-\sum_{l=2}^{3}s(\omega,x_lp^+)-3\int_{\omega}^{\nu}
\frac{d
\overline{\mu}}{\overline{\mu}}\gamma_q(\alpha_{s}(\overline{\mu}))
\right]\Psi(x_{i})\nonumber \\
&&\Phi(x_{i}',b_{i}',p',\nu)=\mathrm{exp}\left[-\sum_{l=1}^{3}s(\omega',x_{l}'p^{-})-3\int_{\omega'}^{\nu}
\frac{d\overline{\mu}}{\overline{\mu}}\gamma_q(\alpha_{s}(\overline{\mu}))\right]
\Phi(x_{i}',\omega'),\nonumber \\
&&\Phi_B(y,b_q,q,\nu)= \exp\left[ -s(\omega_q,q_2^+)
-2\int_{\omega_q}^\nu \frac{d\bar{\mu}}{\bar{\mu}} \gamma
(\alpha_s(\bar{\mu}))\right]\Phi_B(1-y) \label{suda}
\end{eqnarray}
where $\omega = min(1/\tilde b_1,1/\tilde b_2,1/\tilde b_3)$,
$\omega' = min(1/\tilde b'_1,1/\tilde b'_2,1/\tilde b'_3)$ and
$\omega_q = 1/b_q$. $\tilde b^{(')}_1
=|\textbf{b}^{(')}_2-\textbf{b}^{(')}_3|$, $\tilde b^{(')}_2
=|\textbf{b}^{(')}_1-\textbf{b}^{(')}_3|$, $\tilde b^{(')}_3
=|\textbf{b}^{(')}_1-\textbf{b}^{(')}_2|$ and
$b_q=|\textbf{b}_q|$. Here $\textbf{b}$, $\textbf{b}'$ and
$\textbf{b}_q$ are the conjugate variables to $\textbf{k}_T$,
$\textbf{k}'_{T}$ and $\textbf{q}_T$ whose definitions are
collected in Appendix B.

The explicit expressions for the Sudakov factors are given in
\cite{Lip} with
\begin{eqnarray}
&&s(\omega,Q)=\int^Q_\omega{dp\over p}\left[ln({Q\over
p})A[\alpha_s(p)]+B[\alpha_s(p)]\right]\nonumber \\
&&A=C_F{\alpha_s\over \pi}+\left[{67\over 9}-{\pi^2\over
3}-{10\over 27}n_f+{8\over 3}\beta_0ln({e^{\gamma_E}\over
2})\right]({\alpha_s\over
\pi})^2\nonumber \\
&&B={2\over 3}{\alpha_s\over \pi}ln\left({e^{2\gamma_E-1}\over
2}\right)\nonumber \\
&&\gamma_q (\alpha_s(\mu)) = -\alpha_s(\mu)/\pi,\nonumber\\
&&\beta_0={33-2n_f\over 12},
\end{eqnarray}
where $\gamma_E$ is the Euler constant, $n_f$  is the flavor
number and $\gamma_q$ is the anomalous dimension. For $B$ decays,
the typical energy scale is above the charm mass, so we will take
$n_f$  to be 4 in our numerical computations.

Within the PQCD framework, the hadronic matrix element can be
written as:
\begin{eqnarray}
&&M= \int[Dx]\int[Db](\overline{Y}_{\Lambda_c})_{\alpha'
\beta'\gamma'}(x_i',b_i',p',\nu)\nonumber \\
&&H^{\alpha'\beta'\gamma'\rho'\alpha\beta\gamma\rho}(x_i,x_i',y,b_i,b_i',b_q,m_B,m_{\Lambda_c},\nu)
(Y_{\overline{p}})_{\alpha\beta\gamma}(x_i,b_i,p,\nu)(Y_{B})_{\rho\rho'}(y,b_q,q,\nu),\nonumber
\\ \label{integral}
\end{eqnarray}
where
\begin{eqnarray}
&&[Dx] = [dx][dx']dy,\;\;[dx] = dx_1 dx_2 dx_3
\delta(1-\sum_{l=1}^{3}x_l),\;\;[dx'] =
dx_{1}'dx_{2}'dx_3'\delta(1-\sum_{l=1}^{3}x_l').\nonumber \\
\end{eqnarray}
The measures of the transverse parts $[Db]$ are defined in
Appendix A.

To obtain the hard scattering amplitude
$H^{\alpha'\beta'\gamma'\rho'\alpha \beta
\gamma\rho}(x,x',y,b,b',b_q,m_B,m_{\Lambda_c},\nu)$,  one can
derive  the amplitude
$H^{i,\alpha'\beta'\gamma'\rho'\alpha\beta\gamma\rho}(x_i,x_i',y,\textbf{k}_{T},\textbf{k}_{T}'
,\textbf{q}_{T},m_B,m_{\Lambda_c})$ corresponding to the `i'th
diagram in Fig. \ref{diagrams1} and its analytic expression  is
displayed in Appendix B.  Then a Fourier transformation on
$\textbf{k}_T$ and $\textbf{k}'_T$ is carried out to convert them
into the ${\bf b}$ and ${\bf b'}$ space to obtain $\tilde
H^{i,\alpha'\beta'\gamma'\rho'\alpha \beta
\gamma\rho}(x,x',y,b,b',b_q,m_B,m_{\Lambda_c})$. The concrete
procedure for carrying out such transformation is described at the
end of Appendix B.

\begin{figure}[!htb]
\begin{center}
\begin{tabular}{cc}
\includegraphics[width=15cm]{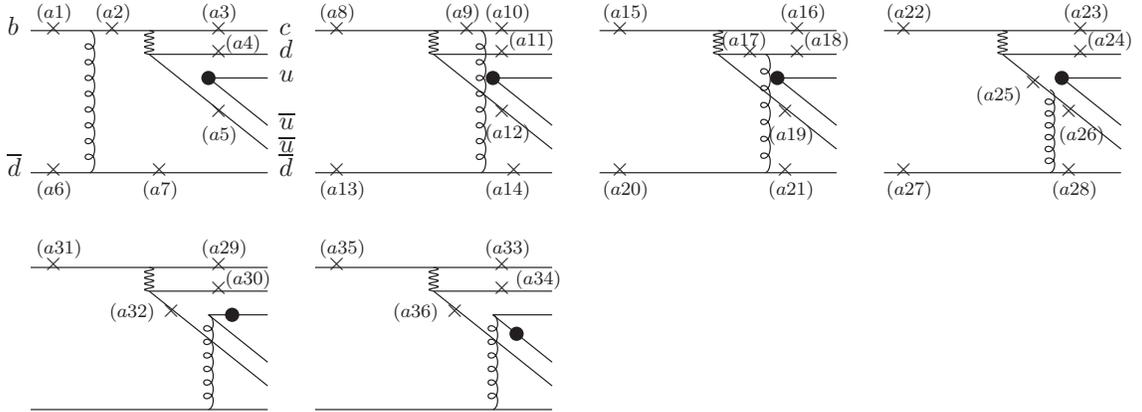}
\end{tabular}
\end{center}
\label{figure} \caption{The lowest order diagrams for the
$\overline{B}^0\to \Lambda_c^+ \overline{p}$ decay according to
the first choice of momenta ($q^1_i$). The wavy line indicates
W-exchange. The solid lines, curly lines denote the quarks and
gluons, respectively. In the diagrams, only one gluon line is
shown. The other gluon line connects the solid black blob and the
cross indicated by $a_i$. There are 36 diagrams. We label each one
by $a_i$.} \label{diagrams1}
\end{figure}

\begin{figure}[!htb]
\begin{center}
\begin{tabular}{cc}
\includegraphics[width=15cm]{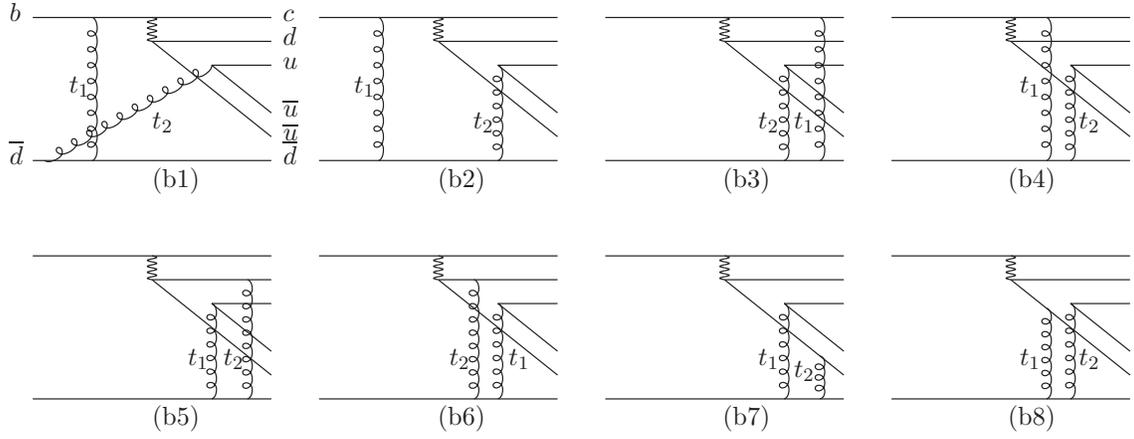}
\end{tabular}
\end{center}
\label{figure} \caption{The lowest order diagrams for the
$\overline{B}^0\to \Lambda_c^+ \overline{p}$ decay according to
the second choice of momenta ($q^2_i$). The wavy line indicates
W-exchange. The solid lines, curly lines denote the quarks and
gluons, respectively. There are 8 diagrams. We label each one by
$b_i$.} \label{diagrams2}
\end{figure}

\begin{figure}[!htb]
\begin{center}
\begin{tabular}{cc}
\includegraphics[width=15cm]{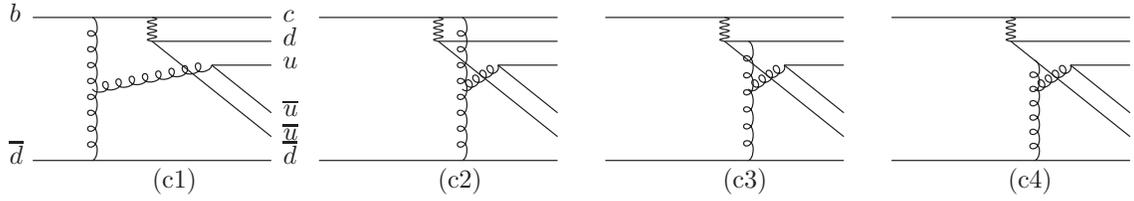}
\end{tabular}
\end{center}
\label{figure} \caption{The lowest order diagrams for the
$\overline{B}^0\to \Lambda_c^+ \overline{p}$ decay including
three-gluon vertices. The wavy line indicates W-exchange. The
solid lines, curly lines denote the quarks and gluons,
respectively. There are 4 diagrams. We label each one by $c_i$.}
\label{diagrams3}
\end{figure}

\begin{figure}[!htb]
\begin{center}
\begin{tabular}{cc}
\includegraphics[width=15cm]{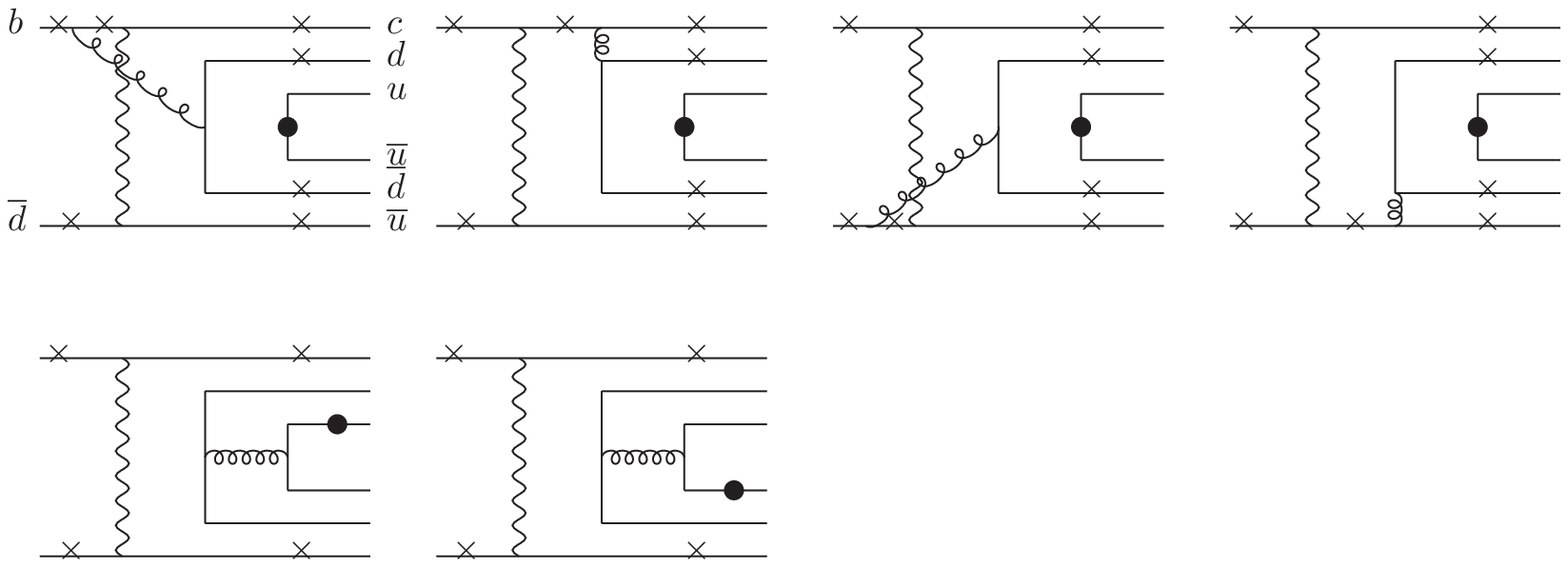}\\
\includegraphics[width=15cm]{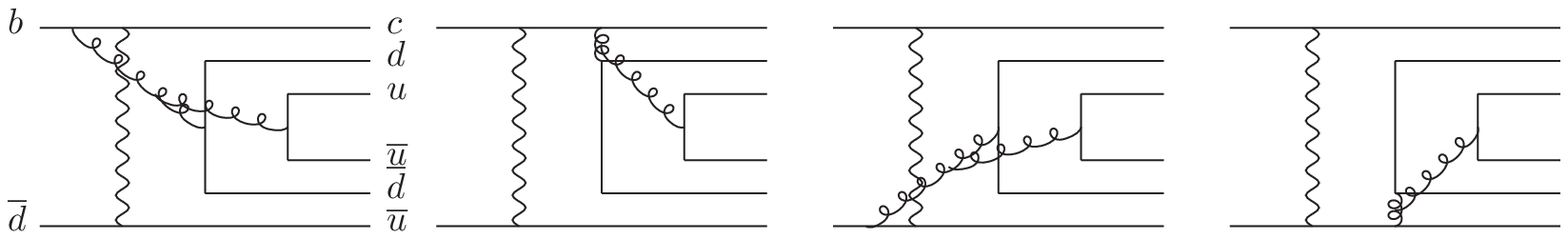}
\end{tabular}
\end{center}
\label{figure} \caption{The W-exchange diagrams for the
$\overline{B}^0\to \Lambda_c^+ \overline{P}$ decay.}
\label{diagrams4}
\end{figure}

We can catalog the diagrams into several types. The first type
includes 36 diagrams, indicated by $a_i$ in Fig. \ref{diagrams1}.
If we adopt the first choice of momenta ($q^1_i$) of valance quark
in $\overline{B^0}$ the total 36 diagrams contribute to the
amplitude in the PQCD frame, where the gluons can be ``hard''
(have large $k^2$). However, one can notice that, if we adopt the
second choice ($q^2_i$), in some of the diagrams the gluon
attached to the spectator quark is always soft, i.e. $k^2\approx
0$. The contributions of such diagrams cannot be counted as PQCD
contributions and are attributed into the wave functions of the
initial or final hadrons. When calculating the PQCD contributions,
these diagrams should not be included at all. In that case, there
are only 8 diagrams, indicated by $b_i$ in Fig. \ref{diagrams2}.
remain where all the exchanged gluons are ``hard''. For another
type of diagrams, 4 of them, indicated by $c_i$ in Fig.
\ref{diagrams3}, three-gluon vertices are involved. We find that
the color factor for two, $c_{3,4}$, of them is zero. There are
also the W-exchange diagrams (where W is exchanged between the
initial b and $\bar d$), 36 of them with gluons connected to quark
lines only and 4 of them with three-gluon vertices shown in Fig.
\ref{diagrams4}. These diagrams are suppressed by the
linear-momentum match and usually ignored\cite{Cheng}. Therefore,
there are 36+8+2=46 diagrams shown in Fig. \ref{diagrams1},
\ref{diagrams2} and \ref{diagrams3}, need to be evaluated to the
leading order.

Multiplying all contributions of the diagrams in
Fig.\ref{diagrams1}, \ref{diagrams2} and \ref{diagrams3} with
corresponding Wilson coefficients, one adds them up to obtain the
full hard scattering amplitude
\begin{eqnarray}
H^{\alpha'\beta'\gamma'\rho'\alpha \beta
\gamma\rho}(x,x',y,b,b',b_q,m_B,m_{\Lambda_c}) = \sum_i C(t)
\tilde H^{i,\alpha'\beta'\gamma'\rho'\alpha \beta
\gamma\rho}(x,x',y,b,b',b_q,m_B,m_{\Lambda_c}).\nonumber
\end{eqnarray}
Here we denote the hard scale as $t$ which is taken to be the
larger of the two variables $t_{1,2}$ associated with the virtual
gluon momentum in Fig.\ref{diagrams1}, i.e.
$t=\textrm{max}(t_{1}^{i},t_{2}^{i})$. The expressions for
$t_{1,2}$ are listed in Appendix C.

Finally a RG running is applied to the hard scattering amplitude
from the scale $\nu$ in the wavefunctions to $t$ and we obtain
\begin{eqnarray}
&&H^{\alpha'\beta'\gamma'\rho'\alpha \beta
\gamma\rho}(x,x',y,b,b',b_q,m_B,m_{\Lambda_c},\nu) =\nonumber \\
&&\textrm{exp}[-8\int_{\nu}^{t}\frac{d\overline{\mu}}
{\overline{\mu}}\gamma_q(\alpha_{s}(\overline{\mu}))]\times
H^{\alpha'\beta'\gamma'\rho'\alpha \beta
\gamma\rho}(x,x',y,b,b',b_q,m_B,m_{\Lambda_b}).
\end{eqnarray}

The form factors are obtained by properly grouping relevant terms
according to the definition in eq.(\ref{formfactor}). Using
eq.(\ref{integral}) we obtain a generic expression for the form
factor corresponding to each diagram as
\begin{eqnarray}
A^i(B^i)&=&\sum_{j=V,A,T}\frac{1}{128N_c^2\sqrt{2N_c}} f_P
f_{_{\Lambda_c}}
\int[Dx]\int[Db]^{i}C^i(t^{i})\nonumber \\
&&\Psi_{\Lambda_c}(x)\Phi_{P}^{j}(x')\phi_B(1-y)\textrm{exp}
[-S^{i}]H_{F}^{ij} \Omega^{i},
\end{eqnarray}
where\begin{eqnarray}
S^i&=&\sum^3_{k=2}s(\omega,x_kp^+)+\sum^3_{k=1}s(\omega',x_k'p'^-)+s(\omega_q,(1-y)q^+)
\nonumber \\
&&+3\int_{\omega}^{t^i}\frac{d\overline{\mu}}
{\overline{\mu}}\gamma_q(\alpha_{s}(\overline{\mu}))+
3\int_{\omega'}^{t^i}\frac{d\overline{\mu}}
{\overline{\mu}}\gamma_q(\alpha_{s}(\overline{\mu}))+
2\int_{\omega_q}^{t^i}\frac{d\overline{\mu}}
{\overline{\mu}}\gamma_q(\alpha_{s}(\overline{\mu}))\nonumber \\
\label{FF}
\end{eqnarray}
and $A^i$ and $B^i$ represent the form factors contributed by the
i-th diagram. $C^i$ is an appropriate combination of the Wilson
coefficients. The superscript $j$ labels different Lorentz
structures $V, A$, and $T$ of anti-proton according to
eq.(\ref{YY}). The explicit expressions of $\Omega^{i}$ are
presented in Appendix D and the functions $H_{F}^{ij}$ are given
in Appendix E. The total form factors are obtained by summing over
the contributions of all the diagrams.

\section{Numerical results}

We are now ready to evaluate the form factors numerically. In our
calculations, we adopt the distribution amplitude $\Psi$ proposed
in Ref.\cite{Lic,lambdacwave} for the $\Lambda_c$ ,
\begin{eqnarray}
&&\Psi(x_1,x_2,x_3) = Nx_1 x_2 x_3
\mathrm{exp}[-{m^2_{\Lambda_c}\over 2\beta^2x_1}-{m^2_q\over
2\beta^2x_2}-{m^2_q\over 2\beta^2x_3}]. \label{WPhi}
\end{eqnarray}
The normalization constant $N$ is fixed by the  condition:
\begin{eqnarray}
&&\int [dx]\Psi(x_1,x_2,x_3) = 1.
\end{eqnarray}
Since the decay constant $f_{\Lambda_c}$ is not experimentally
determined so far, one has to invoke a theoretical evaluation to
fix it. It is assumed that there is a relation between
$f_{\Lambda_c}$ and $f_{\Lambda_c}$ as
$f_{\Lambda_c}=f_{\Lambda_b}{m_{\Lambda_b}\over
m_{\Lambda_c}}$\cite{Lic}. In terms of the PQCD method
$f_{\Lambda_b}$ is determined by fitting $B(\Lambda_b \to
\Lambda_c l \overline{\nu})$ whose central value is 5\% measured
by the DELPHI collaboration\cite{expLic}. In our previous study of
$\Lambda_b\to \Lambda \gamma$\cite{LLr}, we have carried out a new
determination for $f_{\Lambda_b}$. When fitting the data we first
truncate the double logarithmic Sudakov factor in such a way that
we require the factor $\exp(-s)$ to be smaller than 1 following
the prescription of Ref.\cite{suda1}.  It is noted that our
numerical value of $f_{\Lambda_b}$ is different from that obtained
in Ref.\cite{Lic}, where $B(\Lambda_b \to \Lambda_c l
\overline{\nu})$ was taken to be 2\%, whereas we use the newly
measured value instead. In concrete computations, we choose the
cut-offs as $\omega=1.14min(1/\tilde b_1,1/\tilde b_2,1/\tilde
b_3)$ and $\omega'=1.14min(1/\tilde b'_1,1/\tilde b'_2,1/\tilde
b'_3)$. The phenomenological factor 1.14 is adopted according to
Ref.\cite{Lip2} to make the whole picture more realistic.
Moreover, the values of $\beta$ and $m_q$ in the heavy baryon wave
function need to be fixed. In Ref.\cite{Lic,Lip,LiJ}, $\beta=1$
GeV and $m_q=0.3$ GeV were used to estimate the decay rates of
$\Lambda_b \to \Lambda_c l \bar{\nu}$, $\Lambda_b \to p l
\bar{\nu}$ and $\Lambda_b \to \Lambda J/\psi$. It is commonly
expected that $\beta$ should not be too much smaller than 1 GeV if
the form factors are dominated by the perturbative contributions.
To see how different values of $\beta$ and $m_q$ affect the
results, we will let both $\beta$ and $m_q$ vary within ranges of
$0.6\sim 1.2$ GeV and $0.2\sim 0.3$ GeV respectively. Using the
fitted value of $f_{\Lambda_b}$ obtained in Ref.\cite{LLr}, and
the relation: $f_{\Lambda_c}=f_{\Lambda_b}{m_{\Lambda_b}\over
m_{\Lambda_c}}$, we listed $f_{\Lambda_c}$ in Table \ref{flambdab}
for different parameter sets.

\begin{table}[h]
\caption{Decay constant $f_{\Lambda_c}$(GeV) for different choices
of $\beta$ and $m_q$ with Sudkov truncation described in the
text.}
\begin{center}
\begin{tabular}{|c|c|c|c|c|}
  \hline
  $f_{\Lambda_c}$ & $\beta=0.6$GeV & $\beta=0.8$GeV & $\beta=1$GeV & $\beta=1.2$GeV\\
  \hline
  $m_q=0.2$GeV & $1.70\times 10^{-3}$ & $2.50\times 10^{-3}$
  & $3.51\times 10^{-3}$ & $4.71\times 10^{-3}$\\
  \hline
  $m_q=0.3$GeV & $3.12\times 10^{-3}$ & $4.07\times 10^{-3}$
  & $5.22\times 10^{-3}$ & $6.50\times 10^{-3}$\\
  \hline
\end{tabular}
\end{center}\label{flambdab}
\end{table}

The proton distribution amplitudes have been studied using the KS
model. In this work, we adopt the model proposed in
Ref.\cite{protonwave},
\begin{eqnarray}
&&\Phi(x_i,\omega) = \phi_{as}(x_1,x_2,x_3)\sum_{j=0}^5
N_j\left[{\alpha_s(\omega)\over
\alpha_s(\mu_0)}\right]^{b_j/(4\beta_0)}a_jA_j(x_i),\nonumber
\\
&&\phi_{as}(x_1,x_2,x_3) = 120x_1x_2x_3. \label{Wphi}
\end{eqnarray}
with $\mu_0\approx 1$ GeV, the constants $N_j$, $a_j$, $b_j$ and
the Appel polynomials $A_j$ are listed in Table
\ref{proton}\cite{Lip}.

\begin{table}[h]
\caption{The constants and polynomials in eq.(\ref{Wphi}).}
\begin{center}
\begin{tabular}{|c|c|c|c|c|}
  \hline
  $j$ & $a_j$ & $N_j$ & $b_j$ & $A_j(x_i)$\\
  \hline
  0 & 1.00 & 1 & 0 & 1\\
  \hline
  1 & 0.310 & 21/2 & 20/9 & $x_1-x_3$\\
  \hline
  2 & -0.370 & 7/2 & 24/9 & $2-3(x_1+x_3)$\\
  \hline
  3 & 0.630 & 63/10 & 32/9 & $2-7(x_1+x_3)+8(x_1^2+x_3^2)+4x_1x_3$\\
  \hline
  4 & 0.00333 & 567/2 & 40/9 & $x_1-x_3-{4\over 3}(x_1^2-x_3^2)$\\
  \hline
  5 & 0.0632 & 81/5 & 42/9 & $2-7(x_1+x_3)+{14\over 3}(x_1^2+x_3^2)+14x_1x_3$\\
  \hline
\end{tabular}
\end{center}\label{proton}
\end{table}

The constant $f_{P}$ is fixed to be\cite{protonwave}
\begin{eqnarray}
&&f_P(\omega)=f_P(\mu_0)\left[{\alpha_s(\omega)\over
\alpha_s(\mu_0)}\right]^{1/(6\beta_0)},
\end{eqnarray}
with $f_P(\mu_0)=(5.2\pm 0.3)\times 10^{-3}$ GeV$^2$.

The B meson distribution amplitude is given as \cite{B}:
\begin{eqnarray}
&&\phi_B(x,b)=N_Bx^2(1-x)^2exp\left[-{m_B^2x^2\over
2\omega_b^2}-{1\over 2}(\omega_bb)^2\right]
\end{eqnarray}
where $\omega_b=0.4$ GeV and $N_B=91.74$ GeV for $f_B=0.19$ GeV.

Finally, when obtaining the branching ratio of $\overline{B}^0\to
\Lambda_c^+ \overline{p}$, for definitiveness we fix rest of the
parameters as following. The parameter $\Lambda_{QCD}$ which
enters in the strong coupling constant and various Wilson
coefficients is set to be $\Lambda_{QCD} = 0.2$ GeV. For the CKM
mixing parameters, we take their central values\cite{data0}:
$s_{12} = 0.2243$, $s_{23} = 0.0413$, $s_{13} = 0.0037$ and
$\delta_{13} = 1.05$. We take two typical values for $m_q$ as 0.2
and 0.3 GeV and let $\beta$ vary within a reasonable range to
carry out the numerical computations.

The theoretical values of the branching ratio for different
$\beta$ and $m_q$ are listed in Table \ref{BR}. Our numerical
results indicate that BR$(\overline{B}^0\to \Lambda_c^+
\overline{p})=(2.3 \sim 5.1)\times 10^{-5}$. We see that the
results are consistent with experimental data. With the values
$\beta = 1$ GeV and $m_q = 0.3$ GeV used in
Ref.\cite{Lic,Lip,LiJ}, BR$(\overline{B}^0\to \Lambda_c^+
\overline{p})=2.3\times 10^{-5}$.

\begin{table}[h]
\caption{Branching ratio($\times 10^{-5}$) for different choices
of $\beta$ and $m_q$ with Sudakov truncation described in the
text.}
\begin{center}
\begin{tabular}{|c|c|c|c|c|}
  \hline
  BR & $\beta=0.6$GeV & $\beta=0.8$GeV & $\beta=1$GeV & $\beta=1.2$GeV\\
  \hline
  $m_q=0.2$GeV & $4.6$ & $2.6$
  & $3.6$ & $4.2$\\
  \hline
  $m_q=0.3$GeV & $3.2$ & $5.1$
  & $2.3$ & $3.1$\\
  \hline
\end{tabular}
\end{center}\label{BR}
\end{table}

\section{Discussions and conclusions}

In this work, we apply the PQCD method to study the branching
ratio of $\overline{B}^0\rightarrow \Lambda_c^+ \overline{p}$. The
new measurements on the branching ratio by CLEO and Belle indicate
that the previous theoretically estimated branching ratio in
various phenomenological models are obviously larger than the
data, therefore a better treatment is needed. We take the PQCD as
an approach to try since it has been successful in dealing with
mesonic B decays. As we discussed in the introduction, the PQCD
approach should be well applicable in this case, because the
momentum match demands the gluons exchanged between quarks to be
''hard'' and the perturbative contributions dominate. Indeed, our
numerical result is satisfactorily consistent with the measured
value by the Belle collaboration.

There are model-related parameters in various wavefunctions which
may influence the numerical results. We obtain them by fitting
data, thus the theoretical uncertainties would be much reduced. In
fact, except the part involving the wavefunctions of the hadrons,
all the calculations are done in the framework of PQCD. As long as
the perturbative approach is applicable,  the derivations follow
the general principles of quantum field theory. Therefore the
model-dependence is further reduced to minimum.

There is another source of possible uncertainty coming from the
prescription of demanding the Sudakov factor $\exp(-s)$ to be less
than one as proposed in Ref.\cite{suda1}. This truncation seems to
be just a convenient choice to suppress the amplitude. We
therefore have carried out a calculation without the truncation
for a comparison. Indeed, we find in certain kinematic regions,
the Sudakov factor is larger than one. Therefore without Sudakov
truncation the resulting branching ratio is larger, sometimes by a
factor of two. With more accurate data, one can pin down the
detailed differences. Since our results show that the predicted
branching ratio with Sudakov truncation is close to experimental
range, the prescription outlined in Ref.\cite{suda1} is a
reasonable way for calculations.

In conclusion, we have calculated the branching ratio of
$\overline{B}^0 \to \Lambda_c^+ \overline{p}$ in the PQCD
approach. We find that the predicted branching ratio for
BR$(\overline{B}^0 \to \Lambda_c^+ \overline{p})$ in the PQCD
approach can vary over a range of $(2.3\sim 5.1)\times 10^{-5}$
with the largest uncertainty coming from the parameters in the
wave function of $\Lambda_c$. With the favored values  for the
parameters in the $\Lambda_c^+$ wave function,  $\beta = 1$ GeV
and $m_q = 0.3$ GeV, the branching ratio is about $2.3\times
10^{-5}$ which is satisfactorily consistent with the value
measured by experiments.

\noindent {\bf Acknowledgments}:

We thank Prof. Cheng who suggested us to investigate this
interesting topic for fruitful discussions. This work is
supported in part by the NNSF, NSC and NCTS. \\

\noindent{\bf Appendix A: the $b$ measures}\\
The ordinary b measure is defined as
\begin{eqnarray}
&&[d\mathbf{b}]={d^2\mathbf{b}\over (2\pi)^2}.
\end{eqnarray}
The explicit forms of $[D\mathbf{b}]^i$ for $i-$th diagram of the
first type in Fig. \ref{diagrams1} are given by
\begin{eqnarray}
&&[D\mathbf{b}]^{(a1)}=[D\mathbf{b}]^{(a2)}=
[d\mathbf{b}_2][d\mathbf{b}_3'] [d\mathbf{b}_{q}],\nonumber\\
&&[D\mathbf{b}]^{(a3)}=[D\mathbf{b}]^{(a15)}=[D\mathbf{b}]^{(a23)}=
[d\mathbf{b}_2][d\mathbf{b}_3][d\mathbf{b}_3'][d\mathbf{b}_q],\;\;\nonumber\\
&&[D\mathbf{b}]^{(a4)}=[D\mathbf{b}]^{(a8)}=[D\mathbf{b}]^{(a12)}=[D\mathbf{b}]^{(a24)}=
[d\mathbf{b}_1][d\mathbf{b}_2]
[d\mathbf{b}_3'][d\mathbf{b}_{q}],\nonumber \\
&&[D\mathbf{b}]^{(a5)}=[D\mathbf{b}]^{(a22)}=[D\mathbf{b}]^{(a36)}=
[d\mathbf{b}_2][d\mathbf{b}_2'][d\mathbf{b}_3'][d\mathbf{b}_q],\nonumber\\
&&[D\mathbf{b}]^{(a6)}=[D\mathbf{b}]^{(a7)}=
[d\mathbf{b}_2][d\mathbf{b}_1'][d\mathbf{b}_q],\nonumber\\
&&[D\mathbf{b}]^{(a9)}=
[d\mathbf{b}_1][d\mathbf{b}_3'][d\mathbf{b}_q],\nonumber\\
&&[D\mathbf{b}]^{(a10)}=[D\mathbf{b}]^{(a17)}=
[d\mathbf{b}_2][d\mathbf{b}_3][d\mathbf{b}_q],\nonumber\\
&&[D\mathbf{b}]^{(a11)}=
[d\mathbf{b}_1][d\mathbf{b}_2][d\mathbf{b}_2'][d\mathbf{b}_q],\;
\nonumber\\
&&[D\mathbf{b}]^{(a13)}=[D\mathbf{b}]^{(a14)}=[D\mathbf{b}]^{(a29)}=[D\mathbf{b}]^{(a33)}=
[d\mathbf{b}_2][d\mathbf{b}_3][d\mathbf{b}_1'][d\mathbf{b}_q],\;
\nonumber\\
&&[D\mathbf{b}]^{(a16)}=[D\mathbf{b}]^{(a19)}=
[d\mathbf{b}_2][d\mathbf{b}_3][d\mathbf{b}_2'][d\mathbf{b}_q],\nonumber\\
&&[D\mathbf{b}]^{(a18)}=
[d\mathbf{b}_1][d\mathbf{b}_2][d\mathbf{b}_q],\nonumber\\
&&[D\mathbf{b}]^{(a20)}=[D\mathbf{b}]^{(a30)}=[D\mathbf{b}]^{(a34)}=
[d\mathbf{b}_1][d\mathbf{b}_2][d\mathbf{b}_1'][d\mathbf{b}_q],\nonumber\\
&&[D\mathbf{b}]^{(a21)}=
[d\mathbf{b}_1][d\mathbf{b}_1'][d\mathbf{b}_2'][d\mathbf{b}_q],\nonumber\\
&&[D\mathbf{b}]^{(a25)}=[D\mathbf{b}]^{(a26)}=
[d\mathbf{b}_2'][d\mathbf{b}_3'][d\mathbf{b}_q],\nonumber\\
&&[D\mathbf{b}]^{(a27)}=[D\mathbf{b}]^{(a28)}=
[d\mathbf{b}_2][d\mathbf{b}_1'][d\mathbf{b}_2'][d\mathbf{b}_q],\;\;
\nonumber\\
&&[D\mathbf{b}]^{(a31)}=[D\mathbf{b}]^{(a32)}=
[d\mathbf{b}_2][d\mathbf{b}_1'][d\mathbf{b}_3'][d\mathbf{b}_q],\nonumber\\
&&[D\mathbf{b}]^{(a35)}=
[d\mathbf{b}_2][d\mathbf{b}_1'][d\mathbf{b}_3'][d\mathbf{b}_q].
\end{eqnarray}

The explicit forms of $[D\mathbf{b}]^i$ for $i-$th diagram of the
second type in Fig. \ref{diagrams2} are given by
\begin{eqnarray}
&&[D\mathbf{b}]^{(b1)}= [D\mathbf{b}]^{(a6)},\;\;
[D\mathbf{b}]^{(b2)}= [D\mathbf{b}]^{(a7)},\;\;
[D\mathbf{b}]^{(b3)}= [D\mathbf{b}]^{(a13)},\;\;
[D\mathbf{b}]^{(b4)}= [D\mathbf{b}]^{(a14)},\;\;\nonumber\\
&&[D\mathbf{b}]^{(b5)}= [D\mathbf{b}]^{(a20)},\;
[D\mathbf{b}]^{(b6)}= [D\mathbf{b}]^{(a21)}, [D\mathbf{b}]^{(b7)}=
[D\mathbf{b}]^{(a27)},\;\; [D\mathbf{b}]^{(b8)}=
[D\mathbf{b}]^{(a28)}.
\end{eqnarray}

The explicit forms of $[D\mathbf{b}]^i$ for $i-$th diagram of the
third type in Fig. \ref{diagrams3} are given by
\begin{eqnarray}
&&[D\mathbf{b}]^{(c1)}=
[d\mathbf{b}_2][d\mathbf{b}_1'][d\mathbf{b}_2'][d\mathbf{b}_q],\;\;
[D\mathbf{b}]^{(c2)}=
[d\mathbf{b}_2][d\mathbf{b}_3][d\mathbf{b}_1'][d\mathbf{b}_q].
\end{eqnarray}

\noindent{\bf Appendix B: Hard scattering amplitudes
$H^{i,\alpha'\beta'\gamma'\rho'\alpha\beta\gamma\rho}(x_i,x_i',y,\textbf{k}_{T},\textbf{k}_{T}'
,\textbf{q}_{T},m_B,m_{\Lambda_c})$}

Expressions of amplitude
$H^{i,\alpha'\beta'\gamma'\rho'\alpha\beta\gamma\rho}(x_i,x_i',y,\textbf{k}_{T},\textbf{k}_{T}'
,\textbf{q}_{T},m_B,m_{\Lambda_c})$ for each diagram in Fig. 1.
There is only one Lorentz structure for the $\gamma$-matrix in the
effective Hamiltonian, $O_\mu= \gamma_\mu L$.

For the hard amplitude of Fig.1(a1):
\begin{eqnarray}
&&
H^{a1,\alpha'\beta'\gamma'\rho'\alpha\beta\gamma\rho}(x_i,x_i',y,\textbf{k}_{T},\textbf{k}_{T}'
,\textbf{q}_{T},m_B,m_{\Lambda_c})=\nonumber
\\
&&\left[\varepsilon^{abc}\varepsilon^{a'b'c'}\left(C_1(T^j)_{bb'}(T^iT^jT^i)_{ac'}\delta_{a'c}+C_2(T^j)_{bb'}(T^iT^jT^i)_{cc'}\delta_{aa'}\right)\right]g_s^4\nonumber
\\
&&{(\gamma_\lambda)_{\rho'\gamma'}(\gamma_\theta)_{\beta
\beta'}[O_\mu(\rlap /p-\rlap / k_2+\rlap /
k_1'+m_b)\gamma^\lambda(\rlap /q_1-\rlap /k_2-\rlap
/k_2')\gamma^\theta]_{\gamma\rho}(O^\mu)_{\alpha\alpha'}\over
(k_2+k_2')^2(q_2-k_3')^2[(q_1-k_2-k_2')^2-m_b^2][(p-k_2+k_1')^2-m_b^2]}\nonumber
\\
&&={C_N^{a1}
g_s^4(\gamma_\lambda)_{\rho'\gamma'}(\gamma_\theta)_{\beta
\beta'}[O_\mu(\rlap /p-\rlap / k_2+\rlap /
k_1'+m_b)\gamma^\lambda(\rlap /q_1-\rlap /k_2-\rlap
/k_2')\gamma^\theta]_{\gamma\rho}(O^\mu)_{\alpha\alpha'}\over
[A_{a1}+(\textbf{k}_{2T}+\textbf{k}_{2T}')^2][B_{a1}+(\textbf{q}_{T}+\textbf{k}_{3T}')^2][C_{a1}+(-\textbf{q}_{T}
+\textbf{k}_{2T}+\textbf{k}_{2T}')^2][D_{a1}+(\textbf{k}_{2T}+\textbf{k}_{2T}'+\textbf{k}_{3T}')^2]}\nonumber
\\
\end{eqnarray}
with
\begin{eqnarray}
&&A_{a1}=(r^2-1)x_2x_2'm_B^2,\;\;\;\;\;\;\;\;\;\;\;\;\;\;\;\;\;\;\;\;\;\;\;\;\;\;\;\;
B_{a1}=(r^2-1)x_3'(y-1)m_B^2,\nonumber \\
&&C_{a1}=m_b^2+((r^2-1)x_2'+1)(x_2-y)m_B^2,\;
D_{a1}=m_b^2-(r^2(x_1'-1)-x_1')(x_2-1)m_B^2\nonumber \\
\end{eqnarray}
and the color factor
\begin{eqnarray}
&&C_N^{a1}=\varepsilon^{abc}\varepsilon^{a'b'c'}\left(C_1(T^j)_{bb'}(T^iT^jT^i)_{ac'}\delta_{a'c}+C_2(T^j)_{bb'}(T^iT^jT^i)_{cc'}\delta_{aa'}\right)=-{2\over
3}C_1+{2\over 3}C_2.\nonumber \\
\end{eqnarray}

For the hard amplitude of Fig.1(a2):
\begin{eqnarray}
&&
H^{a2,\alpha'\beta'\gamma'\rho'\alpha\beta\gamma\rho}(x_i,x_i',y,\textbf{k}_{T},\textbf{k}_{T}'
,\textbf{q}_{T},m_B,m_{\Lambda_c})=\nonumber
\\
&&\left[\varepsilon^{abc}\varepsilon^{a'b'c'}\left(C_1(T^j)_{bb'}(T^jT^iT^i)_{ac'}\delta_{a'c}+C_2(T^j)_{bb'}(T^jT^iT^i)_{cc'}\delta_{aa'}\right)\right]g_s^4\nonumber
\\
&&{(\gamma_\lambda)_{\rho'\gamma'}(\gamma_\theta)_{\beta
\beta'}[O_\mu(\rlap /p-\rlap / k_2+\rlap /
k_1'+m_b)\gamma^\theta(\rlap /q_-\rlap
/k_3'+m_b)\gamma^\lambda]_{\gamma\rho}(O^\mu)_{\alpha\alpha'}\over
(k_2+k_2')^2(q_2-k_3')^2[(q-k_3')^2-m_b^2][(p-k_2+k_1')^2-m_b^2]}\nonumber
\\
&&={C_N^{a2}
g_s^4(\gamma_\lambda)_{\rho'\gamma'}(\gamma_\theta)_{\beta
\beta'}[O_\mu(\rlap /p-\rlap / k_2+\rlap /
k_1'+m_b)\gamma^\theta(\rlap /q_-\rlap
/k_3'+m_b)\gamma^\lambda]_{\gamma\rho}(O^\mu)_{\alpha\alpha'}\over
[A_{a2}+(\textbf{k}_{2T}+\textbf{k}_{2T}')^2][B_{a2}+(\textbf{q}_{T}+\textbf{k}_{3T}')^2][C_{a2}+\textbf{k}_{3T}'^2]
[D_{a2}+(\textbf{k}_{2T}+\textbf{k}_{2T}'+\textbf{k}_{3T}')^2]}\nonumber
\\
\end{eqnarray}
with
\begin{eqnarray}
&&A_{a2}=(r^2-1)x_2x_2'm_B^2,\;\;\;\;\;\;\;\;\;\;\;\;\;\;\;\;\;\;\;\;\;\;
B_{a2}=(r^2-1)x_3'(y-1)m_B^2,\nonumber \\
&&C_{a2}=m_b^2+(-x_3'r^2+x_3'-1)m_B^2,\;\;\;\;\;
D_{a2}=m_b^2-(r^2(x_1'-1)-x_1')(x_2-1)m_B^2\nonumber \\
\end{eqnarray}
and the color factor
\begin{eqnarray}
&&C_N^{a2}=\varepsilon^{abc}\varepsilon^{a'b'c'}\left(C_1(T^j)_{bb'}(T^jT^iT^i)_{ac'}\delta_{a'c}+C_2(T^j)_{bb'}(T^jT^iT^i)_{cc'}\delta_{aa'}\right)={16\over
3}C_1-{16\over 3}C_2.\nonumber \\
\end{eqnarray}

For the hard amplitude of Fig.1(a3):
\begin{eqnarray}
&&
H^{a3,\alpha'\beta'\gamma'\rho'\alpha\beta\gamma\rho}(x_i,x_i',y,\textbf{k}_{T},\textbf{k}_{T}'
,\textbf{q}_{T},m_B,m_{\Lambda_c})=\nonumber
\\
&&\left[\varepsilon^{abc}\varepsilon^{a'b'c'}\left(C_1(T^j)_{bb'}(T^jT^iT^i)_{ac'}\delta_{a'c}+C_2(T^j)_{bb'}(T^j)_{aa'}(T^iT^i)_{cc'}\right)\right]g_s^4\nonumber
\\
&&{(\gamma_\lambda)_{\rho'\gamma'}(\gamma_\theta)_{\beta
\beta'}[O_\mu(\rlap /q-\rlap /
k_3'+m_b)\gamma^\lambda]_{\gamma\rho}(\gamma^\theta(\rlap /p-\rlap
/k_3+\rlap /k_2'+m_c)O^\mu)_{\alpha\alpha'}\over
(k_2+k_2')^2(q_2-k_3')^2[(q-k_3')^2-m_b^2][(p-k_3+k_2')^2-m_c^2]}\nonumber
\\
&&={C_N^{a3}
g_s^4(\gamma_\lambda)_{\rho'\gamma'}(\gamma_\theta)_{\beta
\beta'}[O_\mu(\rlap /q-\rlap /
k_3'+m_b)\gamma^\lambda]_{\gamma\rho}(\gamma^\theta(\rlap /p-\rlap
/k_3+\rlap /k_2'+m_c)O^\mu)_{\alpha\alpha'}\over
[A_{a3}+(\textbf{k}_{2T}+\textbf{k}_{2T}')^2][B_{a3}+(\textbf{q}_{T}+\textbf{k}_{3T}')^2][C_{a3}+\textbf{k}_{3T}'^2]
[D_{a3}+(\textbf{k}_{3T}-\textbf{k}_{2T}')^2]}\nonumber
\\
\end{eqnarray}
with
\begin{eqnarray}
&&A_{a3}=(r^2-1)x_2x_2'm_B^2,\;\;\;\;\;\;\;\;\;\;\;\;\;\;\;\;\;\;
B_{a3}=(r^2-1)x_3'(y-1)m_B^2,\nonumber \\
&&C_{a3}=m_b^2+(-x_3'r^2+x_3'-1)m_B^2,\;
D_{a3}=m_c^2-(r^2(x_2'-1)-x_2')(x_3-1)m_B^2\nonumber \\
\end{eqnarray}
and the color factor
\begin{eqnarray}
&&C_N^{a3}=\varepsilon^{abc}\varepsilon^{a'b'c'}\left(C_1(T^j)_{bb'}(T^jT^iT^i)_{ac'}\delta_{a'c}+C_2(T^j)_{bb'}(T^j)_{aa'}(T^iT^i)_{cc'}\right)={16\over
3}C_1-{16\over 3}C_2.\nonumber \\
\end{eqnarray}

For the hard amplitude of Fig.1(a4):
\begin{eqnarray}
&&
H^{a4,\alpha'\beta'\gamma'\rho'\alpha\beta\gamma\rho}(x_i,x_i',y,\textbf{k}_{T},\textbf{k}_{T}'
,\textbf{q}_{T},m_B,m_{\Lambda_c})=\nonumber
\\
&&\left[\varepsilon^{abc}\varepsilon^{a'b'c'}\left(C_1(T^j)_{bb'}(T^iT^i)_{ac'}(T^j)_{ca'}+C_2(T^j)_{bb'}(T^jT^iT^i)_{cc'}\delta_{aa'}\right)\right]g_s^4\nonumber
\\
&&{(\gamma_\lambda)_{\rho'\gamma'}(\gamma_\theta)_{\beta
\beta'}[\gamma^\theta(\rlap /p-\rlap / k_1+\rlap /k_2')O_\mu(\rlap
/q-\rlap
/k_3'+m_b)\gamma^\lambda]_{\gamma\rho}(O^\mu)_{\alpha\alpha'}\over
(k_2+k_2')^2(q_2-k_3')^2(p-k_1+k_2')^2[(q-k_3')^2-m_b^2]}\nonumber
\\
&&={C_N^{a4}
g_s^4(\gamma_\lambda)_{\rho'\gamma'}(\gamma_\theta)_{\beta
\beta'}[\gamma^\theta(\rlap /p-\rlap / k_1+\rlap /k_2')O_\mu(\rlap
/q-\rlap
/k_3'+m_b)\gamma^\lambda]_{\gamma\rho}(O^\mu)_{\alpha\alpha'}\over
[A_{a4}+(\textbf{k}_{2T}+\textbf{k}_{2T}')^2][B_{a4}+(\textbf{q}_{T}+\textbf{k}_{3T}')^2][C_{a4}+(\textbf{k}_{1T}-\textbf{k}_{2T}')^2]
[D_{a4}+\textbf{k}_{3T}'^2] }\nonumber
\\
\end{eqnarray}
with
\begin{eqnarray}
&&A_{a4}=(r^2-1)x_2x_2'm_B^2,\;\;\;\;\;\;\;\;\;\;\;\;\;
B_{a4}=(r^2-1)x_3'(y-1)m_B^2,\nonumber \\
&&C_{a4}=-(r^2-1)(x_1-1)x_2'm_B^2,\;
D_{a4}=m_b^2+(-x_3'r^2+x_3'-1)m_B^2\nonumber \\
\end{eqnarray}
and the color factor
\begin{eqnarray}
&&C_N^{a4}=\varepsilon^{abc}\varepsilon^{a'b'c'}\left(C_1(T^j)_{bb'}(T^iT^i)_{ac'}(T^j)_{ca'}+C_2(T^j)_{bb'}(T^jT^iT^i)_{cc'}\delta_{aa'}\right)={16\over
3}C_1-{16\over 3}C_2.\nonumber \\
\end{eqnarray}

For the hard amplitude of Fig.1(a5):
\begin{eqnarray}
&&
H^{a5,\alpha'\beta'\gamma'\rho'\alpha\beta\gamma\rho}(x_i,x_i',y,\textbf{k}_{T},\textbf{k}_{T}'
,\textbf{q}_{T},m_B,m_{\Lambda_c})=\nonumber
\\
&&\left[\varepsilon^{abc}\varepsilon^{a'b'c'}\left(C_1(T^j)_{bb'}(T^iT^i)_{ac'}(T^j)_{ca'}+C_2(T^j)_{bb'}(T^j)_{aa'}(T^iT^i)_{cc'}\right)\right]g_s^4\nonumber
\\
&&{(\gamma_\lambda)_{\rho'\gamma'}(\gamma_\theta)_{\beta
\beta'}[O_\mu(\rlap /q-\rlap /
k_3'+m_b)\gamma^\lambda]_{\gamma\rho}(O^\mu(-\rlap /p'-\rlap
/k_2+\rlap /k_3')\gamma^\theta)_{\alpha\alpha'}\over
(k_2+k_2')^2(q_2-k_3')^2(-p'-k_2+k_3')^2[(q-k_3')^2-m_b^2]}\nonumber
\\
&&={C_N^{a5}
g_s^4(\gamma_\lambda)_{\rho'\gamma'}(\gamma_\theta)_{\beta
\beta'}[O_\mu(\rlap /q-\rlap /
k_3'+m_b)\gamma^\lambda]_{\gamma\rho}(O^\mu(-\rlap /p'-\rlap
/k_2+\rlap /k_3')\gamma^\theta)_{\alpha\alpha'}\over
[A_{a5}+(\textbf{k}_{2T}+\textbf{k}_{2T}')^2][B_{a5}+(\textbf{q}_{T}+\textbf{k}_{3T}')^2][C_{a5}+(\textbf{k}_{2T}-\textbf{k}_{3T}')^2]
[D_{a5}+\textbf{k}_{3T}'^2] }\nonumber
\\
\end{eqnarray}
with
\begin{eqnarray}
&&A_{a5}=(r^2-1)x_2x_2'm_B^2,\;\;\;\;\;\;\;\;\;\;\;\;\;
B_{a5}=(r^2-1)x_3'(y-1)m_B^2,\nonumber \\
&&C_{a5}=-(r^2-1)(x_3'-1)x_2m_B^2,\;
D_{a5}=m_b^2+(-x_3'r^2+x_3'-1)m_B^2\nonumber \\
\end{eqnarray}
and the color factor
\begin{eqnarray}
&&C_N^{a5}=\varepsilon^{abc}\varepsilon^{a'b'c'}\left(C_1(T^j)_{bb'}(T^iT^i)_{ac'}(T^j)_{ca'}+C_2(T^j)_{bb'}(T^j)_{aa'}(T^iT^i)_{cc'}\right)={16\over
3}C_1-{16\over 3}C_2.\nonumber \\
\end{eqnarray}

For the hard amplitude of Fig.1(a6):
\begin{eqnarray}
&&
H^{a6,\alpha'\beta'\gamma'\rho'\alpha\beta\gamma\rho}(x_i,x_i',y,\textbf{k}_{T},\textbf{k}_{T}'
,\textbf{q}_{T},m_B,m_{\Lambda_c})=\nonumber
\\
&&\left[\varepsilon^{abc}\varepsilon^{a'b'c'}\left(C_1(T^j)_{bb'}(T^iT^jT^i)_{ac'}\delta_{ca'}+C_2(T^j)_{bb'}(T^iT^jT^i)_{cc'}\delta_{aa'}\right)\right]g_s^4\nonumber
\\
&&{(\gamma_\theta(\rlap /k_2+\rlap /k_2'-\rlap
/q_2)\gamma_\lambda)_{\rho'\gamma'}(\gamma^\theta)_{\beta
\beta'}[O_\mu(\rlap /p-\rlap /k_2+\rlap /
k_1'+m_b)\gamma^\lambda]_{\gamma\rho}(O^\mu)_{\alpha\alpha'}\over
(k_2+k_2')^2(-p'+q_2-k_2+k_1')^2(k_2+k_2'-q_2)^2[(p-k_2+k_1')^2-m_b^2]}\nonumber
\\
&&={C_N^{a6} g_s^4(\gamma_\theta(\rlap /k_2+\rlap /k_2'-\rlap
/q_2)\gamma_\lambda)_{\rho'\gamma'}(\gamma^\theta)_{\beta
\beta'}[O_\mu(\rlap /p-\rlap /k_2+\rlap /
k_1'+m_b)\gamma^\lambda]_{\gamma\rho}(O^\mu)_{\alpha\alpha'}\over
[A_{a6}+(\textbf{k}_{2T}+\textbf{k}_{2T}')^2][B_{a6}+(\textbf{q}_{T}+\textbf{k}_{2T}-\textbf{k}_{1T}')^2][C_{a6}+(\textbf{k}_{2T}+\textbf{k}_{2T}'+\textbf{q}_{T})^2]
[D_{a6}+(\textbf{k}_{2T}-\textbf{k}_{1T}')^2] }\nonumber
\\
\end{eqnarray}
with
\begin{eqnarray}
&&A_{a6}=(r^2-1)x_2x_2'm_B^2,\;\;\;\;\;\;\;\;\;\;\;\;\;\;\;\;
B_{a6}=(1-r^2)(x_1'-1)(x_2+y-1)m_B^2,\nonumber \\
&&C_{a6}=(r^2-1)(x_2+y-1)x_2'm_B^2,\;
D_{a6}=m_b^2-(r^2(x_1'-1)-x_1')(x_2-1)m_B^2\nonumber \\
\end{eqnarray}
and the color factor
\begin{eqnarray}
&&C_N^{a6}=\varepsilon^{abc}\varepsilon^{a'b'c'}\left(C_1(T^j)_{bb'}(T^iT^jT^i)_{ac'}\delta_{ca'}+C_2(T^j)_{bb'}(T^iT^jT^i)_{cc'}\delta_{aa'}\right)=-{2\over
3}C_1+{2\over 3}C_2.\nonumber \\
\end{eqnarray}

For the hard amplitude of Fig.1(a7):
\begin{eqnarray}
&&
H^{a7,\alpha'\beta'\gamma'\rho'\alpha\beta\gamma\rho}(x_i,x_i',y,\textbf{k}_{T},\textbf{k}_{T}'
,\textbf{q}_{T},m_B,m_{\Lambda_c})=\nonumber
\\
&&\left[\varepsilon^{abc}\varepsilon^{a'b'c'}\left(C_1(T^j)_{bb'}(T^iT^iT^j)_{ac'}\delta_{ca'}+C_2(T^j)_{bb'}(T^iT^iT^j)_{cc'}\delta_{aa'}\right)\right]g_s^4\nonumber
\\
&&{(\gamma_\lambda(-\rlap /p'-\rlap /k_2+\rlap
/k_1')\gamma_\theta)_{\rho'\gamma'}(\gamma^\theta)_{\beta
\beta'}[O_\mu(\rlap /p-\rlap /k_2+\rlap /
k_1'+m_b)\gamma^\lambda]_{\gamma\rho}(O^\mu)_{\alpha\alpha'}\over
(k_2+k_2')^2(-p'+q_2-k_2+k_1')^2(-p'-k_2+k_1')^2[(p-k_2+k_1')^2-m_b^2]}\nonumber
\\
&&={C_N^{a7} g_s^4(\gamma_\lambda(-\rlap /p'-\rlap /k_2+\rlap
/k_1')\gamma_\theta)_{\rho'\gamma'}(\gamma^\theta)_{\beta
\beta'}[O_\mu(\rlap /p-\rlap /k_2+\rlap /
k_1'+m_b)\gamma^\lambda]_{\gamma\rho}(O^\mu)_{\alpha\alpha'}\over
[A_{a7}+(\textbf{k}_{2T}+\textbf{k}_{2T}')^2][B_{a7}+(\textbf{q}_{T}+\textbf{k}_{2T}-\textbf{k}_{1T}')^2][C_{a7}+(\textbf{k}_{2T}-\textbf{k}_{1T}')^2]
[D_{a7}+(\textbf{k}_{2T}-\textbf{k}_{1T}')^2] }\nonumber
\\
\end{eqnarray}
with
\begin{eqnarray}
&&A_{a7}=(r^2-1)x_2x_2'm_B^2,\;\;\;\;\;\;\;\;\;\;
B_{a7}=(1-r^2)(x_1'-1)(x_2+y-1)m_B^2,\nonumber \\
&&C_{a7}=(1-r^2)(x_1'-1)x_2m_B^2,\;
D_{a7}=m_b^2-(r^2(x_1'-1)-x_1')(x_2-1)m_B^2\nonumber \\
\end{eqnarray}
and the color factor
\begin{eqnarray}
&&C_N^{a7}=\varepsilon^{abc}\varepsilon^{a'b'c'}\left(C_1(T^j)_{bb'}(T^iT^iT^j)_{ac'}\delta_{ca'}+C_2(T^j)_{bb'}(T^iT^iT^j)_{cc'}\delta_{aa'}\right)={16\over
3}C_1-{16\over 3}C_2.\nonumber \\
\end{eqnarray}

For the hard amplitude of Fig.1(a8):
\begin{eqnarray}
&&
H^{a8,\alpha'\beta'\gamma'\rho'\alpha\beta\gamma\rho}(x_i,x_i',y,\textbf{k}_{T},\textbf{k}_{T}'
,\textbf{q}_{T},m_B,m_{\Lambda_c})=\nonumber
\\
&&\left[\varepsilon^{abc}\varepsilon^{a'b'c'}\left(C_1(T^j)_{bb'}(T^iT^jT^i)_{ac'}\delta_{ca'}+C_2(T^j)_{bb'}(T^jT^i)_{cc'}(T^i)_{aa'}\right)\right]g_s^4\nonumber
\\
&&{(\gamma_\lambda)_{\rho'\gamma'}(\gamma^\theta)_{\beta
\beta'}[O_\mu(\rlap /q_1-\rlap /k_2-\rlap /
k_2'+m_b)\gamma^\theta]_{\gamma\rho}(\gamma^\lambda(-\rlap
/q_2+\rlap /k_1+\rlap /k_3'+m_c)O^\mu)_{\alpha\alpha'}\over
(k_2+k_2')^2(q_2-k_3)^2[(q_1-k_2-k_2')^2-m_b^2][(-q_2+k_1+k_3')^2-m_c^2]}\nonumber
\\
&&={C_N^{a8}
g_s^4(\gamma_\lambda)_{\rho'\gamma'}(\gamma^\theta)_{\beta
\beta'}[O_\mu(\rlap /q_1-\rlap /k_2-\rlap /
k_2'+m_b)\gamma^\theta]_{\gamma\rho}(\gamma^\lambda(-\rlap
/q_2+\rlap /k_1+\rlap /k_3'+m_c)O^\mu)_{\alpha\alpha'}\over
[A_{a8}+(\textbf{k}_{2T}+\textbf{k}_{2T}')^2][B_{a8}+(\textbf{q}_{T}+\textbf{k}_{3T}')^2][C_{a8}+(\textbf{k}_{2T}+\textbf{k}_{2T}'-\textbf{q}_{T})^2]
[D_{a8}+(\textbf{k}_{1T}+\textbf{k}_{3T}'+\textbf{q}_{T})^2]
}\nonumber
\\
\end{eqnarray}
with
\begin{eqnarray}
&&A_{a8}=(r^2-1)x_2x_2'm_B^2,\;\;\;\;\;\;\;\;\;\;\;\;\;\;\;\;\;\;\;\;\;\;\;\;\;\;\;\;
B_{a8}=(r^2-1)x_3'(y-1)m_B^2,\nonumber \\
&&C_{a8}=m_b^2+((r^2-1)x_2'+1)(x_2-y)m_B^2,\;
D_{a8}=m_c^2+(r^2(x_3'-1)-x_3')(x_1+y-1)m_B^2\nonumber \\
\end{eqnarray}
and the color factor
\begin{eqnarray}
&&C_N^{a8}=\varepsilon^{abc}\varepsilon^{a'b'c'}\left(C_1(T^j)_{bb'}(T^iT^jT^i)_{ac'}\delta_{ca'}+C_2(T^j)_{bb'}(T^jT^i)_{cc'}(T^i)_{aa'}\right)=-{2\over
3}C_1+{8\over 3}C_2.\nonumber \\
\end{eqnarray}

For the hard amplitude of Fig.1(a9):
\begin{eqnarray}
&&
H^{a9,\alpha'\beta'\gamma'\rho'\alpha\beta\gamma\rho}(x_i,x_i',y,\textbf{k}_{T},\textbf{k}_{T}'
,\textbf{q}_{T},m_B,m_{\Lambda_c})=\nonumber
\\
&&\left[\varepsilon^{abc}\varepsilon^{a'b'c'}\left(C_1(T^j)_{bb'}(T^iT^jT^i)_{ac'}\delta_{ca'}+C_2(T^j)_{bb'}(T^i)_{cc'}(T^iT^j)_{aa'}\right)\right]g_s^4\nonumber
\\
&&{(\gamma_\lambda)_{\rho'\gamma'}(\gamma_\theta)_{\beta
\beta'}(O_\mu)_{\gamma \rho}[\gamma^\lambda(-\rlap /q_2+\rlap
/k_1+\rlap / k_3'+m_c)\gamma^\theta(\rlap /q_1-\rlap /k_3-\rlap
/k_1'+m_c)O^\mu)_{\alpha\alpha'}\over
(k_2+k_2')^2(q_2-k_3)^2[(q_1-k_3-k_1')^2-m_c^2][(-q_2+k_1+k_3')^2-m_c^2]}\nonumber
\\
&&={C_N^{a9}
g_s^4(\gamma_\lambda)_{\rho'\gamma'}(\gamma_\theta)_{\beta
\beta'}(O_\mu)_{\gamma \rho}[\gamma^\lambda(-\rlap /q_2+\rlap
/k_1+\rlap / k_3'+m_c)\gamma^\theta(\rlap /q_1-\rlap /k_3-\rlap
/k_1'+m_c)O^\mu)_{\alpha\alpha'}\over
[A_{a9}+(\textbf{k}_{2T}+\textbf{k}_{2T}')^2][B_{a9}+(\textbf{q}_{T}+\textbf{k}_{3T}')^2][C_{a9}+(\textbf{k}_{3T}+\textbf{k}_{1T}'-\textbf{q}_{T})^2]
[D_{a9}+(\textbf{k}_{1T}+\textbf{k}_{3T}'+\textbf{q}_{T})^2]
}\nonumber
\\
\end{eqnarray}
with
\begin{eqnarray}
&&A_{a9}=(r^2-1)x_2x_2'm_B^2,\;\;\;\;\;\;\;\;\;\;\;\;\;\;\;\;\;\;\;\;\;\;\;\;\;\;\;\;
B_{a9}=(r^2-1)x_3'(y-1)m_B^2,\nonumber \\
&&C_{a9}=m_c^2+((r^2-1)x_1'+1)(x_3-y)m_B^2,\;
D_{a9}=m_c^2+(r^2(x_3'-1)-x_3')(x_1+y-1)m_B^2\nonumber \\
\end{eqnarray}
and the color factor
\begin{eqnarray}
&&C_N^{a9}=\varepsilon^{abc}\varepsilon^{a'b'c'}\left(C_1(T^j)_{bb'}(T^iT^jT^i)_{ac'}\delta_{ca'}+C_2(T^j)_{bb'}(T^i)_{cc'}(T^iT^j)_{aa'}\right)=-{2\over
3}C_1+{8\over 3}C_2.\nonumber \\
\end{eqnarray}

For the hard amplitude of Fig.1(a10):
\begin{eqnarray}
&&
H^{a10,\alpha'\beta'\gamma'\rho'\alpha\beta\gamma\rho}(x_i,x_i',y,\textbf{k}_{T},\textbf{k}_{T}'
,\textbf{q}_{T},m_B,m_{\Lambda_c})=\nonumber
\\
&&\left[\varepsilon^{abc}\varepsilon^{a'b'c'}\left(C_1(T^j)_{bb'}(T^jT^iT^i)_{ac'}\delta_{ca'}+C_2(T^j)_{bb'}(T^i)_{cc'}(T^jT^i)_{aa'}\right)\right]g_s^4\nonumber
\\
&&{(\gamma_\lambda)_{\rho'\gamma'}(\gamma_\theta)_{\beta
\beta'}(O_\mu)_{\gamma \rho}[\gamma^\theta(\rlap /p-\rlap
/k_3+\rlap / k_2'+m_c)\gamma^\lambda(\rlap /q_1-\rlap /k_3-\rlap
/k_1'+m_c)O^\mu)_{\alpha\alpha'}\over
(k_2+k_2')^2(q_2-k_3)^2[(q_1-k_3-k_1')^2-m_c^2][(p-k_3+k_2')^2-m_c^2]}\nonumber
\\
&&={C_N^{a10}
g_s^4(\gamma_\lambda)_{\rho'\gamma'}(\gamma_\theta)_{\beta
\beta'}(O_\mu)_{\gamma \rho}[\gamma^\theta(\rlap /p-\rlap
/k_3+\rlap / k_2'+m_c)\gamma^\lambda(\rlap /q_1-\rlap /k_3-\rlap
/k_1'+m_c)O^\mu)_{\alpha\alpha'}\over
[A_{a10}+(\textbf{k}_{2T}+\textbf{k}_{2T}')^2][B_{a10}+(\textbf{q}_{T}+\textbf{k}_{3T}')^2][C_{a10}+(\textbf{k}_{3T}+\textbf{k}_{1T}'-\textbf{q}_{T})^2]
[D_{a10}+(\textbf{k}_{3T}-\textbf{k}_{2T}')^2] }\nonumber
\\
\end{eqnarray}
with
\begin{eqnarray}
&&A_{a10}=(r^2-1)x_2x_2'm_B^2,\;\;\;\;\;\;\;\;\;\;\;\;\;\;\;\;\;\;\;\;\;\;\;\;\;\;\;\;
B_{a10}=(r^2-1)x_3'(y-1)m_B^2,\nonumber \\
&&C_{a10}=m_c^2+((r^2-1)x_1'+1)(x_3-y)m_B^2,\;
D_{a10}=m_c^2-(r^2(x_2'-1)-x_2')(x_3-1)m_B^2\nonumber \\
\end{eqnarray}
and the color factor
\begin{eqnarray}
&&C_N^{a10}=\varepsilon^{abc}\varepsilon^{a'b'c'}\left(C_1(T^j)_{bb'}(T^jT^iT^i)_{ac'}\delta_{ca'}+C_2(T^j)_{bb'}(T^i)_{cc'}(T^jT^i)_{aa'}\right)={16\over
3}C_1+{8\over 3}C_2.\nonumber \\
\end{eqnarray}

For the hard amplitude of Fig.1(a11):
\begin{eqnarray}
&&
H^{a11,\alpha'\beta'\gamma'\rho'\alpha\beta\gamma\rho}(x_i,x_i',y,\textbf{k}_{T},\textbf{k}_{T}'
,\textbf{q}_{T},m_B,m_{\Lambda_c})=\nonumber
\\
&&\left[\varepsilon^{abc}\varepsilon^{a'b'c'}\left(C_1(T^j)_{bb'}(T^j)_{ca'}(T^iT^i)_{ac'}+C_2(T^j)_{bb'}(T^jT^i)_{cc'}(T^i)_{aa'}\right)\right]g_s^4\nonumber
\\
&&{(\gamma_\lambda)_{\rho'\gamma'}(\gamma_\theta)_{\beta
\beta'}[\gamma^\theta(\rlap /p-\rlap /k_1+\rlap / k_2')
O_\mu]_{\gamma \rho}[\gamma^\lambda(-\rlap /q_2+\rlap /k_1+\rlap
/k_3'+m_c)O^\mu)_{\alpha\alpha'}\over
(k_2+k_2')^2(q_2-k_3)^2(p-k_1+k_2')^2[(-q2+k_1+k_3')^2-m_c^2]}\nonumber
\\
&&={C_N^{a11}
g_s^4(\gamma_\lambda)_{\rho'\gamma'}(\gamma_\theta)_{\beta
\beta'}[\gamma^\theta(\rlap /p-\rlap /k_1+\rlap / k_2')
O_\mu]_{\gamma \rho}[\gamma^\lambda(-\rlap /q_2+\rlap /k_1+\rlap
/k_3'+m_c)O^\mu)_{\alpha\alpha'}\over
[A_{a11}+(\textbf{k}_{2T}+\textbf{k}_{2T}')^2][B_{a11}+(\textbf{q}_{T}+\textbf{k}_{3T}')^2][C_{a11}+(\textbf{k}_{1T}-\textbf{k}_{2T}')^2]
[D_{a11}+(\textbf{k}_{1T}+\textbf{k}_{3T}'+\textbf{q}_{T})^2]
}\nonumber
\\
\end{eqnarray}
with
\begin{eqnarray}
&&A_{a11}=(r^2-1)x_2x_2'm_B^2,\;\;\;\;\;\;\;\;\;\;
B_{a11}=(r^2-1)x_3'(y-1)m_B^2,\nonumber \\
&&C_{a11}=(1-r^2)(x_1-1)x_2'm_B^2,\;
D_{a11}=m_c^2+(r^2(x_3'-1)-x_3')(x_1+y-1)m_B^2\nonumber \\
\end{eqnarray}
and the color factor
\begin{eqnarray}
&&C_N^{a11}=\varepsilon^{abc}\varepsilon^{a'b'c'}\left(C_1(T^j)_{bb'}(T^j)_{ca'}(T^iT^i)_{ac'}+C_2(T^j)_{bb'}(T^jT^i)_{cc'}(T^i)_{aa'}\right)={16\over
3}C_1+{8\over 3}C_2.\nonumber \\
\end{eqnarray}

For the hard amplitude of Fig.1(a12):
\begin{eqnarray}
&&
H^{a12,\alpha'\beta'\gamma'\rho'\alpha\beta\gamma\rho}(x_i,x_i',y,\textbf{k}_{T},\textbf{k}_{T}'
,\textbf{q}_{T},m_B,m_{\Lambda_c})=\nonumber
\\
&&\left[\varepsilon^{abc}\varepsilon^{a'b'c'}\left(C_1(T^j)_{bb'}(T^j)_{ca'}(T^iT^i)_{ac'}+C_2(T^j)_{bb'}(T^i)_{cc'}(T^iT^j)_{aa'}\right)\right]g_s^4\nonumber
\\
&&{(\gamma_\lambda)_{\rho'\gamma'}(\gamma_\theta)_{\beta
\beta'}(O_\mu)_{\gamma \rho}[\gamma^\lambda(-\rlap /q_2+\rlap
/k_1+\rlap /k_3'+m_c)O^\mu(-\rlap p'-\rlap /k_2+\rlap
/k_3')\gamma^\theta]_{\alpha\alpha'}\over
(k_2+k_2')^2(q_2-k_3)^2(-p'-k_2+k_3')^2[(-q2+k_1+k_3')^2-m_c^2]}\nonumber
\\
&&={C_N^{a12}
g_s^4(\gamma_\lambda)_{\rho'\gamma'}(\gamma_\theta)_{\beta
\beta'}(O_\mu)_{\gamma \rho}[\gamma^\lambda(-\rlap /q_2+\rlap
/k_1+\rlap /k_3'+m_c)O^\mu(-\rlap p'-\rlap /k_2+\rlap
/k_3')\gamma^\theta]_{\alpha\alpha'}\over
[A_{a12}+(\textbf{k}_{2T}+\textbf{k}_{2T}')^2][B_{a12}+(\textbf{q}_{T}+\textbf{k}_{3T}')^2][C_{a12}+(\textbf{k}_{2T}-\textbf{k}_{3T}')^2]
[D_{a12}+(\textbf{k}_{1T}+\textbf{k}_{3T}'+\textbf{q}_{T})^2]
}\nonumber
\\
\end{eqnarray}
with
\begin{eqnarray}
&&A_{a12}=(r^2-1)x_2x_2'm_B^2,\;\;\;\;\;\;\;\;\;\;
B_{a12}=(r^2-1)x_3'(y-1)m_B^2,\nonumber \\
&&C_{a12}=(1-r^2)(x_3'-1)x_2m_B^2,\;
D_{a12}=m_c^2+(r^2(x_3'-1)-x_3')(x_1+y-1)m_B^2\nonumber \\
\end{eqnarray}
and the color factor
\begin{eqnarray}
&&C_N^{a12}=\varepsilon^{abc}\varepsilon^{a'b'c'}\left(C_1(T^j)_{bb'}(T^j)_{ca'}(T^iT^i)_{ac'}+C_2(T^j)_{bb'}(T^i)_{cc'}(T^iT^j)_{aa'}\right)={16\over
3}C_1+{8\over 3}C_2.\nonumber \\
\end{eqnarray}

For the hard amplitude of Fig.1(a13):
\begin{eqnarray}
&&
H^{a13,\alpha'\beta'\gamma'\rho'\alpha\beta\gamma\rho}(x_i,x_i',y,\textbf{k}_{T},\textbf{k}_{T}'
,\textbf{q}_{T},m_B,m_{\Lambda_c})=\nonumber
\\
&&\left[\varepsilon^{abc}\varepsilon^{a'b'c'}\left(C_1(T^j)_{bb'}(T^iT^jT^i)_{ac'}\delta_{ca'}+C_2(T^j)_{bb'}(T^i)_{aa'}(T^jT^i)_{cc'}\right)\right]g_s^4\nonumber
\\
&&{[\gamma_\theta(\rlap /k_2+\rlap /k_2'-\rlap
/q_2)\gamma_\lambda]_{\rho'\gamma'}(\gamma^\theta)_{\beta
\beta'}(O_\mu)_{\gamma \rho}[\gamma^\lambda(\rlap /q_1-\rlap
/k_3-\rlap /k_1'+m_c)O^\mu]_{\alpha\alpha'}\over
(k_2+k_2')^2(p-q_1-k_2+k_1')^2(k_2+k_2'-q_2)^2[(q1-k_3-k_1')^2-m_c^2]}\nonumber
\\
&&={C_N^{a13} g_s^4[\gamma_\theta(\rlap /k_2+\rlap /k_2'-\rlap
/q_2)\gamma_\lambda]_{\rho'\gamma'}(\gamma^\theta)_{\beta
\beta'}(O_\mu)_{\gamma \rho}[\gamma^\lambda(\rlap /q_1-\rlap
/k_3-\rlap /k_1'+m_c)O^\mu]_{\alpha\alpha'}\over
[A_{a13}+(\textbf{k}_{2T}+\textbf{k}_{2T}')^2][B_{a13}+(\textbf{q}_{T}+\textbf{k}_{2T}-\textbf{k}_{1T}')^2][C_{a13}+(\textbf{k}_{2T}+\textbf{k}_{2T}'+\textbf{q}_{T})^2]
[D_{a13}+(\textbf{k}_{1T}'+\textbf{k}_{3T}-\textbf{q}_{T})^2]
}\nonumber
\\
\end{eqnarray}
with
\begin{eqnarray}
&&A_{a13}=(r^2-1)x_2x_2'm_B^2,\;\;\;\;\;\;\;\;\;\;\;\;\;\;\;\;
B_{a13}=(1-r^2)(x_1'-1)(x_2+y-1)m_B^2,\nonumber \\
&&C_{a13}=(r^2-1)(x_2+y-1)x_2'm_B^2,\;
D_{a13}=m_c^2+(x_1'(r^2-1)+1)(x_3-y)m_B^2\nonumber \\
\end{eqnarray}
and the color factor
\begin{eqnarray}
&&C_N^{a13}=\varepsilon^{abc}\varepsilon^{a'b'c'}\left(C_1(T^j)_{bb'}(T^iT^jT^i)_{ac'}\delta_{ca'}+C_2(T^j)_{bb'}(T^i)_{aa'}(T^jT^i)_{cc'}\right)=-{2\over
3}C_1+{8\over 3}C_2.\nonumber \\
\end{eqnarray}

For the hard amplitude of Fig.1(a14):
\begin{eqnarray}
&&
H^{a14,\alpha'\beta'\gamma'\rho'\alpha\beta\gamma\rho}(x_i,x_i',y,\textbf{k}_{T},\textbf{k}_{T}'
,\textbf{q}_{T},m_B,m_{\Lambda_c})=\nonumber
\\
&&\left[\varepsilon^{abc}\varepsilon^{a'b'c'}\left(C_1(T^j)_{bb'}(T^iT^iT^j)_{ac'}\delta_{ca'}+C_2(T^j)_{bb'}(T^i)_{aa'}(T^iT^j)_{cc'}\right)\right]g_s^4\nonumber
\\
&&{[\gamma_\lambda(-\rlap /p'-\rlap /k_2+\rlap
/k_1')\gamma_\theta]_{\rho'\gamma'}(\gamma^\theta)_{\beta
\beta'}(O_\mu)_{\gamma \rho}[\gamma^\lambda(\rlap /q_1-\rlap
/k_3-\rlap /k_1'+m_c)O^\mu]_{\alpha\alpha'}\over
(k_2+k_2')^2(q_2-p'-k_2+k_1')^2(-p'-k_2+k_1')^2[(q1-k_3-k_1')^2-m_c^2]}\nonumber
\\
&&={C_N^{a14} g_s^4[\gamma_\lambda(-\rlap /p'-\rlap /k_2+\rlap
/k_1')\gamma_\theta]_{\rho'\gamma'}(\gamma^\theta)_{\beta
\beta'}(O_\mu)_{\gamma \rho}[\gamma^\lambda(\rlap /q_1-\rlap
/k_3-\rlap /k_1'+m_c)O^\mu]_{\alpha\alpha'}\over
[A_{a14}+(\textbf{k}_{2T}+\textbf{k}_{2T}')^2][B_{a14}+(\textbf{q}_{T}+\textbf{k}_{2T}-\textbf{k}_{1T}')^2][C_{a14}+(\textbf{k}_{2T}-\textbf{k}_{1T}')^2]
[D_{a14}+(\textbf{k}_{1T}'+\textbf{k}_{3T}-\textbf{q}_{T})^2]
}\nonumber
\\
\end{eqnarray}
with
\begin{eqnarray}
&&A_{a14}=(r^2-1)x_2x_2'm_B^2,\;\;\;\;\;\;\;\;\;\;
B_{a14}=(1-r^2)(x_1'-1)(x_2+y-1)m_B^2,\nonumber \\
&&C_{a14}=(1-r^2)(x_1'-1)x_2m_B^2,\;
D_{a14}=m_c^2+(x_1'(r^2-1)+1)(x_3-y)m_B^2\nonumber \\
\end{eqnarray}
and the color factor
\begin{eqnarray}
&&C_N^{a14}=\varepsilon^{abc}\varepsilon^{a'b'c'}\left(C_1(T^j)_{bb'}(T^iT^iT^j)_{ac'}\delta_{ca'}+C_2(T^j)_{bb'}(T^i)_{aa'}(T^iT^j)_{cc'}\right)={16\over
3}C_1+{8\over 3}C_2.\nonumber \\
\end{eqnarray}

For the hard amplitude of Fig.1(a15):
\begin{eqnarray}
&&
H^{a15,\alpha'\beta'\gamma'\rho'\alpha\beta\gamma\rho}(x_i,x_i',y,\textbf{k}_{T},\textbf{k}_{T}'
,\textbf{q}_{T},m_B,m_{\Lambda_c})=\nonumber
\\
&&\left[\varepsilon^{abc}\varepsilon^{a'b'c'}\left(C_1(T^j)_{bb'}(T^jT^i)_{ac'}(T^i)_{ca'}+C_2(T^j)_{bb'}(T^iT^jT^i)_{cc'}\delta_{aa'}\right)\right]g_s^4\nonumber
\\
&&{(\gamma_\lambda)_{\rho'\gamma'}(\gamma_\theta)_{\beta
\beta'}[\gamma^\lambda(\rlap /k_3+\rlap /k_3'-\rlap
/q_2)O_\mu(\rlap /q_1-\rlap /k_2-\rlap
/k_2'+m_b)\gamma^\theta]_{\gamma \rho}(O^\mu)_{\alpha\alpha'}\over
(k_2+k_2')^2(q_2-k_3')^2(k_3+k_3'-q_2)^2[(q1-k_2-k_2')^2-m_b^2]}\nonumber
\\
&&={C_N^{a15}
g_s^4(\gamma_\lambda)_{\rho'\gamma'}(\gamma_\theta)_{\beta
\beta'}[\gamma^\lambda(\rlap /k_3+\rlap /k_3'-\rlap
/q_2)O_\mu(\rlap /q_1-\rlap /k_2-\rlap
/k_2'+m_b)\gamma^\theta]_{\gamma \rho}(O^\mu)_{\alpha\alpha'}\over
[A_{a15}+(\textbf{k}_{2T}+\textbf{k}_{2T}')^2][B_{a15}+(\textbf{q}_{T}+\textbf{k}_{3T}')^2][C_{a15}+(\textbf{k}_{3T}+\textbf{k}_{3T}'+\textbf{q}_{T})^2]
[D_{a15}+(\textbf{k}_{2T}'+\textbf{k}_{2T}-\textbf{q}_{T})^2]
}\nonumber
\\
\end{eqnarray}
with
\begin{eqnarray}
&&A_{a15}=(r^2-1)x_2x_2'm_B^2,\;\;\;\;\;\;\;\;\;\;\;\;\;\;\;\;
B_{a15}=(r^2-1)x_3'(y-1)m_B^2,\nonumber \\
&&C_{a15}=(r^2-1)(x_3+y-1)x_3'm_B^2,\;
D_{a15}=m_b^2+(x_2'(r^2-1)+1)(x_2-y)m_B^2\nonumber \\
\end{eqnarray}
and the color factor
\begin{eqnarray}
&&C_N^{a15}=\varepsilon^{abc}\varepsilon^{a'b'c'}\left(C_1(T^j)_{bb'}(T^jT^i)_{ac'}(T^i)_{ca'}+C_2(T^j)_{bb'}(T^iT^jT^i)_{cc'}\delta_{aa'}\right)=-{8\over
3}C_1+{2\over 3}C_2.\nonumber \\
\end{eqnarray}

For the hard amplitude of Fig.1(a16):
\begin{eqnarray}
&&
H^{a16,\alpha'\beta'\gamma'\rho'\alpha\beta\gamma\rho}(x_i,x_i',y,\textbf{k}_{T},\textbf{k}_{T}'
,\textbf{q}_{T},m_B,m_{\Lambda_c})=\nonumber
\\
&&\left[\varepsilon^{abc}\varepsilon^{a'b'c'}\left(C_1(T^j)_{bb'}(T^jT^i)_{ac'}(T^i)_{ca'}+C_2(T^j)_{bb'}(T^j)_{aa'}(T^iT^i)_{cc'}\right)\right]g_s^4\nonumber
\\
&&{(\gamma_\lambda)_{\rho'\gamma'}(\gamma_\theta)_{\beta
\beta'}[\gamma^\lambda(\rlap /k_3+\rlap /k_3'-\rlap
/q_2)O_\mu]_{\gamma \rho}[\gamma^\theta(\rlap /p-\rlap /k_3+\rlap
/k_2'+m_c)O^\mu]_{\alpha\alpha'}\over
(k_2+k_2')^2(q_2-k_3')^2(k_3+k_3'-q_2)^2[(p-k_3+k_2')^2-m_c^2]}\nonumber
\\
&&={C_N^{a16}
g_s^4(\gamma_\lambda)_{\rho'\gamma'}(\gamma_\theta)_{\beta
\beta'}[\gamma^\lambda(\rlap /k_3+\rlap /k_3'-\rlap
/q_2)O_\mu]_{\gamma \rho}[\gamma^\theta(\rlap /p-\rlap /k_3+\rlap
/k_2'+m_c)O^\mu]_{\alpha\alpha'}\over
[A_{a16}+(\textbf{k}_{2T}+\textbf{k}_{2T}')^2][B_{a16}+(\textbf{q}_{T}+\textbf{k}_{3T}')^2][C_{a16}+(\textbf{k}_{3T}+\textbf{k}_{3T}'+\textbf{q}_{T})^2]
[D_{a16}+(\textbf{k}_{3T}-\textbf{k}_{2T}')^2] }\nonumber
\\
\end{eqnarray}
with
\begin{eqnarray}
&&A_{a16}=(r^2-1)x_2x_2'm_B^2,\;\;\;\;\;\;\;\;\;\;\;\;\;\;\;\;
B_{a16}=(r^2-1)x_3'(y-1)m_B^2,\nonumber \\
&&C_{a16}=(r^2-1)(x_3+y-1)x_3'm_B^2,\;
D_{a16}=m_c^2-(r^2(x_2'-1)-x_2')(x_3-1)m_B^2\nonumber \\
\end{eqnarray}
and the color factor
\begin{eqnarray}
&&C_N^{a16}=\varepsilon^{abc}\varepsilon^{a'b'c'}\left(C_1(T^j)_{bb'}(T^jT^i)_{ac'}(T^i)_{ca'}+C_2(T^j)_{bb'}(T^j)_{aa'}(T^iT^i)_{cc'}\delta_{aa'}\right)=-{8\over
3}C_1-{16\over 3}C_2.\nonumber \\
\end{eqnarray}

For the hard amplitude of Fig.1(a17):
\begin{eqnarray}
&&
H^{a17,\alpha'\beta'\gamma'\rho'\alpha\beta\gamma\rho}(x_i,x_i',y,\textbf{k}_{T},\textbf{k}_{T}'
,\textbf{q}_{T},m_B,m_{\Lambda_c})=\nonumber
\\
&&\left[\varepsilon^{abc}\varepsilon^{a'b'c'}\left(C_1(T^j)_{bb'}(T^i)_{ac'}(T^iT^j)_{ca'}+C_2(T^j)_{bb'}(T^iT^jT^i)_{cc'}\delta_{aa'}\right)\right]g_s^4\nonumber
\\
&&{(\gamma_\lambda)_{\rho'\gamma'}(\gamma_\theta)_{\beta
\beta'}[\gamma^\lambda(\rlap /k_3+\rlap /k_3'-\rlap
/q_2)\gamma^\theta(\rlap /q_1-\rlap /k_1-\rlap
/k_1')O_\mu]_{\gamma \rho}(O^\mu)_{\alpha\alpha'}\over
(k_2+k_2')^2(q_2-k_3')^2(k_3+k_3'-q_2)^2(q_1-k_1-k_1')^2}\nonumber
\\
&&={C_N^{a17}
g_s^4(\gamma_\lambda)_{\rho'\gamma'}(\gamma_\theta)_{\beta
\beta'}[\gamma^\lambda(\rlap /k_3+\rlap /k_3'-\rlap
/q_2)\gamma^\theta(\rlap /q_1-\rlap /k_1-\rlap
/k_1')O_\mu]_{\gamma \rho}(O^\mu)_{\alpha\alpha'}\over
[A_{a17}+(\textbf{k}_{2T}+\textbf{k}_{2T}')^2][B_{a17}+(\textbf{q}_{T}+\textbf{k}_{3T}')^2][C_{a17}+(\textbf{k}_{3T}+\textbf{k}_{3T}'+\textbf{q}_{T})^2]
[D_{a17}+(\textbf{k}_{1T}+\textbf{k}_{1T}'-\textbf{q}_{T})^2]
}\nonumber
\\
\end{eqnarray}
with
\begin{eqnarray}
&&A_{a17}=(r^2-1)x_2x_2'm_B^2,\;\;\;\;\;\;\;\;\;\;\;\;\;\;\;\;
B_{a17}=(r^2-1)x_3'(y-1)m_B^2,\nonumber \\
&&C_{a17}=(r^2-1)(x_3+y-1)x_3'm_B^2,\;
D_{a17}=(r^2-1)(x_1'-1)(x_1-y)m_B^2\nonumber \\
\end{eqnarray}
and the color factor
\begin{eqnarray}
&&C_N^{a17}=\varepsilon^{abc}\varepsilon^{a'b'c'}\left(C_1(T^j)_{bb'}(T^i)_{ac'}(T^iT^j)_{ca'}+C_2(T^j)_{bb'}(T^iT^jT^i)_{cc'}\delta_{aa'}\right)=-{8\over
3}C_1+{2\over 3}C_2.\nonumber \\
\end{eqnarray}

For the hard amplitude of Fig.1(a18):
\begin{eqnarray}
&&
H^{a18,\alpha'\beta'\gamma'\rho'\alpha\beta\gamma\rho}(x_i,x_i',y,\textbf{k}_{T},\textbf{k}_{T}'
,\textbf{q}_{T},m_B,m_{\Lambda_c})=\nonumber
\\
&&\left[\varepsilon^{abc}\varepsilon^{a'b'c'}\left(C_1(T^j)_{bb'}(T^i)_{ac'}(T^jT^i)_{ca'}+C_2(T^j)_{bb'}(T^jT^iT^i)_{cc'}\delta_{aa'}\right)\right]g_s^4\nonumber
\\
&&{(\gamma_\lambda)_{\rho'\gamma'}(\gamma_\theta)_{\beta
\beta'}[\gamma^\theta(\rlap /p-\rlap /k_1+\rlap
/k_2')\gamma^\lambda(\rlap /q_1-\rlap /k_1-\rlap
/k_1')O_\mu]_{\gamma \rho}(O^\mu)_{\alpha\alpha'}\over
(k_2+k_2')^2(q_2-k_3')^2(p-k_1+k_2')^2(q_1-k_1-k_1')^2}\nonumber
\\
&&={C_N^{a18}
g_s^4(\gamma_\lambda)_{\rho'\gamma'}(\gamma_\theta)_{\beta
\beta'}[\gamma^\theta(\rlap /p-\rlap /k_1+\rlap
/k_2')\gamma^\lambda(\rlap /q_1-\rlap /k_1-\rlap
/k_1')O_\mu]_{\gamma \rho}(O^\mu)_{\alpha\alpha'}\over
[A_{a18}+(\textbf{k}_{2T}+\textbf{k}_{2T}')^2][B_{a18}+(\textbf{q}_{T}+\textbf{k}_{3T}')^2][C_{a18}+(\textbf{k}_{1T}-\textbf{k}_{2T}')^2]
[D_{a18}+(\textbf{k}_{1T}+\textbf{k}_{1T}'-\textbf{q}_{T})^2]
}\nonumber
\\
\end{eqnarray}
with
\begin{eqnarray}
&&A_{a18}=(r^2-1)x_2x_2'm_B^2,\;\;\;\;\;\;\;\;\;\;
B_{a18}=(r^2-1)x_3'(y-1)m_B^2,\nonumber \\
&&C_{a18}=(1-r^2)(x_1-1)x_2'm_B^2,\;
D_{a18}=(r^2-1)(x_1'-1)(x_1-y)m_B^2\nonumber \\
\end{eqnarray}
and the color factor
\begin{eqnarray}
&&C_N^{a18}=\varepsilon^{abc}\varepsilon^{a'b'c'}\left(C_1(T^j)_{bb'}(T^i)_{ac'}(T^jT^i)_{ca'}+C_2(T^j)_{bb'}(T^jT^iT^i)_{cc'}\delta_{aa'}\right)=-{8\over
3}C_1-{16\over 3}C_2.\nonumber \\
\end{eqnarray}

For the hard amplitude of Fig.1(a19):
\begin{eqnarray}
&&
H^{a19,\alpha'\beta'\gamma'\rho'\alpha\beta\gamma\rho}(x_i,x_i',y,\textbf{k}_{T},\textbf{k}_{T}'
,\textbf{q}_{T},m_B,m_{\Lambda_c})=\nonumber
\\
&&\left[\varepsilon^{abc}\varepsilon^{a'b'c'}\left(C_1(T^j)_{bb'}(T^i)_{ac'}(T^iT^j)_{ca'}+C_2(T^j)_{bb'}(T^j)_{aa'}(T^iT^i)_{cc'}\right)\right]g_s^4\nonumber
\\
&&{(\gamma_\lambda)_{\rho'\gamma'}(\gamma_\theta)_{\beta
\beta'}[\gamma^\lambda(\rlap /k_3+\rlap /k_3'-\rlap
/q_2)O_\mu]_{\gamma \rho}(O^\mu(-\rlap /p'-\rlap /k_2+\rlap
/k_3')\gamma^\theta)_{\alpha\alpha'}\over
(k_2+k_2')^2(q_2-k_3')^2(k_3+k_3'-q_2)^2(-p'-k_2+k_3')^2}\nonumber
\\
&&={C_N^{a19}
g_s^4(\gamma_\lambda)_{\rho'\gamma'}(\gamma_\theta)_{\beta
\beta'}[\gamma^\lambda(\rlap /k_3+\rlap /k_3'-\rlap
/q_2)O_\mu]_{\gamma \rho}(O^\mu(-\rlap /p'-\rlap /k_2+\rlap
/k_3')\gamma^\theta)_{\alpha\alpha'}\over
[A_{a19}+(\textbf{k}_{2T}+\textbf{k}_{2T}')^2][B_{a19}+(\textbf{q}_{T}+\textbf{k}_{3T}')^2][C_{a19}+(\textbf{k}_{3T}+\textbf{k}_{3T}'+\textbf{q}_{T})^2]
[D_{a19}+(\textbf{k}_{2T}-\textbf{k}_{3T}')^2] }\nonumber
\\
\end{eqnarray}
with
\begin{eqnarray}
&&A_{a19}=(r^2-1)x_2x_2'm_B^2,\;\;\;\;\;\;\;\;\;\;\;\;\;\;\;\;
B_{a19}=(r^2-1)x_3'(y-1)m_B^2,\nonumber \\
&&C_{a19}=(r^2-1)(x_3+y-1)x_3'm_B^2,\;
D_{a19}=(1-r^2)(x_3'-1)x_2m_B^2\nonumber \\
\end{eqnarray}
and the color factor
\begin{eqnarray}
&&C_N^{a19}=\varepsilon^{abc}\varepsilon^{a'b'c'}\left(C_1(T^j)_{bb'}(T^i)_{ac'}(T^iT^j)_{ca'}+C_2(T^j)_{bb'}(T^j)_{aa'}(T^iT^i)_{cc'}\right)=-{8\over
3}C_1-{16\over 3}C_2.\nonumber \\
\end{eqnarray}

For the hard amplitude of Fig.1(a20):
\begin{eqnarray}
&&
H^{a20,\alpha'\beta'\gamma'\rho'\alpha\beta\gamma\rho}(x_i,x_i',y,\textbf{k}_{T},\textbf{k}_{T}'
,\textbf{q}_{T},m_B,m_{\Lambda_c})=\nonumber
\\
&&\left[\varepsilon^{abc}\varepsilon^{a'b'c'}\left(C_1(T^j)_{bb'}(T^jT^i)_{ac'}(T^i)_{ca'}+C_2(T^j)_{bb'}(T^iT^jT^i)_{cc'}\delta_{aa'}\right)\right]g_s^4\nonumber
\\
&&{[\gamma_\theta(\rlap /k_2+\rlap /k_2'-\rlap
/q_2)\gamma_\lambda]_{\rho'\gamma'}(\gamma^\theta)_{\beta
\beta'}[\gamma^\lambda(\rlap /q_1-\rlap /k_1-\rlap
/k_1')O^\mu]_{\gamma\rho}(O_\mu)_{\alpha\alpha'}\over
(k_2+k_2')^2(q_1-p+k_2-k_1')^2(k_2+k_2'-q_2)^2(q_1-k_1-k_1')^2}\nonumber
\\
&&={C_N^{a20} g_s^4[\gamma_\theta(\rlap /k_2+\rlap /k_2'-\rlap
/q_2)\gamma_\lambda]_{\rho'\gamma'}(\gamma^\theta)_{\beta
\beta'}[\gamma^\lambda(\rlap /q_1-\rlap /k_1-\rlap
/k_1')O^\mu]_{\gamma\rho}(O_\mu)_{\alpha\alpha'}\over
[A_{a20}+(\textbf{k}_{2T}
+\textbf{k}_{2T}')^2][B_{a20}+(\textbf{k}_{2T}-\textbf{k}_{1T}'+\textbf{q}_{T})^2][C_{a20}+(\textbf{k}_{2T}+\textbf{k}_{2T}'+\textbf{q}_{T})^2]
[D_{a20}+(-\textbf{q}_{T}+\textbf{k}_{1T}+\textbf{k}_{1T}')^2]}\nonumber
\\
\end{eqnarray}
with
\begin{eqnarray}
&&A_{a20}=x_2x_2'(r^2-1)m_B^2,\;\;\;\;\;\;\;\;\;\;\;\;\;\;\;\;
B_{a20}=(1-r^2)(x_1'-1)(x_2+y-1)m_B^2\nonumber \\
&&C_{a20}=x_2'(r^2-1)(x_2+y-1)m_B^2,\;
D_{a20}=(r^2-1)(x_1'-1)(x_1-y)m_B^2\nonumber \\
\end{eqnarray}
and the color factor
\begin{eqnarray}
&&C_N^{a20}=\varepsilon^{abc}\varepsilon^{a'b'c'}\left(C_1(T^j)_{bb'}(T^jT^i)_{ac'}(T^i)_{ca'}+C_2(T^j)_{bb'}(T^iT^jT^i)_{cc'}\delta_{aa'}\right)=-{8\over
3}C_1+{2\over 3}C_2.\nonumber \\
\end{eqnarray}

For the hard amplitude of Fig.1(a21):
\begin{eqnarray}
&&
H^{a21,\alpha'\beta'\gamma'\rho'\alpha\beta\gamma\rho}(x_i,x_i',y,\textbf{k}_{T},\textbf{k}_{T}'
,\textbf{q}_{T},m_B,m_{\Lambda_c})=\nonumber
\\
&&\left[\varepsilon^{abc}\varepsilon^{a'b'c'}\left(C_1(T^j)_{bb'}(T^iT^j)_{ac'}(T^i)_{ca'}+C_2(T^j)_{bb'}(T^iT^iT^j)_{cc'}\delta_{aa'}\right)\right]g_s^4\nonumber
\\
&&{[\gamma_\lambda(-\rlap /p'-\rlap /k_2+\rlap
/k_1')\gamma_\theta]_{\rho'\gamma'}(\gamma^\theta)_{\beta
\beta'}[\gamma^\lambda(\rlap /q_1-\rlap /k_1-\rlap
/k_1')O^\mu]_{\gamma\rho}(O_\mu)_{\alpha\alpha'}\over
(k_2+k_2')^2(q_1-p+k_2-k_1')^2(-p'-k_2+k_1')^2(q_1-k_1-k_1')^2}\nonumber
\\
&&={C_N^{a21} g_s^4[\gamma_\lambda(-\rlap /p'-\rlap /k_2+\rlap
/k_1')\gamma_\theta]_{\rho'\gamma'}(\gamma^\theta)_{\beta
\beta'}[\gamma^\lambda(\rlap /q_1-\rlap /k_1-\rlap
/k_1')O^\mu]_{\gamma\rho}(O_\mu)_{\alpha\alpha'}\over
[A_{a21}+(\textbf{k}_{2T}+\textbf{k}_{2T}')^2][B_{a21}+(\textbf{q}_{T}+\textbf{k}_{2T}-\textbf{k}_{1T}')^2]
[C_{a21}+(\textbf{k}_{2T}
-\textbf{k}_{1T}')^2][D_{a21}+(\textbf{k}_{1T}+\textbf{k}_{1T}'-\textbf{q}_{T})^2]}\nonumber
\\
\end{eqnarray}
with
\begin{eqnarray}
&&A_{a21}=x_2x_2'(r^2-1)m_B^2,\;\;\;\;\;\;\;\;\;\;\;\;\;\;\;\;
B_{a21}=(1-r^2)(x_1'-1)(x_2+y-1)m_B^2\nonumber \\
&&C_{a21}=(r^2-1)x_2'(x_2+y-1)m_B^2,\;
D_{a21}=(r^2-1)(x_1'-1)(x_1-y)m_B^2,\nonumber \\
\end{eqnarray}
and the color factor
\begin{eqnarray}
&&C_N^{a21}=\varepsilon^{abc}\varepsilon^{a'b'c'}\left(C_1(T^j)_{bb'}(T^iT^j)_{ac'}(T^i)_{ca'}+C_2(T^j)_{bb'}(T^iT^iT^j)_{cc'}\delta_{aa'}\right)=-{8\over
3}C_1-{16\over 3}C_2.\nonumber \\
\end{eqnarray}

For the hard amplitude of Fig.1(a22):
\begin{eqnarray}
&&
H^{a22,\alpha'\beta'\gamma'\rho'\alpha\beta\gamma\rho}(x_i,x_i',y,\textbf{k}_{T},\textbf{k}_{T}'
,\textbf{q}_{T},m_B,m_{\Lambda_c})=\nonumber
\\
&&\left[\varepsilon^{abc}\varepsilon^{a'b'c'}\left(C_1(T^j)_{bb'}(T^jT^i)_{ac'}(T^i)_{ca'}+C_2(T^j)_{bb'}(T^jT^i)_{cc'}(T^i)_{aa'}\right)\right]g_s^4\nonumber
\\
&&{(\gamma_\lambda)_{\rho'\gamma'}(\gamma_\theta)_{\beta
\beta'}[O_\mu(\rlap /q_1-\rlap /k_2-\rlap
/k_2'+m_b)\gamma^\theta]_{\gamma\rho}[O^\mu(\rlap /q_2-\rlap
/p'+\rlap /k_2')\gamma^\lambda]_{\alpha\alpha'}\over
(k_2+k_2')^2(q_2-k_3')^2(q_2-p'+k_2')^2[(q_1-k_2-k_2')^2-m_b^2]}\nonumber
\\
&&={C_N^{a22}
g_s^4(\gamma_\lambda)_{\rho'\gamma'}(\gamma_\theta)_{\beta
\beta'}[O_\mu(\rlap /q_1-\rlap /k_2-\rlap
/k_2'+m_b)\gamma^\theta]_{\gamma\rho}[O^\mu(\rlap /q_2-\rlap
/p'+\rlap /k_2')\gamma^\lambda]_{\alpha\alpha'}\over
[A_{a22}+(\textbf{k}_{2T}+\textbf{k}_{2T}')^2][B_{a22}+(\textbf{q}_{T}+\textbf{k}_{3T}')^2]
[C_{a22}+(\textbf{k}_{2T}'
-\textbf{q}_{T})^2][D_{a22}+(\textbf{k}_{2T}+\textbf{k}_{2T}'-\textbf{q}_{T})^2]}\nonumber
\\
\end{eqnarray}
with
\begin{eqnarray}
&&A_{a22}=x_2x_2'(r^2-1)m_B^2,\;\;\;\;\;\;\;\;\;\;\;\;\;\;\;\;\;
B_{a22}=(r^2-1)x_3'(y-1)m_B^2\nonumber \\
&&C_{a22}=(1-r^2)(x_2'-1)(y-1)m_B^2,\;
D_{a22}=m_b^2+((r^2-1)x_2'+1)(x_2-y)m_B^2,\nonumber \\
\end{eqnarray}
and the color factor
\begin{eqnarray}
&&C_N^{a22}=\varepsilon^{abc}\varepsilon^{a'b'c'}\left(C_1(T^j)_{bb'}(T^jT^i)_{ac'}(T^i)_{ca'}+C_2(T^j)_{bb'}(T^jT^i)_{cc'}(T^i)_{aa'}\right)=-{8\over
3}C_1+{8\over 3}C_2.\nonumber \\
\end{eqnarray}

For the hard amplitude of Fig.1(a23):
\begin{eqnarray}
&&
H^{a23,\alpha'\beta'\gamma'\rho'\alpha\beta\gamma\rho}(x_i,x_i',y,\textbf{k}_{T},\textbf{k}_{T}'
,\textbf{q}_{T},m_B,m_{\Lambda_c})=\nonumber
\\
&&\left[\varepsilon^{abc}\varepsilon^{a'b'c'}\left(C_1(T^j)_{bb'}(T^jT^i)_{ac'}(T^i)_{ca'}+C_2(T^j)_{bb'}(T^jT^i)_{aa'}(T^i)_{cc'}\right)\right]g_s^4\nonumber
\\
&&{(\gamma_\lambda)_{\rho'\gamma'}(\gamma_\theta)_{\beta
\beta'}(O_\mu)_{\gamma\rho}[\gamma^\theta(\rlap /p-\rlap
/k_3+\rlap /k_2'+m_c) O^\mu(\rlap /q_2-\rlap /p'+\rlap
/k_2')\gamma^\lambda]_{\alpha\alpha'}\over
(k_2+k_2')^2(q_2-k_3')^2(q_2-p'+k_2')^2[(p-k_3+k_2')^2-m_c^2]}\nonumber
\\
&&={C_N^{a23}
g_s^4(\gamma_\lambda)_{\rho'\gamma'}(\gamma_\theta)_{\beta
\beta'}(O_\mu)_{\gamma\rho}[\gamma^\theta(\rlap /p-\rlap
/k_3+\rlap /k_2'+m_c) O^\mu(\rlap /q_2-\rlap /p'+\rlap
/k_2')\gamma^\lambda]_{\alpha\alpha'}\over
[A_{a23}+(\textbf{k}_{2T}+\textbf{k}_{2T}')^2][B_{a23}+(\textbf{q}_{T}+\textbf{k}_{3T}')^2]
[C_{a23}+(\textbf{k}_{2T}'
-\textbf{q}_{T})^2][D_{a23}+(\textbf{k}_{3T}-\textbf{k}_{2T}')^2]}\nonumber
\\
\end{eqnarray}
with
\begin{eqnarray}
&&A_{a23}=x_2x_2'(r^2-1)m_B^2,\;\;\;\;\;\;\;\;\;\;\;\;\;\;\;\;\;
B_{a23}=(r^2-1)x_3'(y-1)m_B^2\nonumber \\
&&C_{a23}=(1-r^2)(x_2'-1)(y-1)m_B^2,\;
D_{a23}=m_c^2-(r^2(x_2'-1)-x_2')(x_3-1)m_B^2,\nonumber \\
\end{eqnarray}
and the color factor
\begin{eqnarray}
&&C_N^{a23}=\varepsilon^{abc}\varepsilon^{a'b'c'}\left(C_1(T^j)_{bb'}(T^jT^i)_{ac'}(T^i)_{ca'}+C_2(T^j)_{bb'}(T^jT^i)_{aa'}(T^i)_{cc'}\right)=-{8\over
3}C_1+{8\over 3}C_2.\nonumber \\
\end{eqnarray}

For the hard amplitude of Fig.1(a24):
\begin{eqnarray}
&&
H^{a24,\alpha'\beta'\gamma'\rho'\alpha\beta\gamma\rho}(x_i,x_i',y,\textbf{k}_{T},\textbf{k}_{T}'
,\textbf{q}_{T},m_B,m_{\Lambda_c})=\nonumber
\\
&&\left[\varepsilon^{abc}\varepsilon^{a'b'c'}\left(C_1(T^j)_{bb'}(T^i)_{ac'}(T^jT^i)_{ca'}+C_2(T^j)_{bb'}(T^jT^i)_{cc'}(T^i)_{aa'}\right)\right]g_s^4\nonumber
\\
&&{(\gamma_\lambda)_{\rho'\gamma'}(\gamma_\theta)_{\beta
\beta'}[\gamma^\theta(\rlap /p-\rlap /k_1+\rlap
/k_2')O_\mu]_{\gamma\rho}[O^\mu(\rlap /q_2-\rlap /p'+\rlap /k_2')
\gamma^\lambda]_{\alpha\alpha'}\over
(k_2+k_2')^2(q_2-k_3')^2(q_2-p'+k_2')^2(p-k_1+k_2')^2}\nonumber
\\
&&={C_N^{a24}
g_s^4(\gamma_\lambda)_{\rho'\gamma'}(\gamma_\theta)_{\beta
\beta'}[\gamma^\theta(\rlap /p-\rlap /k_1+\rlap
/k_2')O_\mu]_{\gamma\rho}[O^\mu(\rlap /q_2-\rlap /p'+\rlap /k_2')
\gamma^\lambda]_{\alpha\alpha'}\over
[A_{a24}+(\textbf{k}_{2T}+\textbf{k}_{2T}')^2][B_{a24}+(\textbf{q}_{T}+\textbf{k}_{3T}')^2]
[C_{a24}+(\textbf{k}_{2T}'
-\textbf{q}_{T})^2][D_{a24}+(\textbf{k}_{1T}-\textbf{k}_{2T}')^2]}\nonumber
\\
\end{eqnarray}
with
\begin{eqnarray}
&&A_{a24}=x_2x_2'(r^2-1)m_B^2,\;\;\;\;\;\;\;\;\;\;\;\;\;\;\;\;\;
B_{a24}=(r^2-1)x_3'(y-1)m_B^2\nonumber \\
&&C_{a24}=(1-r^2)(x_2'-1)(y-1)m_B^2,\;
D_{a24}=(1-r^2)(x_1-1)x_2'm_B^2,\nonumber \\
\end{eqnarray}
and the color factor
\begin{eqnarray}
&&C_N^{a24}=\varepsilon^{abc}\varepsilon^{a'b'c'}\left(C_1(T^j)_{bb'}(T^i)_{ac'}(T^jT^i)_{ca'}+C_2(T^j)_{bb'}(T^jT^i)_{cc'}(T^i)_{aa'}\right)=-{8\over
3}C_1+{8\over 3}C_2.\nonumber \\
\end{eqnarray}

For the hard amplitude of Fig.1(a25):
\begin{eqnarray}
&&
H^{a25,\alpha'\beta'\gamma'\rho'\alpha\beta\gamma\rho}(x_i,x_i',y,\textbf{k}_{T},\textbf{k}_{T}'
,\textbf{q}_{T},m_B,m_{\Lambda_c})=\nonumber
\\
&&\left[\varepsilon^{abc}\varepsilon^{a'b'c'}\left(C_1(T^j)_{bb'}(T^i)_{ac'}(T^jT^i)_{ca'}+C_2(T^j)_{bb'}(T^jT^i)_{aa'}(T^i)_{cc'}\right)\right]g_s^4\nonumber
\\
&&{(\gamma_\lambda)_{\rho'\gamma'}(\gamma_\theta)_{\beta
\beta'}(O_\mu)_{\gamma\rho}[O^\mu(\rlap /q_2-\rlap /p'-\rlap /k_2)
\gamma^\theta(\rlap /q_2-\rlap /p'+\rlap
/k_2')\gamma^\lambda]_{\alpha\alpha'}\over
(k_2+k_2')^2(q_2-k_3')^2(q_2-p'+k_2')^2(q_2-p'-k_2)^2}\nonumber
\\
&&={C_N^{a25}
g_s^4(\gamma_\lambda)_{\rho'\gamma'}(\gamma_\theta)_{\beta
\beta'}(O_\mu)_{\gamma\rho}[O^\mu(\rlap /q_2-\rlap /p'-\rlap /k_2)
\gamma^\theta(\rlap /q_2-\rlap /p'+\rlap
/k_2')\gamma^\lambda]_{\alpha\alpha'}\over
[A_{a25}+(\textbf{k}_{2T}+\textbf{k}_{2T}')^2][B_{a25}+(\textbf{q}_{T}+\textbf{k}_{3T}')^2]
[C_{a25}+(\textbf{k}_{2T}'
-\textbf{q}_{T})^2][D_{a25}+(\textbf{k}_{2T}+\textbf{q}_{T})^2]}\nonumber
\\
\end{eqnarray}
with
\begin{eqnarray}
&&A_{a25}=x_2x_2'(r^2-1)m_B^2,\;\;\;\;\;\;\;\;\;\;\;\;\;\;\;\;\;
B_{a25}=(r^2-1)x_3'(y-1)m_B^2\nonumber \\
&&C_{a25}=(1-r^2)(x_2'-1)(y-1)m_B^2,\;
D_{a25}=(r^2-1)(x_2+y-1)m_B^2,\nonumber \\
\end{eqnarray}
and the color factor
\begin{eqnarray}
&&C_N^{a25}=\varepsilon^{abc}\varepsilon^{a'b'c'}\left(C_1(T^j)_{bb'}(T^i)_{ac'}(T^jT^i)_{ca'}+C_2(T^j)_{bb'}(T^jT^i)_{aa'}(T^i)_{cc'}\right)=-{8\over
3}C_1+{8\over 3}C_2.\nonumber \\
\end{eqnarray}

For the hard amplitude of Fig.1(a26):
\begin{eqnarray}
&&
H^{a26,\alpha'\beta'\gamma'\rho'\alpha\beta\gamma\rho}(x_i,x_i',y,\textbf{k}_{T},\textbf{k}_{T}'
,\textbf{q}_{T},m_B,m_{\Lambda_c})=\nonumber
\\
&&\left[\varepsilon^{abc}\varepsilon^{a'b'c'}\left(C_1(T^j)_{bb'}(T^i)_{ac'}(T^iT^j)_{ca'}+C_2(T^j)_{bb'}(T^iT^j)_{aa'}(T^i)_{cc'}\right)\right]g_s^4\nonumber
\\
&&{(\gamma_\lambda)_{\rho'\gamma'}(\gamma_\theta)_{\beta
\beta'}(O_\mu)_{\gamma\rho}[O^\mu(\rlap /q_2-\rlap /p'-\rlap /k_2)
\gamma^\lambda(-\rlap /p'-\rlap /k_2+\rlap
/k_3')\gamma^\theta]_{\alpha\alpha'}\over
(k_2+k_2')^2(q_2-k_3')^2(q_2-p'+k_2')^2(-p'-k_2+k_3')^2}\nonumber
\\
&&={C_N^{a26}
g_s^4(\gamma_\lambda)_{\rho'\gamma'}(\gamma_\theta)_{\beta
\beta'}(O_\mu)_{\gamma\rho}[O^\mu(\rlap /q_2-\rlap /p'-\rlap /k_2)
\gamma^\lambda(-\rlap /p'-\rlap /k_2+\rlap
/k_3')\gamma^\theta]_{\alpha\alpha'}\over
[A_{a26}+(\textbf{k}_{2T}+\textbf{k}_{2T}')^2][B_{a26}+(\textbf{q}_{T}+\textbf{k}_{3T}')^2]
[C_{a26}+(\textbf{k}_{2T}
+\textbf{q}_{T})^2][D_{a26}+(\textbf{k}_{2T}-\textbf{k}_{3T})^2]}\nonumber
\\
\end{eqnarray}
with
\begin{eqnarray}
&&A_{a26}=x_2x_2'(r^2-1)m_B^2,\;\;\;\;\;\;\;\;\;\;\;\;\;
B_{a26}=(r^2-1)x_3'(y-1)m_B^2\nonumber \\
&&C_{a26}=(r^2-1)(x_2+y-1)m_B^2,\;
D_{a26}=(1-r^2)x_2(x_3'-1)m_B^2,\nonumber \\
\end{eqnarray}
and the color factor
\begin{eqnarray}
&&C_N^{a26}=\varepsilon^{abc}\varepsilon^{a'b'c'}\left(C_1(T^j)_{bb'}(T^i)_{ac'}(T^iT^j)_{ca'}+C_2(T^j)_{bb'}(T^iT^j)_{aa'}(T^i)_{cc'}\right)=-{8\over
3}C_1+{8\over 3}C_2.\nonumber \\
\end{eqnarray}

For the hard amplitude of Fig.1(a27):
\begin{eqnarray}
&&
H^{a27,\alpha'\beta'\gamma'\rho'\alpha\beta\gamma\rho}(x_i,x_i',y,\textbf{k}_{T},\textbf{k}_{T}'
,\textbf{q}_{T},m_B,m_{\Lambda_c})=\nonumber
\\
&&\left[\varepsilon^{abc}\varepsilon^{a'b'c'}\left(C_1(T^j)_{bb'}(T^jT^i)_{ac'}(T^i)_{ca'}+C_2(T^j)_{bb'}(T^jT^i)_{cc'}(T^i)_{aa'}\right)\right]g_s^4\nonumber
\\
&&{[\gamma_\theta(\rlap /k_2+\rlap /k_2'-\rlap
/q_2)\gamma_\lambda]_{\rho'\gamma'}(\gamma^\theta)_{\beta
\beta'}(O^\mu)_{\gamma\rho}[O^\mu(-\rlap /p'+\rlap /q_2-\rlap
/k_2)\gamma_\lambda]_{\alpha\alpha'}\over
(k_2+k_2')^2(q_2-p'-k_2+k_1')^2(q_2-p'-k_2)^2(-q_2+k_2+k_2')^2}\nonumber
\\
&&={C_N^{a27} g_s^4[\gamma_\theta(\rlap /k_2+\rlap /k_2'-\rlap
/q_2)\gamma_\lambda]_{\rho'\gamma'}(\gamma^\theta)_{\beta
\beta'}(O^\mu)_{\gamma\rho}[O^\mu(-\rlap /p'+\rlap /q_2-\rlap
/k_2)\gamma_\lambda]_{\alpha\alpha'}\over
[A_{a27}+(\textbf{k}_{2T}+\textbf{k}_{2T}')^2][B_{a27}+(\textbf{q}_{T}+\textbf{k}_{2T}-\textbf{k}_{1T}')^2]
[C_{a27}+(\textbf{k}_{2T}
+\textbf{q}_{T})^2][D_{a27}+(\textbf{k}_{2T}+\textbf{k}_{2T}'+\textbf{q}_{T})^2]
}\nonumber
\\
\end{eqnarray}
with
\begin{eqnarray}
&&A_{a27}=x_2x_2'(r^2-1)m_B^2,\;\;\;\;\;\;\;\;\;\;\;\;
B_{a27}=(1-r^2)(x_1'-1)(x_2+y-1)m_B^2\nonumber \\
&&C_{a27}=(r^2-1)(x_2+y-1)m_B^2,\;
D_{a27}=(r^2-1)x_2'(x_2+y-1)m_B^2,\nonumber \\
\end{eqnarray}
and the color factor
\begin{eqnarray}
&&C_N^{a27}=\varepsilon^{abc}\varepsilon^{a'b'c'}\left(C_1(T^j)_{bb'}(T^jT^i)_{ac'}(T^i)_{ca'}+C_2(T^j)_{bb'}(T^jT^i)_{cc'}(T^i)_{aa'}\right)=-{8\over
3}C_1+{8\over 3}C_2.\nonumber \\
\end{eqnarray}

For the hard amplitude of Fig.1(a28):
\begin{eqnarray}
&&
H^{a28,\alpha'\beta'\gamma'\rho'\alpha\beta\gamma\rho}(x_i,x_i',y,\textbf{k}_{T},\textbf{k}_{T}'
,\textbf{q}_{T},m_B,m_{\Lambda_c})=\nonumber
\\
&&\left[\varepsilon^{abc}\varepsilon^{a'b'c'}\left(C_1(T^j)_{bb'}(T^iT^j)_{ac'}(T^i)_{ca'}+C_2(T^j)_{bb'}(T^iT^j)_{cc'}(T^i)_{aa'}\right)\right]g_s^4\nonumber
\\
&&{[\gamma_\lambda(-\rlap /p'-\rlap /k_2+\rlap
/k_1')\gamma_\lambda]_{\rho'\gamma'}(\gamma^\theta)_{\beta\beta'}(O^\mu)_{\gamma\rho}[O^\mu(-\rlap
/p'+\rlap /q_2-\rlap /k_2)\gamma_\lambda]_{\alpha\alpha'}\over
(k_2+k_2')^2(q_2-p'-k_2+k_1')^2(q_2-p'-k_2)^2(-p'-k_2+k_1')^2}\nonumber
\\
&&={C_N^{a28} g_s^4[\gamma_\lambda(-\rlap /p'-\rlap /k_2+\rlap
/k_1')\gamma_\lambda]_{\rho'\gamma'}(\gamma^\theta)_{\beta\beta'}(O^\mu)_{\gamma\rho}[O^\mu(-\rlap
/p'+\rlap /q_2-\rlap /k_2)\gamma_\lambda]_{\alpha\alpha'}\over
[A_{a28}+(\textbf{k}_{2T}+\textbf{k}_{2T}')^2][B_{a28}+(\textbf{q}_{T}+\textbf{k}_{2T}-\textbf{k}_{1T}')^2]
[C_{a28}+(\textbf{k}_{2T}
+\textbf{q}_{T})^2][D_{a28}+(\textbf{k}_{2T}-\textbf{k}_{1T}')^2]
}\nonumber
\\
\end{eqnarray}
with
\begin{eqnarray}
&&A_{a28}=x_2x_2'(r^2-1)m_B^2,\;\;\;\;\;\;\;\;\;\;\;\;\;\;\;
B_{a28}=(1-r^2)(x_1'-1)(-1+x_2+y)m_B^2\nonumber \\
&&C_{a28}=(r^2-1)(-1+x_2+y)m_B^2,\;
D_{a28}=(1-r^2)(x_1'-1)x_2m_B^2,\nonumber \\
\end{eqnarray}
and the color factor
\begin{eqnarray}
&&C_N^{a28}=\varepsilon^{abc}\varepsilon^{a'b'c'}\left(C_1(T^j)_{bb'}(T^iT^j)_{ac'}(T^i)_{ca'}+C_2(T^j)_{bb'}(T^iT^j)_{cc'}(T^i)_{aa'}\right)=-{8\over
3}C_1+{8\over 3}C_2.\nonumber \\
\end{eqnarray}

For the hard amplitude of Fig.1(a29):
\begin{eqnarray}
&&
H^{a29,\alpha'\beta'\gamma'\rho'\alpha\beta\gamma\rho}(x_i,x_i',y,\textbf{k}_{T},\textbf{k}_{T}'
,\textbf{q}_{T},m_B,m_{\Lambda_c})=\nonumber
\\
&&\left[\varepsilon^{abc}\varepsilon^{a'b'c'}\left(C_1(T^jT^i)_{bb'}(T^jT^i)_{ac'}\delta_{ca'}+C_2(T^jT^i)_{bb'}(T^j)_{aa'}(T^i)_{cc'}\right)\right]g_s^4\nonumber
\\
&&{(\gamma_\lambda)_{\rho'\gamma'}[\gamma_\theta(\rlap /p-\rlap
/q_1+\rlap
/k_1')\gamma^\lambda]_{\beta\beta'}(O^\mu)_{\gamma\rho}[\gamma^\theta
(\rlap /q_1-\rlap /k_3-\rlap /k_1'+m_c)O^\mu]_{\alpha\alpha'}\over
(p-q_1-k_2+k_1')^2(q_2-k_3')^2(p-q_1+k_1')^2[(q_1-k_3-k_1')^2-m_c^2}\nonumber
\\
&&={C_N^{a29}
g_s^4(\gamma_\lambda)_{\rho'\gamma'}[\gamma_\theta(\rlap /p-\rlap
/q_1+\rlap
/k_1')\gamma^\lambda]_{\beta\beta'}(O^\mu)_{\gamma\rho}[\gamma^\theta
(\rlap /q_1-\rlap /k_3-\rlap /k_1'+m_c)O^\mu]_{\alpha\alpha'}\over
[A_{a29}+(\textbf{k}_{2T}-\textbf{k}_{1T}'+\textbf{q}_{T})^2][B_{a29}+(\textbf{q}_{T}+\textbf{k}_{3T}')^2]
[C_{a29}+(\textbf{k}_{1T}'
-\textbf{q}_{T})^2][D_{a29}+(\textbf{k}_{3T}+\textbf{k}_{1T}'-\textbf{q}_{T})^2]
}\nonumber
\\
\end{eqnarray}
with
\begin{eqnarray}
&&A_{a29}=(1-r^2)(x_1'-1)(x_2+y-1)m_B^2,\;\;
B_{a29}=(r^2-1)x_3'(y-1)m_B^2\nonumber \\
&&C_{a29}=(1-r^2)(x_1'-1)(-1+y)m_B^2,\;\;\;\;\;\;
D_{a29}=m_c^2+(1+(r^2-1)x_1')(x_3-y)m_B^2,\nonumber \\
\end{eqnarray}
and the color factor
\begin{eqnarray}
&&C_N^{a29}=\varepsilon^{abc}\varepsilon^{a'b'c'}\left(C_1(T^jT^i)_{bb'}(T^jT^i)_{ac'}\delta_{ca'}+C_2(T^jT^i)_{bb'}(T^j)_{aa'}(T^i)_{cc'}\right)=-{8\over
3}C_1+{8\over 3}C_2.\nonumber \\
\end{eqnarray}

For the hard amplitude of Fig.1(a30):
\begin{eqnarray}
&&
H^{a30,\alpha'\beta'\gamma'\rho'\alpha\beta\gamma\rho}(x_i,x_i',y,\textbf{k}_{T},\textbf{k}_{T}'
,\textbf{q}_{T},m_B,m_{\Lambda_c})=\nonumber
\\
&&\left[\varepsilon^{abc}\varepsilon^{a'b'c'}\left(C_1(T^jT^i)_{bb'}(T^i)_{ac'}(T^j)_{ca'}+C_2(T^jT^i)_{bb'}(T^jT^i)_{cc'}\delta_{aa'}\right)\right]g_s^4\nonumber
\\
&&{(\gamma_\lambda)_{\rho'\gamma'}[\gamma_\theta(\rlap /p-\rlap
/q_1+\rlap /k_1')\gamma^\lambda]_{\beta\beta'}[\gamma^\theta(\rlap
/q_1-\rlap /k_1-\rlap
/k_1')O_\mu]_{\gamma\rho}(O^\mu)_{\alpha\alpha'}\over
(p-q_1-k_2+k_1')^2(q_2-k_3')^2(p-q_1+k_1')^2(q_1-k_1-k_1')^2}\nonumber
\\
&&={C_N^{a30}
g_s^4(\gamma_\lambda)_{\rho'\gamma'}[\gamma_\theta(\rlap /p-\rlap
/q_1+\rlap /k_1')\gamma^\lambda]_{\beta\beta'}[\gamma^\theta(\rlap
/q_1-\rlap /k_1-\rlap
/k_1')O_\mu]_{\gamma\rho}(O^\mu)_{\alpha\alpha'}\over
[A_{a30}+(\textbf{k}_{2T}-\textbf{k}_{1T}'+\textbf{q}_{T})^2][B_{a30}+(\textbf{q}_{T}+\textbf{k}_{3T}')^2]
[C_{a30}+(\textbf{k}_{1T}'
-\textbf{q}_{T})^2][D_{a30}+(\textbf{k}_{1T}+\textbf{k}_{1T}'-\textbf{q}_{T})^2]
}\nonumber
\\
\end{eqnarray}
with
\begin{eqnarray}
&&A_{a30}=(1-r^2)(x_1'-1)(x_2+y-1)m_B^2,\;\;
B_{a30}=(r^2-1)x_3'(y-1)m_B^2\nonumber \\
&&C_{a30}=(1-r^2)(x_1'-1)(-1+y)m_B^2,\;\;\;\;\;\;
D_{a30}=(r^2-1)(x_1'-1)(x_1-y)m_B^2,\nonumber \\
\end{eqnarray}
and the color factor
\begin{eqnarray}
&&C_N^{a30}=\varepsilon^{abc}\varepsilon^{a'b'c'}\left(C_1(T^jT^i)_{bb'}(T^i)_{ac'}(T^j)_{ca'}+C_2(T^jT^i)_{bb'}(T^jT^i)_{cc'}\delta_{aa'}\right)=-{8\over
3}C_1+{8\over 3}C_2.\nonumber \\
\end{eqnarray}

For the hard amplitude of Fig.1(a31):
\begin{eqnarray}
&&
H^{a31,\alpha'\beta'\gamma'\rho'\alpha\beta\gamma\rho}(x_i,x_i',y,\textbf{k}_{T},\textbf{k}_{T}'
,\textbf{q}_{T},m_B,m_{\Lambda_c})=\nonumber
\\
&&\left[\varepsilon^{abc}\varepsilon^{a'b'c'}\left(C_1(T^jT^i)_{bb'}(T^jT^i)_{ac'}\delta_{ca'}+C_2(T^jT^i)_{bb'}(T^jT^i)_{cc'}\delta_{aa'}\right)\right]g_s^4\nonumber
\\
&&{(\gamma_\lambda)_{\rho'\gamma'}[\gamma_\theta(\rlap /p-\rlap
/q_1+\rlap
/k_1')\gamma^\lambda]_{\beta\beta'}[O_\mu\gamma^\theta(\rlap
/p-\rlap /k_2+\rlap
/k_1'+m_b)\gamma^\theta]_{\gamma\rho}(O^\mu)_{\alpha\alpha'}\over
(p-q_1-k_2+k_1')^2(q_2-k_3')^2(p-q_1+k_1')^2[(p-k_2+k_1')^2-m_b^2]}\nonumber
\\
&&={C_N^{a31}
g_s^4(\gamma_\lambda)_{\rho'\gamma'}[\gamma_\theta(\rlap /p-\rlap
/q_1+\rlap
/k_1')\gamma^\lambda]_{\beta\beta'}[O_\mu\gamma^\theta(\rlap
/p-\rlap /k_2+\rlap
/k_1'+m_b)\gamma^\theta]_{\gamma\rho}(O^\mu)_{\alpha\alpha'}\over
[A_{a31}+(\textbf{k}_{2T}-\textbf{k}_{1T}'+\textbf{q}_{T})^2][B_{a31}+(\textbf{q}_{T}+\textbf{k}_{3T}')^2]
[C_{a31}+(\textbf{k}_{1T}'
-\textbf{q}_{T})^2][D_{a31}+(\textbf{k}_{2T}-\textbf{k}_{1T}')^2]
}\nonumber
\\
\end{eqnarray}
with
\begin{eqnarray}
&&A_{a31}=(1-r^2)(x_1'-1)(x_2+y-1)m_B^2,\;\;
B_{a31}=(r^2-1)x_3'(y-1)m_B^2\nonumber \\
&&C_{a31}=(1-r^2)(x_1'-1)(-1+y)m_B^2,\;\;\;\;\;\;\;
D_{a31}=(1 +
x_1'(-1 + x_2) + r^2(-1 + x_1' + x_2 - x_1'x_2))m_B^2,\nonumber \\
\end{eqnarray}
and the color factor
\begin{eqnarray}
&&C_N^{a31}=\varepsilon^{abc}\varepsilon^{a'b'c'}\left(C_1(T^jT^i)_{bb'}(T^jT^i)_{ac'}\delta_{ca'}+C_2(T^jT^i)_{bb'}(T^jT^i)_{cc'}\delta_{aa'}\right)=-{8\over
3}C_1+{8\over 3}C_2.\nonumber \\
\end{eqnarray}

For the hard amplitude of Fig.1(a32):
\begin{eqnarray}
&&
H^{a32,\alpha'\beta'\gamma'\rho'\alpha\beta\gamma\rho}(x_i,x_i',y,\textbf{k}_{T},\textbf{k}_{T}'
,\textbf{q}_{T},m_B,m_{\Lambda_c})=\nonumber
\\
&&\left[\varepsilon^{abc}\varepsilon^{a'b'c'}\left(C_1(T^jT^i)_{bb'}(T^i)_{ac'}(T^j)_{ca'}+C_2(T^jT^i)_{bb'}(T^i)_{cc'}(T^j)_{aa'}\right)\right]g_s^4\nonumber
\\
&&{(\gamma_\lambda)_{\rho'\gamma'}[\gamma_\theta(\rlap /q_2-\rlap
/p'+\rlap
/k_1')\gamma^\lambda]_{\beta\beta'}(O_\mu)_{\gamma\rho}[O^\mu(\rlap
/p-\rlap /q_1-\rlap /k_2)\gamma^\theta]_{\alpha\alpha'}\over
(p-q_1-k_2+k_1')^2(q_2-k_3')^2(p-q_1-k_2)^2(q_2-p'+k_1')^2}\nonumber
\\
&&={C_N^{a32}
g_s^4(\gamma_\lambda)_{\rho'\gamma'}[\gamma_\theta(\rlap
/q_2-\rlap /p'+\rlap
/k_1')\gamma^\lambda]_{\beta\beta'}(O_\mu)_{\gamma\rho}[O^\mu(\rlap
/p-\rlap /q_1-\rlap /k_2)\gamma^\theta]_{\alpha\alpha'}\over
[A_{a32}+(\textbf{k}_{2T}-\textbf{k}_{1T}'+\textbf{q}_{T})^2][B_{a32}+(\textbf{q}_{T}+\textbf{k}_{3T}')^2]
[C_{a32}+(\textbf{k}_{2T}+\textbf{q}_{T})^2][D_{a32}+(-\textbf{q}_{T}+\textbf{k}_{1T}')^2]
}\nonumber
\\
\end{eqnarray}
with
\begin{eqnarray}
&&A_{a32}=(1-r^2)(x_1'-1)(x_2+y-1)m_B^2,\;
B_{a32}=(r^2-1)x_3'(y-1)m_B^2\nonumber \\
&&C_{a32}=(r^2-1)(x_2+y-1)m_B^2,\;\;\;\;\;\;\;\;\;\;\;\;\;
D_{a32}=(1-r^2)(x_1'-1)(y-1)m_B^2,\nonumber \\
\end{eqnarray}
and the color factor
\begin{eqnarray}
&&C_N^{a32}=\varepsilon^{abc}\varepsilon^{a'b'c'}\left(C_1(T^jT^i)_{bb'}(T^i)_{ac'}(T^j)_{ca'}+C_2(T^jT^i)_{bb'}(T^i)_{cc'}(T^j)_{aa'}\right)=-{8\over
3}C_1+{8\over 3}C_2.\nonumber \\
\end{eqnarray}

For the hard amplitude of Fig.1(a33):
\begin{eqnarray}
&&
H^{a33,\alpha'\beta'\gamma'\rho'\alpha\beta\gamma\rho}(x_i,x_i',y,\textbf{k}_{T},\textbf{k}_{T}'
,\textbf{q}_{T},m_B,m_{\Lambda_c})=\nonumber
\\
&&\left[\varepsilon^{abc}\varepsilon^{a'b'c'}\left(C_1(T^jT^i)_{bb'}(T^iT^j)_{ac'}\delta_{ca'}+C_2(T^jT^i)_{bb'}(T^j)_{cc'}(T^i)_{aa'}\right)\right]g_s^4\nonumber
\\
&&{(\gamma_\theta)_{\rho'\gamma'}[\gamma^\theta(-\rlap /q_2+\rlap
/k_2+\rlap
/k_3')\gamma_\lambda]_{\beta\beta'}(O_\mu)_{\gamma\rho}[\gamma^\lambda(\rlap
/q_1-\rlap /k_3-\rlap /k_1'+m_c)O^\mu]_{\alpha\alpha'}\over
(q_2-k_3')^2(q_2-p'-k_2+k_1')^2(-q_2+k_2+k_3')^2[(q_1-k_3-k_1')^2-m_c^2]}\nonumber
\\
&&={C_N^{a33}
g_s^4(\gamma_\theta)_{\rho'\gamma'}[\gamma^\theta(-\rlap
/q_2+\rlap /k_2+\rlap
/k_3')\gamma_\lambda]_{\beta\beta'}(O_\mu)_{\gamma\rho}[\gamma^\lambda(\rlap
/q_1-\rlap /k_3-\rlap /k_1'+m_c)O^\mu]_{\alpha\alpha'}\over
[A_{a33}+(\textbf{k}_{3T}'+\textbf{q}_{T})^2][B_{a33}+(\textbf{k}_{2T}-\textbf{k}_{1T}'+\textbf{q}_{T})^2]
[C_{a33}+(\textbf{k}_{2T}+\textbf{k}_{3T}'+\textbf{q}_{T})^2][D_{a33}+(-\textbf{q}_{T}+\textbf{k}_{1T}'+\textbf{k}_{3T})^2]
}\nonumber
\\
\end{eqnarray}
with
\begin{eqnarray}
&&A_{a33}=(r^2-1)x_3'(y-1)m_B^2,\;\;\;\;\;\;\;\;\;\;
B_{a33}=(1-r^2)(x_1'-1)(x_2+y-1)m_B^2\nonumber \\
&&C_{a33}=(r^2-1)(x_2+y-1)x_3'm_B^2,\;
D_{a33}=m_c^2+(1+(r^2-1)x_1')(x_3-y)m_B^2,\nonumber \\
\end{eqnarray}
and the color factor
\begin{eqnarray}
&&C_N^{a33}=\varepsilon^{abc}\varepsilon^{a'b'c'}\left(C_1(T^jT^i)_{bb'}(T^iT^j)_{ac'}\delta_{ca'}+C_2(T^jT^i)_{bb'}(T^j)_{cc'}(T^i)_{aa'}\right)={10\over
3}C_1+{8\over 3}C_2.\nonumber \\
\end{eqnarray}

For the hard amplitude of Fig.1(a34):
\begin{eqnarray}
&&
H^{a34,\alpha'\beta'\gamma'\rho'\alpha\beta\gamma\rho}(x_i,x_i',y,\textbf{k}_{T},\textbf{k}_{T}'
,\textbf{q}_{T},m_B,m_{\Lambda_c})=\nonumber
\\
&&\left[\varepsilon^{abc}\varepsilon^{a'b'c'}\left(C_1(T^jT^i)_{bb'}(T^j)_{ac'}(T^i)_{ca'}+C_2(T^jT^i)_{bb'}(T^iT^j)_{cc'}\delta_{aa'}\right)\right]g_s^4\nonumber
\\
&&{(\gamma_\theta)_{\rho'\gamma'}[\gamma^\theta(-\rlap /q_2+\rlap
/k_2+\rlap
/k_3')\gamma_\lambda]_{\beta\beta'}[\gamma^\lambda(\rlap
/q_1-\rlap /k_1-\rlap
/k_1')O_\mu]_{\gamma\rho}(O^\mu)_{\alpha\alpha'}\over
(q_2-k_3')^2(p-q_1-k_2+k_1')^2(-q_2+k_2+k_3')^2(q_1-k_1-k_1')^2}\nonumber
\\
&&={C_N^{a34}
g_s^4(\gamma_\theta)_{\rho'\gamma'}[\gamma^\theta(-\rlap
/q_2+\rlap /k_2+\rlap
/k_3')\gamma_\lambda]_{\beta\beta'}[\gamma^\lambda(\rlap
/q_1-\rlap /k_1-\rlap
/k_1')O_\mu]_{\gamma\rho}(O^\mu)_{\alpha\alpha'}\over
[A_{a34}+(\textbf{k}_{3T}'+\textbf{q}_{T})^2][B_{a34}+(\textbf{k}_{2T}-\textbf{k}_{1T}'+\textbf{q}_{T})^2]
[C_{a34}+(\textbf{k}_{2T}+\textbf{k}_{3T}'+\textbf{q}_{T})^2][D_{a34}+(-\textbf{q}_{T}+\textbf{k}_{1T}'+\textbf{k}_{1T})^2]
}\nonumber
\\
\end{eqnarray}
with
\begin{eqnarray}
&&A_{a34}=(r^2-1)x_3'(y-1)m_B^2,\;\;\;\;\;\;\;\;\;\;
B_{a34}=(1-r^2)(x_1'-1)(x_2+y-1)m_B^2\nonumber \\
&&C_{a34}=(r^2-1)(x_2+y-1)x_3'm_B^2,\;
D_{a34}=(r^2-1)(x_1'-1)(x_1-y)m_B^2,\nonumber \\
\end{eqnarray}
and the color factor
\begin{eqnarray}
&&C_N^{a34}=\varepsilon^{abc}\varepsilon^{a'b'c'}\left(C_1(T^jT^i)_{bb'}(T^j)_{ac'}(T^i)_{ca'}+C_2(T^jT^i)_{bb'}(T^iT^j)_{cc'}\delta_{aa'}\right)=-{8\over
3}C_1-{10\over 3}C_2.\nonumber \\
\end{eqnarray}

For the hard amplitude of Fig.1(a35):
\begin{eqnarray}
&&
H^{a35,\alpha'\beta'\gamma'\rho'\alpha\beta\gamma\rho}(x_i,x_i',y,\textbf{k}_{T},\textbf{k}_{T}'
,\textbf{q}_{T},m_B,m_{\Lambda_c})=\nonumber
\\
&&\left[\varepsilon^{abc}\varepsilon^{a'b'c'}\left(C_1(T^jT^i)_{bb'}(T^iT^j)_{ac'}\delta_{ca'}+C_2(T^jT^i)_{bb'}(T^iT^j)_{cc'}\delta_{aa'}\right)\right]g_s^4\nonumber
\\
&&{(\gamma_\theta)_{\rho'\gamma'}[\gamma^\theta(-\rlap /q_2+\rlap
/k_2+\rlap /k_3')\gamma_\lambda]_{\beta\beta'}[O_\mu(\rlap
/p-\rlap /k_2+\rlap
/k_1'+m_b)\gamma^\lambda]_{\gamma\rho}(O^\mu)_{\alpha\alpha'}\over
(q_2-k_3')^2(q_2-p'-k_2+k_1')^2(-q_2+k_2+k_3')^2[(p-k_2+k_1')^2-m_b^2]}\nonumber
\\
&&={C_N^{a35}
g_s^4(\gamma_\theta)_{\rho'\gamma'}[\gamma^\theta(-\rlap
/q_2+\rlap /k_2+\rlap
/k_3')\gamma_\lambda]_{\beta\beta'}[O_\mu(\rlap /p-\rlap
/k_2+\rlap
/k_1'+m_b)\gamma^\lambda]_{\gamma\rho}(O^\mu)_{\alpha\alpha'}\over
[A_{a35}+(\textbf{k}_{3T}'+\textbf{q}_{T})^2][B_{a35}+(\textbf{k}_{2T}-\textbf{k}_{1T}'+\textbf{q}_{T})^2]
[C_{a35}+(\textbf{k}_{2T}+\textbf{k}_{3T}'+\textbf{q}_{T})^2][D_{a35}+(-\textbf{k}_{1T}'+\textbf{k}_{2T})^2]
}\nonumber
\\
\end{eqnarray}
with
\begin{eqnarray}
&&A_{a35}=(r^2-1)x_3'(y-1)m_B^2,\;\;\;\;\;\;\;\;\;\;
B_{a35}=(1-r^2)(x_1'-1)(x_2+y-1)m_B^2\nonumber \\
&&C_{a35}=(r^2-1)(x_2+y-1)x_3'm_B^2,\;
D_{a35}=(1 + x_1'(-1 + x_2)
+r^2(-1 + x_1' + x_2 - x_1'x_2))m_B^2,\nonumber \\
\end{eqnarray}
and the color factor
\begin{eqnarray}
&&C_N^{a35}=\varepsilon^{abc}\varepsilon^{a'b'c'}\left(C_1(T^jT^i)_{bb'}(T^j)_{ac'}(T^i)_{ca'}+C_2(T^jT^i)_{bb'}(T^iT^j)_{cc'}\delta_{aa'}\right)={10\over
3}C_1-{10\over 3}C_2.\nonumber \\
\end{eqnarray}

For the hard amplitude of Fig.1(a36):
\begin{eqnarray}
&&
H^{a36,\alpha'\beta'\gamma'\rho'\alpha\beta\gamma\rho}(x_i,x_i',y,\textbf{k}_{T},\textbf{k}_{T}'
,\textbf{q}_{T},m_B,m_{\Lambda_c})=\nonumber
\\
&&\left[\varepsilon^{abc}\varepsilon^{a'b'c'}\left(C_1(T^jT^i)_{bb'}(T^j)_{ac'}(T^i)_{ca'}+C_2(T^jT^i)_{bb'}(T^j)_{cc'}(T^i)_{aa'}\right)\right]g_s^4\nonumber
\\
&&{(\gamma_\theta)_{\rho'\gamma'}[\gamma^\theta(-\rlap /q_2+\rlap
/k_2+\rlap
/k_3')\gamma_\lambda]_{\beta\beta'}(O_\mu)_{\gamma\rho}[O^\mu(\rlap
/p-\rlap /q_1-\rlap /k_2+m_b)\gamma^\lambda]_{\alpha\alpha'}\over
(q_2-k_3')^2(-p+q_1+k_2-k_1')^2(-q_2+k_2+k_3')^2(p-q_1-k_2)^2}\nonumber
\\
&&={C_N^{a36}
g_s^4(\gamma_\theta)_{\rho'\gamma'}[\gamma^\theta(-\rlap
/q_2+\rlap /k_2+\rlap
/k_3')\gamma_\lambda]_{\beta\beta'}(O_\mu)_{\gamma\rho}[O^\mu(\rlap
/p-\rlap /q_1-\rlap /k_2+m_b)\gamma^\lambda]_{\alpha\alpha'}\over
[A_{a36}+(\textbf{k}_{3T}'+\textbf{q}_{T})^2][B_{a36}+(-\textbf{k}_{2T}+\textbf{k}_{1T}'-\textbf{q}_{T})^2]
[C_{a36}+(\textbf{k}_{2T}+\textbf{k}_{3T}'+\textbf{q}_{T})^2][D_{a36}+(\textbf{k}_{2T}+\textbf{q}_{T})^2]
}\nonumber
\\
\end{eqnarray}
with
\begin{eqnarray}
&&A_{a36}=(r^2-1)x_3'(y-1)m_B^2,\;\;\;\;\;\;\;\;\;\;
B_{a36}=(1-r^2)(x_1'-1)(x_2+y-1)m_B^2\nonumber \\
&&C_{a36}=(r^2-1)(x_2+y-1)x_3'm_B^2,\;
D_{a36}=(r^2-1)(x_2+y-1)m_B^2,\nonumber \\
\end{eqnarray}
and the color factor
\begin{eqnarray}
&&C_N^{a36}=\varepsilon^{abc}\varepsilon^{a'b'c'}\left(C_1(T^jT^i)_{bb'}(T^j)_{ac'}(T^i)_{ca'}+C_2(T^jT^i)_{bb'}(T^j)_{cc'}(T^i)_{aa'}\right)=-{8\over
3}C_1+{8\over 3}C_2.\nonumber \\
\end{eqnarray}

For the hard amplitude of Fig.2(b1):
\begin{eqnarray}
&&
H^{b1,\alpha'\beta'\gamma'\rho'\alpha\beta\gamma\rho}(x_i,x_i',y,\textbf{k}_{T},\textbf{k}_{T}'
,\textbf{q}_{T},m_B,m_{\Lambda_c})=H^{a6,\alpha'\beta'\gamma'\rho'\alpha\beta\gamma\rho}(x_i,x_i',y,\textbf{k}_{T},\textbf{k}_{T}'
,\textbf{q}_{T},m_B,m_{\Lambda_c})\nonumber
\\
&&={C_N^{b1} g_s^4(\gamma_\theta(\rlap /k_2+\rlap /k_2'-\rlap
/q_2)\gamma_\lambda)_{\rho'\gamma'}(\gamma^\theta)_{\beta
\beta'}[O_\mu(\rlap /p-\rlap /k_2+\rlap /
k_1'+m_b)\gamma^\lambda]_{\gamma\rho}(O^\mu)_{\alpha\alpha'}\over
[A_{b1}+(\textbf{k}_{2T}+\textbf{k}_{2T}')^2][B_{b1}+(\textbf{q}_{T}+\textbf{k}_{2T}-\textbf{k}_{1T}')^2][C_{b1}+(\textbf{k}_{2T}+\textbf{k}_{2T}'+\textbf{q}_{T})^2]
[D_{b1}+(\textbf{k}_{2T}-\textbf{k}_{1T}')^2] }\nonumber
\\
\end{eqnarray}
with
\begin{eqnarray}
&&A_{b1}=x_2x_2'(r^2-1)m_B^2,\;\;\;\;\;\;\;\;\;\;\;\;\;\;\;\;\;
B_{b1}=x_2(-(x_1'-1)r^2+x_1'-y)m_B^2,\nonumber \\
&&C_{b1}=x_2((r^2-1)x_2'-y+1)m_B^2,\;
D_{b1}=m_b^2-(r^2(x_1'-1)-x_1')(x_2-1)m_B^2\nonumber \\
\end{eqnarray}
and the color factor
\begin{eqnarray}
&&C_N^{b1}=C_N^{a6}.
\end{eqnarray}

For the hard amplitude of Fig.2(b2):
\begin{eqnarray}
&&
H^{b2,\alpha'\beta'\gamma'\rho'\alpha\beta\gamma\rho}(x_i,x_i',y,\textbf{k}_{T},\textbf{k}_{T}'
,\textbf{q}_{T},m_B,m_{\Lambda_c})=H^{a7,\alpha'\beta'\gamma'\rho'\alpha\beta\gamma\rho}(x_i,x_i',y,\textbf{k}_{T},\textbf{k}_{T}'
,\textbf{q}_{T},m_B,m_{\Lambda_c})\nonumber
\\
&&={C_N^{b2} g_s^4(\gamma_\lambda(-\rlap /p'-\rlap /k_2+\rlap
/k_1')\gamma_\theta)_{\rho'\gamma'}(\gamma^\theta)_{\beta
\beta'}[O_\mu(\rlap /p-\rlap /k_2+\rlap /
k_1'+m_b)\gamma^\lambda]_{\gamma\rho}(O^\mu)_{\alpha\alpha'}\over
[A_{b2}+(\textbf{k}_{2T}+\textbf{k}_{2T}')^2][B_{b2}+(\textbf{q}_{T}+\textbf{k}_{2T}-\textbf{k}_{1T}')^2][C_{b2}+(\textbf{k}_{2T}-\textbf{k}_{1T}')^2]
[D_{b2}+(\textbf{k}_{2T}-\textbf{k}_{1T}')^2] }\nonumber
\\
\end{eqnarray}
with
\begin{eqnarray}
&&A_{b2}=x_2x_2'(r^2-1)m_B^2,\;\;\;\;\;\;\;\;\;\;\;\;\;
B_{b2}=x_2(-(x'_1-1)r^2+x_1'-y)m_B^2,\nonumber \\
&&C_{b2}=-x_2(r^2-1)(x_1'-1)m_B^2,\;
D_{b2}=m_b^2-(x_2-1)(r^2(x_1'-1)-x_1')m_B^2\nonumber \\
\end{eqnarray}
and the color factor
\begin{eqnarray}
&&C_N^{b2}=C_N^{a7}.\nonumber \\
\end{eqnarray}

For the hard amplitude of Fig.2(b3):
\begin{eqnarray}
&&
H^{b3,\alpha'\beta'\gamma'\rho'\alpha\beta\gamma\rho}(x_i,x_i',y,\textbf{k}_{T},\textbf{k}_{T}'
,\textbf{q}_{T},m_B,m_{\Lambda_c})=H^{a13,\alpha'\beta'\gamma'\rho'\alpha\beta\gamma\rho}(x_i,x_i',y,\textbf{k}_{T},\textbf{k}_{T}'
,\textbf{q}_{T},m_B,m_{\Lambda_c})\nonumber
\\
&&={C_N^{b3} g_s^4[\gamma_\theta(\rlap /k_2+\rlap /k_2'-\rlap
/q_2)\gamma_\lambda]_{\rho'\gamma'}(\gamma^\theta)_{\beta
\beta'}(O_\mu)_{\gamma \rho}[\gamma^\lambda(\rlap /q_1-\rlap
/k_3-\rlap /k_1'+m_c)O^\mu]_{\alpha\alpha'}\over
[A_{b3}+(\textbf{k}_{2T}+\textbf{k}_{2T}')^2][B_{b3}+(\textbf{q}_{T}+\textbf{k}_{2T}-\textbf{k}_{1T}')^2][C_{b3}+(\textbf{k}_{2T}+\textbf{k}_{2T}'+\textbf{q}_{T})^2]
[D_{b3}+(\textbf{k}_{1T}'+\textbf{k}_{3T}-\textbf{q}_{T})^2]
}\nonumber
\\
\end{eqnarray}
with
\begin{eqnarray}
&&A_{b3}=x_2x_2'(r^2-1)m_B^2,\;\;\;\;\;\;\;\;\;\;\;\;\;\;\;\;
B_{b3}=x_2(-(x_1'-1)r^2+x_1'-y)m_B^2,\nonumber \\
&&C_{b3}=x_2((r^2-1)x_2'-y+1)m_B^2,\;
D_{b3}=m_c^2+(x_3-1)((r^2-1)x_1'+y)m_B^2\nonumber \\
\end{eqnarray}
and the color factor
\begin{eqnarray}
&&C_N^{b3}=C_N^{a13}.
\end{eqnarray}

For the hard amplitude of Fig.2(b4):
\begin{eqnarray}
&&
H^{b4,\alpha'\beta'\gamma'\rho'\alpha\beta\gamma\rho}(x_i,x_i',y,\textbf{k}_{T},\textbf{k}_{T}'
,\textbf{q}_{T},m_B,m_{\Lambda_c})=H^{a14,\alpha'\beta'\gamma'\rho'\alpha\beta\gamma\rho}(x_i,x_i',y,\textbf{k}_{T},\textbf{k}_{T}'
,\textbf{q}_{T},m_B,m_{\Lambda_c})\nonumber
\\
&&={C_N^{b4} g_s^4[\gamma_\lambda(-\rlap /p'-\rlap /k_2+\rlap
/k_1')\gamma_\theta]_{\rho'\gamma'}(\gamma^\theta)_{\beta
\beta'}(O_\mu)_{\gamma \rho}[\gamma^\lambda(\rlap /q_1-\rlap
/k_3-\rlap /k_1'+m_c)O^\mu]_{\alpha\alpha'}\over
[A_{b4}+(\textbf{k}_{2T}+\textbf{k}_{2T}')^2][B_{b4}+(\textbf{q}_{T}+\textbf{k}_{2T}-\textbf{k}_{1T}')^2][C_{b4}+(\textbf{k}_{2T}-\textbf{k}_{1T}')^2]
[D_{b4}+(\textbf{k}_{1T}'+\textbf{k}_{3T}-\textbf{q}_{T})^2]
}\nonumber
\\
\end{eqnarray}
with
\begin{eqnarray}
&&A_{b4}=x_2x_2'(r^2-1)m_B^2,\;\;\;\;\;\;\;\;\;\;\;\;\;
B_{b4}=x_2(-(x_1'-1)r^2+x_1'-y)m_B^2,\nonumber \\
&&C_{b4}=-x_2(r^2-1)(x_1'-1)m_B^2,\;
D_{b4}=m_c^2+(x_3-1)((r^2-1)x_1'+y)m_B^2\nonumber \\
\end{eqnarray}
and the color factor
\begin{eqnarray}
&&C_N^{b4}=C_N^{a14}.
\end{eqnarray}

For the hard amplitude of Fig.2(b5):
\begin{eqnarray}
&&
H^{b5,\alpha'\beta'\gamma'\rho'\alpha\beta\gamma\rho}(x_i,x_i',y,\textbf{k}_{T},\textbf{k}_{T}'
,\textbf{q}_{T},m_B,m_{\Lambda_c})=H^{a20,\alpha'\beta'\gamma'\rho'\alpha\beta\gamma\rho}(x_i,x_i',y,\textbf{k}_{T},\textbf{k}_{T}'
,\textbf{q}_{T},m_B,m_{\Lambda_c})\nonumber
\\
&&={C_N^{b5} g_s^4[\gamma_\theta(\rlap /k_2+\rlap /k_2'-\rlap
/q_2)\gamma_\lambda]_{\rho'\gamma'}(\gamma^\theta)_{\beta
\beta'}[\gamma^\lambda(\rlap /q_1-\rlap /k_1-\rlap
/k_1')O^\mu]_{\gamma\rho}(O_\mu)_{\alpha\alpha'}\over
[A_{b5}+(\textbf{k}_{2T}
+\textbf{k}_{2T}')^2][B_{b5}+(\textbf{k}_{2T}-\textbf{k}_{1T}'+\textbf{q}_{T})^2][C_{b5}+(\textbf{k}_{2T}+\textbf{k}_{2T}'+\textbf{q}_{T})^2]
[D_{b5}+(-\textbf{q}_{T}+\textbf{k}_{1T}+\textbf{k}_{1T}')^2]}\nonumber
\\
\end{eqnarray}
with
\begin{eqnarray}
&&A_{b5}=x_2x_2'(r^2-1)m_B^2,\;\;\;\;\;\;\;\;\;\;\;\;\;\;\;\;\;\;\;\;
B_{b5}=m_B^2(r^2(x_2-x_1'x_2)+x_2(x_1'-y)),\nonumber \\
&&C_{b5}=m_B^2x_2(1+(-1+r^2)x_2'-y),\;
D_{b5}=m_B^2(r^2(1-x_1+(-1+x_1)x_1')-(-1+x_1)(x_1'-y)),\nonumber \\
\end{eqnarray}
and the color factor
\begin{eqnarray}
&&C_N^{b5}=C_N^{a20}.
\end{eqnarray}

For the hard amplitude of Fig.2(b6):
\begin{eqnarray}
&&
H^{b6,\alpha'\beta'\gamma'\rho'\alpha\beta\gamma\rho}(x_i,x_i',y,\textbf{k}_{T},\textbf{k}_{T}'
,\textbf{q}_{T},m_B,m_{\Lambda_c})=H^{a21,\alpha'\beta'\gamma'\rho'\alpha\beta\gamma\rho}(x_i,x_i',y,\textbf{k}_{T},\textbf{k}_{T}'
,\textbf{q}_{T},m_B,m_{\Lambda_c})\nonumber
\\
&&={C_N^{b6} g_s^4[\gamma_\lambda(-\rlap /p'-\rlap /k_2+\rlap
/k_1')\gamma_\theta]_{\rho'\gamma'}(\gamma^\theta)_{\beta
\beta'}[\gamma^\lambda(\rlap /q_1-\rlap /k_1-\rlap
/k_1')O^\mu]_{\gamma\rho}(O_\mu)_{\alpha\alpha'}\over
[A_{b6}+(\textbf{k}_{2T}+\textbf{k}_{2T}')^2][B_{b6}+(\textbf{q}_{T}+\textbf{k}_{2T}-\textbf{k}_{1T}')^2]
[C_{b6}+(\textbf{k}_{2T}
-\textbf{k}_{1T}')^2][D_{b6}+(\textbf{k}_{1T}+\textbf{k}_{1T}'-\textbf{q}_{T})^2]}\nonumber
\\
\end{eqnarray}
with
\begin{eqnarray}
&&A_{b6}=x_2x_2'(r^2-1)m_B^2,\;\;\;\;\;\;\;\;\;\;\;\;\;\;\;\;\;\;\;
B_{b6}=m_B^2(r^2(x_2-x_1'x_2)+x_2(x_1'-y))\nonumber \\
&&C_{b6}=-m_B^2x_2(-1+r^2)(-1+x_1'),\;
D_{b6}=m_B^2(r^2(1-x_1+(-1+x_1)x_1')-(-1+x_1)(x_1'-y)),\nonumber \\
\end{eqnarray}
and the color factor
\begin{eqnarray}
&&C_N^{b6}=C_N^{a21}.\nonumber \\
\end{eqnarray}

For the hard amplitude of Fig.2(b7):
\begin{eqnarray}
&&
H^{b7,\alpha'\beta'\gamma'\rho'\alpha\beta\gamma\rho}(x_i,x_i',y,\textbf{k}_{T},\textbf{k}_{T}'
,\textbf{q}_{T},m_B,m_{\Lambda_c})=H^{a27,\alpha'\beta'\gamma'\rho'\alpha\beta\gamma\rho}(x_i,x_i',y,\textbf{k}_{T},\textbf{k}_{T}'
,\textbf{q}_{T},m_B,m_{\Lambda_c})\nonumber
\\
&&={C_N^{b7} g_s^4[\gamma_\theta(\rlap /k_2+\rlap /k_2'-\rlap
/q_2)\gamma_\lambda]_{\rho'\gamma'}(\gamma^\theta)_{\beta
\beta'}(O^\mu)_{\gamma\rho}[O^\mu(-\rlap /p'+\rlap /q_2-\rlap
/k_2)\gamma_\lambda]_{\alpha\alpha'}\over
[A_{b7}+(\textbf{k}_{2T}+\textbf{k}_{2T}')^2][B_{b7}+(\textbf{q}_{T}+\textbf{k}_{2T}-\textbf{k}_{1T}')^2]
[C_{b7}+(\textbf{k}_{2T}
+\textbf{q}_{T})^2][D_{b7}+(\textbf{k}_{2T}+\textbf{k}_{2T}'+\textbf{q}_{T})^2]
}\nonumber
\\
\end{eqnarray}
with
\begin{eqnarray}
&&A_{b7}=x_2x_2'(r^2-1)m_B^2,\;\;\;\;
B_{b7}=m_B^2x_2(-r^2(-1+x_1')+x_1'-y)\nonumber \\
&&C_{b7}=m_B^2x_2(-y+r^2),\;\;\;\;
D_{b7}=m_B^2x_2(1+x_2'(-1+r^2)-y),\nonumber \\
\end{eqnarray}
and the color factor
\begin{eqnarray}
&&C_N^{b7}=C_N^{a27}.\nonumber \\
\end{eqnarray}

For the hard amplitude of Fig.2(b8):
\begin{eqnarray}
&&
H^{b8,\alpha'\beta'\gamma'\rho'\alpha\beta\gamma\rho}(x_i,x_i',y,\textbf{k}_{T},\textbf{k}_{T}'
,\textbf{q}_{T},m_B,m_{\Lambda_c})=H^{a28,\alpha'\beta'\gamma'\rho'\alpha\beta\gamma\rho}(x_i,x_i',y,\textbf{k}_{T},\textbf{k}_{T}'
,\textbf{q}_{T},m_B,m_{\Lambda_c})\nonumber
\\
&&={C_N^{a28} g_s^4[\gamma_\lambda(-\rlap /p'-\rlap /k_2+\rlap
/k_1')\gamma_\lambda]_{\rho'\gamma'}(\gamma^\theta)_{\beta\beta'}(O^\mu)_{\gamma\rho}[O^\mu(-\rlap
/p'+\rlap /q_2-\rlap /k_2)\gamma_\lambda]_{\alpha\alpha'}\over
[A_{b8}+(\textbf{k}_{2T}+\textbf{k}_{2T}')^2][B_{b8}+(\textbf{q}_{T}+\textbf{k}_{2T}-\textbf{k}_{1T}')^2]
[C_{b8}+(\textbf{k}_{2T}
+\textbf{q}_{T})^2][D_{b8}+(\textbf{k}_{2T}-\textbf{k}_{1T}')^2]
}\nonumber
\\
\end{eqnarray}
with
\begin{eqnarray}
&&A_{b8}=x_2x_2'(r^2-1)m_B^2,\;\;\;\;
B_{b8}=m_B^2x_2(-r^2(-1+x_1')+x_1'-y)\nonumber \\
&&C_{b8}=m_B^2x_2(-y+r^2),\;\;\;\;
D_{b8}=-m_B^2x_2(-1+r^2)(-1+x_1'),\nonumber \\
\end{eqnarray}
and the color factor
\begin{eqnarray}
&&C_N^{b8}=C_N^{a28}.
\end{eqnarray}

For the hard amplitude of Fig.3(c1):
\begin{eqnarray}
&&
H^{c1,\alpha'\beta'\gamma'\rho'\alpha\beta\gamma\rho}(x_i,x_i',y,\textbf{k}_{T},\textbf{k}_{T}'
,\textbf{q}_{T},m_B,m_{\Lambda_c})=\nonumber
\\
&&i\left[\varepsilon^{abc}\varepsilon^{a'b'c'}f_{jil}\left(C_1(T^jT^i)_{ac'}(T^l)_{bb'}\delta_{ca'}+C_2(T^jT^i)_{cc'}(T^l)_{bb'}\delta_{aa'}\right)\right]g_s^4\nonumber
\\
&&{(\gamma_\lambda)_{\rho'\gamma'}(\gamma_\eta)_{\beta\beta'}[O_\mu(\rlap
/p-\rlap /k_2-\rlap
/k_1'+m_b)\gamma_\theta]_{\gamma\rho}(O^\mu)_{\alpha\alpha'}\over
(k_2+k_2')^2(-q_2+k_3')^2(p-q_1-k_2-k_1')^2[(p-k_2-k_1')^2-m_b^2]}\nonumber
\\
&&[g^{\theta\lambda}(p-q_1-k_2-k_1'-(-q_2+k_3'))^\eta+
g^{\lambda\eta}(-q_2+k_3'-(k_2+k_2'))^\theta+g^{\eta\theta}(k_2+k_2'-(p-k_2-k_1'))^\lambda]\nonumber
\\
&&={iC_N^{c1}
g_s^4(\gamma_\lambda)_{\rho'\gamma'}(\gamma_\eta)_{\beta\beta'}[O_\mu(\rlap
/p-\rlap /k_2-\rlap
/k_1'+m_b)\gamma_\theta]_{\gamma\rho}(O^\mu)_{\alpha\alpha'}\over
[A_{c1}+(\textbf{k}_{2T}'+\textbf{k}_{2T}')^2][B_{c1}+(\textbf{k}_{3T}'+\textbf{q}_{T})^2]
[C_{c1}+(\textbf{k}_{2T}+\textbf{k}_{1T}'+\textbf{q}_{T})^2][D_{c1}+(\textbf{k}_{2T}+\textbf{k}_{1T}')^2]}\nonumber
\\
&&[g^{\theta\lambda}(p-q_1-k_2-k_1'-(-q_2+k_3'))^\eta+
g^{\lambda\eta}(-q_2+k_3'-(k_2+k_2'))^\theta+g^{\eta\theta}(k_2+k_2'-(p-k_2-k_1'))^\lambda]
\nonumber
\\
\end{eqnarray}
with
\begin{eqnarray}
&&A_{c1}=x_2x_2'(r^2-1)m_B^2,\;\;\;\;\;\;\;\;\;\;\;\;\;\;\;\;\;\;\;\;\;\;\;\;\;
B_{c1}=(r^2-1)x_3'(y-1)m_B^2\nonumber \\
&&C_{c1}=(r^2-1)(x_2+y-1)(x_1'+1)m_B^2,\;
D_{c1}=m_b^2+(r^2(x_1'+1)-x_1')(x_2-1)m_B^2,\nonumber \\
\end{eqnarray}
and the color factor
\begin{eqnarray}
&&C_N^{c1}=\varepsilon^{abc}\varepsilon^{a'b'c'}f_{jil}\left(C_1(T^jT^i)_{ac'}(T^l)_{bb'}\delta_{ca'}+C_2(T^jT^i)_{cc'}(T^l)_{bb'}\delta_{aa'}\right)=6iC_1-6iC_2.\nonumber \\
\end{eqnarray}

For the hard amplitude of Fig.3(c2):
\begin{eqnarray}
&&
H^{c2,\alpha'\beta'\gamma'\rho'\alpha\beta\gamma\rho}(x_i,x_i',y,\textbf{k}_{T},\textbf{k}_{T}'
,\textbf{q}_{T},m_B,m_{\Lambda_c})=\nonumber
\\
&&i\left[\varepsilon^{abc}\varepsilon^{a'b'c'}f_{jil}\left(C_1(T^jT^i)_{ac'}(T^l)_{bb'}\delta_{ca'}+C_2(T^i)_{cc'}(T^l)_{bb'}(T^j)_{aa'}\right)\right]g_s^4\nonumber
\\
&&{(\gamma_\lambda)_{\rho'\gamma'}(\gamma_\eta)_{\beta\beta'}(O_\mu)_{\gamma\rho}[\gamma_\theta(\rlap
/q_1-\rlap /k_3-\rlap /k_1'+m_c)O^\mu]_{\alpha\alpha'}\over
(k_2+k_2')^2(-q_2+k_3')^2(p-q_1-k_2-k_1')^2[(q_1-k_3-k_1')^2-m_c^2]}\nonumber
\\
&&[g^{\theta\lambda}(p-q_1-k_2-k_1'-(-q_2+k_3'))^\eta+
g^{\lambda\eta}(-q_2+k_3'-(k_2+k_2'))^\theta+g^{\eta\theta}(k_2+k_2'-(p-k_2-k_1'))^\lambda]\nonumber
\\
&&={iC_N^{c2}
g_s^4(\gamma_\lambda)_{\rho'\gamma'}(\gamma_\eta)_{\beta\beta'}(O_\mu)_{\gamma\rho}[\gamma_\theta(\rlap
/q_1-\rlap /k_3-\rlap /k_1'+m_c)O^\mu]_{\alpha\alpha'}\over
[A_{c2}+(\textbf{k}_{2T}'+\textbf{k}_{2T}')^2][B_{c2}+(\textbf{k}_{3T}'+\textbf{q}_{T})^2]
[C_{c2}+(\textbf{k}_{2T}+\textbf{k}_{1T}'+\textbf{q}_{T})^2][D_{c2}+(\textbf{k}_{3T}+\textbf{k}_{1T}'-\textbf{q}_{T})^2]}\nonumber
\\
&&[g^{\theta\lambda}(p-q_1-k_2-k_1'-(-q_2+k_3'))^\eta+
g^{\lambda\eta}(-q_2+k_3'-(k_2+k_2'))^\theta+g^{\eta\theta}(k_2+k_2'-(p-k_2-k_1'))^\lambda]
\nonumber
\\
\end{eqnarray}
with
\begin{eqnarray}
&&A_{c2}=x_2x_2'(r^2-1)m_B^2,\;\;\;\;\;\;\;\;\;\;\;\;\;\;\;\;\;\;\;\;\;\;\;\;\;
B_{c2}=(r^2-1)x_3'(y-1)m_B^2\nonumber \\
&&C_{c2}=(r^2-1)(x_2+y-1)(x_1'+1)m_B^2,\;
D_{c2}=m_c^2+((r^2-1)x_1'+1)(x_3-y)m_B^2,\nonumber \\
\end{eqnarray}
and the color factor
\begin{eqnarray}
&&C_N^{c2}=\varepsilon^{abc}\varepsilon^{a'b'c'}f_{jil}\left(C_1(T^jT^i)_{ac'}(T^l)_{bb'}\delta_{ca'}+C_2(T^i)_{cc'}(T^l)_{bb'}(T^j)_{aa'}\right)=6iC_1.\nonumber \\
\end{eqnarray}

\noindent{\bf Appendix C: The maximal one of $t_{1,2}$}

The  maximal one of the hard scales $t_1^i$ and $t_2^i$ for
diagrams (a1)$\sim$(a36) in Fig.~1, (b1)$\sim$(b8) in Fig.~2 and
(c1), (c2) in Fig.~3 are given below in the table.

$t_1^{a1}=max\{\sqrt{|A_{a1}|},{1\over
|\mathbf{b_2}-\mathbf{b_3'}+\mathbf{b_q}|},\omega,\omega'\},\;\;\;\;\;\;\;\;\;
t_2^{a1}=max\{\sqrt{|B_{a1}|},{1\over
|\mathbf{b_3'}|},\omega,\omega'\}$

$t_1^{a2}=max\{\sqrt{|A_{a2}|},{1\over
|\mathbf{b_2}|},\omega,\omega'\},\;\;\;\;\;\;\;\;\;\;\;\;\;\;\;\;\;\;\;\;
t_2^{a2}=max\{\sqrt{|B_{a2}|},{1\over
|\mathbf{b_q}|},\omega,\omega'\}$

$t_1^{a3}=max\{\sqrt{|A_{a3}|},{1\over
|\mathbf{b_2}|},\omega,\omega'\},\;\;\;\;\;\;\;\;\;\;\;\;\;\;\;\;\;\;\;\;
t_2^{a3}=max\{\sqrt{|B_{a3}|},{1\over
|\mathbf{b_q}|},\omega,\omega'\}$

$t_1^{a4}=max\{\sqrt{|A_{a4}|},{1\over
|\mathbf{b_2}|},\omega,\omega'\},\;\;\;\;\;\;\;\;\;\;\;\;\;\;\;\;\;\;\;\;
t_2^{a4}=max\{\sqrt{|B_{a4}|},{1\over
|\mathbf{b_q}|},\omega,\omega'\}$

$t_1^{a5}=max\{\sqrt{|A_{a5}|},{1\over
|\mathbf{b_2'}|},\omega,\omega'\},\;\;\;\;\;\;\;\;\;\;\;\;\;\;\;\;\;\;\;\;
t_2^{a5}=max\{\sqrt{|B_{a5}|},{1\over
|\mathbf{b_q}|},\omega,\omega'\}$

$t_1^{a6}=max\{\sqrt{|A_{a6}|},{1\over
|\mathbf{b_2}-\mathbf{b_q}|},\omega,\omega'\},\;\;\;\;\;\;\;\;\;\;\;\;\;\;
t_2^{a6}=max\{\sqrt{|B_{a6}|},{1\over
|\mathbf{b_1'}|},\omega,\omega'\}$

$t_1^{a7}=max\{\sqrt{|A_{a7}|},{1\over
|\mathbf{b_2}+\mathbf{b_1'}|},\omega,\omega'\},\;\;\;\;\;\;\;\;\;\;\;\;\;\;
t_2^{a7}=max\{\sqrt{|B_{a7}|},{1\over
|\mathbf{b_q}|},\omega,\omega'\}$

$t_1^{a8}=max\{\sqrt{|A_{a8}|},{1\over
|\mathbf{b_2}-\mathbf{b_3'}+\mathbf{b_q}|},\omega,\omega'\},\;\;\;\;\;\;\;\;
t_2^{a8}=max\{\sqrt{|B_{a8}|},{1\over
|\mathbf{b_1}-\mathbf{b_3'}|},\omega,\omega'\}$

$t_1^{a9}=max\{\sqrt{|A_{a9}|},{1\over
|\mathbf{b_1}|},\omega,\omega'\},\;\;\;\;\;\;\;\;\;\;\;\;\;\;\;\;\;\;\;\;
t_2^{a9}=max\{\sqrt{|B_{a9}|},{1\over
|\mathbf{b_1}-\mathbf{b_3'}|},\omega,\omega'\}$

$t_1^{a10}=max\{\sqrt{|A_{a10}|},{1\over
|\mathbf{b_2}|},\omega,\omega'\},\;\;\;\;\;\;\;\;\;\;\;\;\;\;\;\;\;
t_2^{a10}=max\{\sqrt{|B_{a10}|},{1\over
|\mathbf{b_3}+\mathbf{b_q}|},\omega,\omega'\}$

$t_1^{a11}=max\{\sqrt{|A_{a11}|},{1\over
|\mathbf{b_2}|},\omega,\omega'\},\;\;\;\;\;\;\;\;\;\;\;\;\;\;\;\;\;
t_2^{a11}=max\{\sqrt{|B_{a11}|},{1\over
|\mathbf{b_1}-\mathbf{b_2}+\mathbf{b_2'}-\mathbf{b_q}|},\omega,\omega'\}$

$t_1^{a12}=max\{\sqrt{|A_{a12}|},{1\over
|\mathbf{b_2}+\mathbf{b_3'}-\mathbf{b_q}|},\omega,\omega'\},\;\;\;\;\;\;
t_2^{a12}=max\{\sqrt{|B_{a12}|},{1\over
|\mathbf{b_1}-\mathbf{b_q}|},\omega,\omega'\}$

$t_1^{a13}=max\{\sqrt{|A_{a13}|},{1\over
|\mathbf{b_2}-\mathbf{b_3}-\mathbf{b_q}|},\omega,\omega'\},\;\;\;\;\;\;
t_2^{a13}=max\{\sqrt{|B_{a13}|},{1\over
|\mathbf{b_3}-\mathbf{b_1'}|},\omega,\omega'\}$

$t_1^{a14}=max\{\sqrt{|A_{a14}|},{1\over
|\mathbf{b_2}-\mathbf{b_3}+\mathbf{b_1'}|},\omega,\omega'\},\;\;\;\;\;\;
t_2^{a14}=max\{\sqrt{|B_{a14}|},{1\over
|\mathbf{b_3}+\mathbf{b_q}|},\omega,\omega'\}$

$t_1^{a15}=max\{\sqrt{|A_{a15}|},{1\over
|\mathbf{b_2}-\mathbf{b_3'}+\mathbf{b_q}|},\omega,\omega'\},\;\;\;\;\;\;
t_2^{a15}=max\{\sqrt{|B_{a15}|},{1\over
|\mathbf{b_3}-\mathbf{b_3'}|},\omega,\omega'\}$

$t_1^{a16}=max\{\sqrt{|A_{a16}|},{1\over
|\mathbf{b_2}|},\omega,\omega'\},\;\;\;\;\;\;\;\;\;\;\;\;\;\;\;\;\;
t_2^{a16}=max\{\sqrt{|B_{a16}|},{1\over
|\mathbf{b_2}-\mathbf{b_3}-\mathbf{b_2'}+\mathbf{b_q}|},\omega,\omega'\}$

$t_1^{a17}=max\{\sqrt{|A_{a17}|},{1\over
|\mathbf{b_2}|},\omega,\omega'\},\;\;\;\;\;\;\;\;\;\;\;\;\;\;\;\;\;
t_2^{a17}=max\{\sqrt{|B_{a17}|},{1\over
|\mathbf{b_3}-\mathbf{b_q}|},\omega,\omega'\}$

$t_1^{a18}=max\{\sqrt{|A_{a18}|},{1\over
|\mathbf{b_2}|},\omega,\omega'\},\;\;\;\;\;\;\;\;\;\;\;\;\;\;\;\;\;
t_2^{a18}=max\{\sqrt{|B_{a18}|},{1\over
|\mathbf{b_q}|},\omega,\omega'\}$

$t_1^{a19}=max\{\sqrt{|A_{a19}|},{1\over
|\mathbf{b_2'}|},\omega,\omega'\},\;\;\;\;\;\;\;\;\;\;\;\;\;\;\;\;\;
t_2^{a19}=max\{\sqrt{|B_{a19}|},{1\over
|\mathbf{b_3}-\mathbf{b_q}|},\omega,\omega'\}$

$t_1^{a20}=max\{\sqrt{|A_{a20}|},{1\over
|\mathbf{b_1}-\mathbf{b_2}+\mathbf{b_q}|},\omega,\omega'\},\;\;\;\;\;\;
t_2^{a20}=max\{\sqrt{|B_{a20}|},{1\over
|\mathbf{b_1}-\mathbf{b_1'}|},\omega,\omega'\}$

$t_1^{a21}=max\{\sqrt{|A_{a21}|},{1\over
|\mathbf{b_2'}|},\omega,\omega'\},\;\;\;\;\;\;\;\;\;\;\;\;\;\;\;\;\;
t_2^{a21}=max\{\sqrt{|B_{a21}|},{1\over
|\mathbf{b_1}+\mathbf{b_q}|},\omega,\omega'\}$

$t_1^{a22}=max\{\sqrt{|A_{a22}|},{1\over
|\mathbf{b_2'}-\mathbf{b_3'}+\mathbf{b_q}|},\omega,\omega'\},\;\;\;\;\;\;
t_2^{a22}=max\{\sqrt{|B_{a22}|},{1\over
|\mathbf{b_3'}|},\omega,\omega'\}$

$t_1^{a23}=max\{\sqrt{|A_{a23}|},{1\over
|\mathbf{b_2}|},\omega,\omega'\},\;\;\;\;\;\;\;\;\;\;\;\;\;\;\;\;\;
t_2^{a23}=max\{\sqrt{|B_{a23}|},{1\over
|\mathbf{b_3'}|},\omega,\omega'\}$

$t_1^{a24}=max\{\sqrt{|A_{a24}|},{1\over
|\mathbf{b_2}|},\omega,\omega'\},\;\;\;\;\;\;\;\;\;\;\;\;\;\;\;\;\;
t_2^{a24}=max\{\sqrt{|B_{a24}|},{1\over
|\mathbf{b_3'}|},\omega,\omega'\}$

$t_1^{a25}=max\{\sqrt{|A_{a25}|},{1\over
|\mathbf{b_2'}|},\omega,\omega'\},\;\;\;\;\;\;\;\;\;\;\;\;\;\;\;\;\;
t_2^{a25}=max\{\sqrt{|B_{a25}|},{1\over
|\mathbf{b_3'}|},\omega,\omega'\}$

$t_1^{a26}=max\{\sqrt{|A_{a26}|},{1\over
|\mathbf{b_2'}|},\omega,\omega'\},\;\;\;\;\;\;\;\;\;\;\;\;\;\;\;\;\;
t_2^{a26}=max\{\sqrt{|B_{a26}|},{1\over
|\mathbf{b_q}|},\omega,\omega'\}$

$t_1^{a27}=max\{\sqrt{|A_{a27}|},{1\over
|\mathbf{b_2}-\mathbf{b_q}|},\omega,\omega'\},\;\;\;\;\;\;\;\;\;\;\;
t_2^{a27}=max\{\sqrt{|B_{a27}|},{1\over
|\mathbf{b_1'}|},\omega,\omega'\}$

$t_1^{a28}=max\{\sqrt{|A_{a28}|},{1\over
|\mathbf{b_2'}|},\omega,\omega'\},\;\;\;\;\;\;\;\;\;\;\;\;\;\;\;\;\;
t_2^{a28}=max\{\sqrt{|B_{a28}|},{1\over
|\mathbf{b_2}+\mathbf{b_1'}-\mathbf{b_2'}-\mathbf{b_q}|},\omega,\omega'\}$

$t_1^{a29}=max\{\sqrt{|A_{a29}|},{1\over
|\mathbf{b_2}|},\omega,\omega'\},\;\;\;\;\;\;\;\;\;\;\;\;\;\;\;\;\;
t_2^{a29}=max\{\sqrt{|B_{a29}|},{1\over
|\mathbf{b_1'}+\mathbf{b_q}|},\omega,\omega'\}$

$t_1^{a30}=max\{\sqrt{|A_{a30}|},{1\over
|\mathbf{b_2}|},\omega,\omega'\},\;\;\;\;\;\;\;\;\;\;\;\;\;\;\;\;\;
t_2^{a30}=max\{\sqrt{|B_{a30}|},{1\over
|\mathbf{b_1'}+\mathbf{b_q}|},\omega,\omega'\}$

$t_1^{a31}=max\{\sqrt{|A_{a31}|},{1\over
|\mathbf{b_2}+\mathbf{b_1'}-\mathbf{b_3'}+\mathbf{b_q}|},\omega,\omega'\},\;
t_2^{a31}=max\{\sqrt{|B_{a31}|},{1\over
|\mathbf{b_3'}|},\omega,\omega'\}$

$t_1^{a32}=max\{\sqrt{|A_{a32}|},{1\over
|\mathbf{b_2}-\mathbf{b_1'}+\mathbf{b_3'}-\mathbf{b_q}|},\omega,\omega'\},\;
t_2^{a32}=max\{\sqrt{|B_{a32}|},{1\over
|\mathbf{b_3'}|},\omega,\omega'\}$

$t_1^{a33}=max\{\sqrt{|A_{a33}|},{1\over
|\mathbf{b_2}-\mathbf{b_3}-\mathbf{b_q}|},\omega,\omega'\},\;\;\;\;\;\;
t_2^{a33}=max\{\sqrt{|B_{a33}|},{1\over
|\mathbf{b_3}-\mathbf{b_1'}|},\omega,\omega'\}$

$t_1^{a34}=max\{\sqrt{|A_{a34}|},{1\over
|\mathbf{b_1}-\mathbf{b_2}+\mathbf{b_q}|},\omega,\omega'\},\;\;\;\;\;\;
t_2^{a34}=max\{\sqrt{|B_{a34}|},{1\over
|\mathbf{b_1}-\mathbf{b_1'}|},\omega,\omega'\}$

$t_1^{a35}=max\{\sqrt{|A_{a35}|},{1\over
|\mathbf{b_2}+\mathbf{b_1'}-\mathbf{b_3'}|},\omega,\omega'\},\;\;\;\;\;\;
t_2^{a35}=max\{\sqrt{|B_{a35}|},{1\over
|\mathbf{b_3'}-\mathbf{b_q}|},\omega,\omega'\}$

$t_1^{a36}=max\{\sqrt{|A_{a36}|},{1\over
|\mathbf{b_2}-\mathbf{b_q}|},\omega,\omega'\},\;\;\;\;\;\;\;\;\;\;\;\;
t_2^{a36}=max\{\sqrt{|B_{a36}|},{1\over
|\mathbf{b_2'}|},\omega,\omega'\}$

$t_1^{b1}=max\{\sqrt{|A_{b1}|},{1\over
|\mathbf{b_2}-\mathbf{b_q}|},\omega,\omega'\},\;\;\;\;\;\;\;\;\;\;\;\;\;\;\;
t_2^{b1}=max\{\sqrt{|B_{b1}|},{1\over
|\mathbf{b_1'}|},\omega,\omega'\}$

$t_1^{b2}=max\{\sqrt{|A_{b2}|},{1\over
|\mathbf{b_2}+\mathbf{b_1'}|},\omega,\omega'\},\;\;\;\;\;\;\;\;\;\;\;\;\;\;\;
t_2^{b2}=max\{\sqrt{|B_{b2}|},{1\over
|\mathbf{b_q}|},\omega,\omega'\}$

$t_1^{b3}=max\{\sqrt{|A_{b3}|},{1\over
|\mathbf{b_2}-\mathbf{b_3}-\mathbf{b_q}|},\omega,\omega'\},\;\;\;\;\;\;\;\;\;\;
t_2^{b3}=max\{\sqrt{|B_{b3}|},{1\over
|\mathbf{b_3}-\mathbf{b_1'}|},\omega,\omega'\}$

$t_1^{b4}=max\{\sqrt{|A_{b4}|},{1\over
|\mathbf{b_2}-\mathbf{b_3}+\mathbf{b_1'}|},\omega,\omega'\},\;\;\;\;\;\;\;\;\;\;
t_2^{b4}=max\{\sqrt{|B_{b4}|},{1\over
|\mathbf{b_3}+\mathbf{b_q}|},\omega,\omega'\}$

$t_1^{b5}=max\{\sqrt{|A_{b5}|},{1\over
|\mathbf{b_1}-\mathbf{b_2}+\mathbf{b_q}|},\omega,\omega'\},\;\;\;\;\;\;\;\;\;\;
t_2^{b5}=max\{\sqrt{|B_{b5}|},{1\over
|\mathbf{b_1}-\mathbf{b_1'}|},\omega,\omega'\}$

$t_1^{b6}=max\{\sqrt{|A_{b6}|},{1\over
|\mathbf{b_2'}|},\omega,\omega'\},\;\;\;\;\;\;\;\;\;\;\;\;\;\;\;\;\;\;\;\;\;
t_2^{b6}=max\{\sqrt{|B_{b6}|},{1\over
|\mathbf{b_1}+\mathbf{b_q}|},\omega,\omega'\}$

$t_1^{b7}=max\{\sqrt{|A_{b7}|},{1\over
|\mathbf{b_2}-\mathbf{b_q}|},\omega,\omega'\},\;\;\;\;\;\;\;\;\;\;\;\;\;\;\;
t_2^{b7}=max\{\sqrt{|B_{b7}|},{1\over
|\mathbf{b_1'}|},\omega,\omega'\}$

$t_1^{b8}=max\{\sqrt{|A_{b8}|},{1\over
|\mathbf{b_2'}|},\omega,\omega'\},\;\;\;\;\;\;\;\;\;\;\;\;\;\;\;\;\;\;\;\;\;
t_2^{b8}=max\{\sqrt{|B_{b8}|},{1\over
|\mathbf{b_2}+\mathbf{b_1'}-\mathbf{b_2'}-\mathbf{b_q}|},\omega,\omega'\}$

$t_1^{c1}=max\{\sqrt{|A_{c1}|},{2\over
|\mathbf{b_2}-\mathbf{b_1'}+\mathbf{b_2'}|},\omega,\omega'\},\;\;\;\;\;\;\;\;\;\;
t_2^{c1}=max\{\sqrt{|B_{c1}|},{2\over
|\mathbf{b_2}-\mathbf{b_1'}-\mathbf{b_2'}|},\omega,\omega'\},$

$t_3^{c1}=max\{\sqrt{|C_{c1}|},{2\over
|\mathbf{b_2}-\mathbf{b_1'}-2\mathbf{b_q}-\mathbf{b_2'}|},\omega,\omega'\}$

$t_1^{c2}=max\{\sqrt{|A_{c2}|},{2\over
|2\mathbf{b_2}-\mathbf{b_1'}-\mathbf{b_q}|},\omega,\omega'\},\;\;\;\;\;\;\;\;\;\;
t_2^{c2}=max\{\sqrt{|B_{c2}|},{2\over
|2\mathbf{b_3}-\mathbf{b_1'}+\mathbf{b_q}|},\omega,\omega'\},$

$t_3^{c2}=max\{\sqrt{|C_{c2}|},{2\over
|\mathbf{b_1'}+\mathbf{b_q}|},\omega,\omega'\}$

\noindent{\bf Appendix D: Expressions of $\Omega^i$}\\
The  expression of $\Omega^i$ for diagrams (a1)$\sim$(a36) in
Fig.~1, (b1)$\sim$(b8) in Fig.~2 and (c1), (c2) in Fig.~3 are
presented below where $K_0$, $K_1$, $N_0$, $N_1$, $J_0$ and $J_1$
are Bessel functions and $\theta(x)$ is $\theta$-function.

\begin{eqnarray}
&&\Omega^{a1} =\nonumber \\
&&\int_0^1 {dz_1 dz_2 \over {z_1
(1-z_1)}}\frac{\sqrt{X_2}}{\sqrt{|Z_2|}}\left[\pi^2K_1(\sqrt{X_2Z_2})\theta(Z_2)+\frac{\pi^3}{2}[N_1(\sqrt{X_2|Z_2|})-iJ_1(\sqrt{X_2|Z_2|})]\theta(-Z_2)\right]\nonumber \\
&&\left[2\pi
K_0(\sqrt{A_{a1}}|\mathbf{b_2}-\mathbf{b_3'}+\mathbf{b_q}|)\theta(A_{a1})+
\pi^2[-N_0(\sqrt{|A_{a1}|}|\mathbf{b_2}-\mathbf{b_3'}+\mathbf{b_q}|)+iJ_0(\sqrt{|A_{a1}|}|\mathbf{b_2}-\mathbf{b_3'}+\mathbf{b_q}|)]\theta(-A_{a1})\right]\nonumber
\\
\end{eqnarray}
with
\begin{eqnarray}
&&Z_2=B_{a1}(1-z_2)+{z_2 \over z_1(1-z_1)}[C_{a1}(1-z_1)+D_{a1}z_1]\nonumber \\
&&X_2=[\mathbf{b_3'}-z_1(\mathbf{b_3'}-\mathbf{b_q})]^2+{z_1(1-z_1)\over
z_2}(\mathbf{b_3'}-\mathbf{b_q})^2.
\end{eqnarray}

\begin{eqnarray}
&&\Omega^{a2} =\nonumber \\
&&\int_0^1 {dz_1 dz_2 \over {z_1
(1-z_1)}}\frac{\sqrt{X_2}}{\sqrt{|Z_2|}}\left[\pi^2K_1(\sqrt{X_2Z_2})\theta(Z_2)+\frac{\pi^3}{2}[N_1(\sqrt{X_2|Z_2|})-iJ_1(\sqrt{X_2|Z_2|})]\theta(-Z_2)\right]\nonumber \\
&&\left[2\pi K_0(\sqrt{B_{a2}}|\mathbf{b_q}|)\theta(B_{a2})+
\pi^2[-N_0(\sqrt{|B_{a2}|}|\mathbf{b_q}|)+iJ_0(\sqrt{|B_{a2}|}|\mathbf{b_q}|)]\theta(-B_{a2})\right]\nonumber
\\
\end{eqnarray}
with
\begin{eqnarray}
&&Z_2=C_{a2}(1-z_2)+{z_2 \over z_1(1-z_1)}[A_{a2}(1-z_1)+D_{a2}z_1]\nonumber \\
&&X_2=[\mathbf{b_3'}-\mathbf{b_q}-z_1\mathbf{b_2}]^2+{z_1(1-z_1)\over
z_2}\mathbf{b_2}^2.
\end{eqnarray}

\begin{eqnarray}
&&\Omega^{a3} =\nonumber \\
&&\left[2\pi
K_0(\sqrt{A_{a3}}|\mathbf{b_2}|)\theta(A_{a3})+\pi^2\left(-N_0(\sqrt{|A_{a3}|}|\mathbf{b_2}|)+iJ_0(\sqrt{|A_{a3}|}|\mathbf{b_2}|)
\right)\theta(-A_{a3})\right]\nonumber \\
&&\left[2\pi
K_0(\sqrt{B_{a3}}|\mathbf{b_q}|)\theta(B_{a3})+\pi^2\left(-N_0(\sqrt{|B_{a3}|}|\mathbf{b_q}|)+iJ_0(\sqrt{|B_{a3}|}|\mathbf{b_q}|)
\right)\theta(-B_{a3})\right]\nonumber \\
&&\left[2\pi
K_0(\sqrt{C_{a3}}|\mathbf{b_3'}-\mathbf{b_q}|)\theta(C_{a3})+\pi^2\left(-N_0(\sqrt{|C_{a3}|}|\mathbf{b_3'}-\mathbf{b_q}|)+iJ_0(\sqrt{|C_{a3}|}|\mathbf{b_3'}-\mathbf{b_q}|)
\right)\theta(-C_{a3})\right]\nonumber \\
&&\left[2\pi
K_0(\sqrt{D_{a3}}|\mathbf{b_3}|)\theta(D_{a3})+\pi^2\left(-N_0(\sqrt{|D_{a3}|}|\mathbf{b_3}|)+iJ_0(\sqrt{|D_{a3}|}|\mathbf{b_3}|)
\right)\theta(-D_{a3})\right]\nonumber \\
\end{eqnarray}

\begin{eqnarray}
&&\Omega^{a4} =\nonumber \\
&&\left[2\pi
K_0(\sqrt{A_{a4}}|\mathbf{b_2}|)\theta(A_{a4})+\pi^2\left(-N_0(\sqrt{|A_{a4}|}|\mathbf{b_2}|)+iJ_0(\sqrt{|A_{a4}|}|\mathbf{b_2}|)
\right)\theta(-A_{a4})\right]\nonumber \\
&&\left[2\pi
K_0(\sqrt{B_{a4}}|\mathbf{b_q}|)\theta(B_{a4})+\pi^2\left(-N_0(\sqrt{|B_{a4}|}|\mathbf{b_q}|)+iJ_0(\sqrt{|B_{a4}|}|\mathbf{b_q}|)
\right)\theta(-B_{a4})\right]\nonumber \\
&&\left[2\pi
K_0(\sqrt{C_{a4}}|\mathbf{b_1}|)\theta(C_{a4})+\pi^2\left(-N_0(\sqrt{|C_{a4}|}|\mathbf{b_1}|)+iJ_0(\sqrt{|C_{a4}|}|\mathbf{b_1}|)
\right)\theta(-C_{a4})\right]\nonumber \\
&&\left[2\pi
K_0(\sqrt{D_{a4}}|\mathbf{b_3'}-\mathbf{b_q}|)\theta(D_{a4})+\pi^2\left(-N_0(\sqrt{|D_{a4}|}|\mathbf{b_3'}-\mathbf{b_q}|)+iJ_0(\sqrt{|D_{a4}|}|\mathbf{b_3'}-\mathbf{b_q}|)
\right)\theta(-D_{a4})\right]\nonumber \\
\end{eqnarray}

\begin{eqnarray}
&&\Omega^{a5} =\nonumber \\
&&\left[2\pi
K_0(\sqrt{A_{a5}}|\mathbf{b_2'}|)\theta(A_{a5})+\pi^2\left(-N_0(\sqrt{|A_{a5}|}|\mathbf{b_2'}|)+iJ_0(\sqrt{|A_{a5}|}|\mathbf{b_2'}|)
\right)\theta(-A_{a5})\right]\nonumber \\
&&\left[2\pi
K_0(\sqrt{B_{a5}}|\mathbf{b_q}|)\theta(B_{a5})+\pi^2\left(-N_0(\sqrt{|B_{a5}|}|\mathbf{b_q}|)+iJ_0(\sqrt{|B_{a5}|}|\mathbf{b_q}|)
\right)\theta(-B_{a5})\right]\nonumber \\
&&\left[2\pi
K_0(\sqrt{C_{a5}}|\mathbf{b_2}-\mathbf{b_2'}|)\theta(C_{a5})+\pi^2\left(-N_0(\sqrt{|C_{a5}|}|\mathbf{b_2}-\mathbf{b_2'}|)+iJ_0(\sqrt{|C_{a5}|}|\mathbf{b_2}-\mathbf{b_2'}|)
\right)\theta(-C_{a5})\right]\nonumber \\
&&[2\pi
K_0(\sqrt{D_{a5}}|\mathbf{b_2}-\mathbf{b_2'}+\mathbf{b_3'}-\mathbf{b_q}|)\theta(D_{a5})+\nonumber
\\
&&\pi^2\left(-N_0(\sqrt{|D_{a5}|}|\mathbf{b_2}-\mathbf{b_2'}+\mathbf{b_3'}-\mathbf{b_q}|)+iJ_0(\sqrt{|D_{a5}|}|\mathbf{b_2}-\mathbf{b_2'}+\mathbf{b_3'}-\mathbf{b_q}|)
\right)\theta(-D_{a5})]\nonumber \\
\end{eqnarray}

\begin{eqnarray}
&&\Omega^{a6} =\nonumber \\
&&\int_0^1 {dz_1 dz_2 dz_3 \over z_1
(1-z_1)z_2(1-z_2)}\frac{\sqrt{X_3}}{\sqrt{|Z_3|}}\left[\pi^3K_1(\sqrt{X_3Z_3})\theta(Z_3)+\frac{\pi^4}{2}[N_1(\sqrt{X_3|Z_3|})-iJ_1(\sqrt{X_3|Z_3|})]\theta(-Z_3)\right]\nonumber \\
\end{eqnarray}
with
\begin{eqnarray}
&&Z_3=B_{a6}(1-z_3)+{z_3 \over z_2(1-z_2)}\left[C_{a6}(1-z_2)+{z_2\over z_1(1-z_1)}[A_{a6}(1-z_1)+D_{a6}z_1]\right]\nonumber \\
&&X_3=[-\mathbf{b_1'}+z_2(\mathbf{b_1'}+\mathbf{b_q})-(\mathbf{b_2}-\mathbf{b_q})z_2(1-z_1)]^2+{z_2(1-z_2)\over
z_3}[-\mathbf{b_1'}-\mathbf{b_q}-(\mathbf{b_2}-\mathbf{b_q})z_1]^2.
\end{eqnarray}

\begin{eqnarray}
&&\Omega^{a7} =\nonumber \\
&&\int_0^1 dz\frac{|\mathbf{b_1'}+\mathbf{b_q}|}{\sqrt{|Z_1|}}\left[\pi K_1(\sqrt{Z_1}|\mathbf{b_1'}+\mathbf{b_q}|)\theta(Z_1)+
\frac{\pi^2}{2}[N_1(\sqrt{|Z_1|}|\mathbf{b_1'}+\mathbf{b_q}|)-iJ_1(\sqrt{|Z_1|}|\mathbf{b_1'}+\mathbf{b_q}|)]\theta(-Z_1)\right]\nonumber \\
&&\left[2\pi
K_0(\sqrt{A_{a7}}|\mathbf{b_2}+\mathbf{b_1'}|)\theta(A_{a7})+\pi^2\left(-N_0(\sqrt{|A_{a7}|}|\mathbf{b_2}+\mathbf{b_1'}|)+
iJ_0(\sqrt{|A_{a7}|}|\mathbf{b_2}+\mathbf{b_1'}|)
\right)\theta(-A_{a7})\right]\nonumber \\
&&\left[2\pi
K_0(\sqrt{B_{a7}}|\mathbf{b_q}|)\theta(B_{a7})+\pi^2\left(-N_0(\sqrt{|B_{a7}|}|\mathbf{b_q}|)+
iJ_0(\sqrt{|B_{a7}|}|\mathbf{b_q}|)
\right)\theta(-B_{a7})\right]\nonumber \\
\end{eqnarray}
with
\begin{eqnarray}
&&Z_1=C_{a7}z+D_{a7}(1-z).
\end{eqnarray}

\begin{eqnarray}
&&\Omega^{a8} =\nonumber \\
&&\left[2\pi
K_0(\sqrt{A_{a8}}|\mathbf{b_2}-\mathbf{b_3'}+\mathbf{b_q}|)\theta(A_{a8})+\pi^2\left(-N_0(\sqrt{|A_{a8}|}|\mathbf{b_2}-\mathbf{b_3'}+\mathbf{b_q}|)+
iJ_0(\sqrt{|A_{a8}|}|\mathbf{b_2}-\mathbf{b_3'}+\mathbf{b_q}|)
\right)\theta(-A_{a8})\right]\nonumber \\
&&\left[2\pi
K_0(\sqrt{B_{a8}}|\mathbf{b_1}-\mathbf{b_3'}|)\theta(B_{a8})+\pi^2\left(-N_0(\sqrt{|B_{a8}|}|\mathbf{b_1}-\mathbf{b_3'}|)+iJ_0(\sqrt{|B_{a8}|}|\mathbf{b_1}-\mathbf{b_3'}|)
\right)\theta(-B_{a8})\right]\nonumber \\
&&\left[2\pi
K_0(\sqrt{C_{a8}}|\mathbf{b_3'}-\mathbf{b_q}|)\theta(C_{a8})+\pi^2\left(-N_0(\sqrt{|C_{a8}|}|\mathbf{b_3'}-\mathbf{b_q}|)+iJ_0(\sqrt{|C_{a8}|}|\mathbf{b_3'}-\mathbf{b_q}|)
\right)\theta(-C_{a8})\right]\nonumber \\
&&\left[2\pi
K_0(\sqrt{D_{a8}}|\mathbf{b_1}|)\theta(D_{a8})+\pi^2\left(-N_0(\sqrt{|D_{a8}|}|\mathbf{b_1}|)+iJ_0(\sqrt{|D_{a8}|}|\mathbf{b_1}|)
\right)\theta(-D_{a8})\right]\nonumber \\
\end{eqnarray}

\begin{eqnarray}
&&\Omega^{a9} =\nonumber \\
&&\int_0^1 {dz_1 dz_2 \over {z_1
(1-z_1)}}\frac{\sqrt{X_2}}{\sqrt{|Z_2|}}\left[\pi^2K_1(\sqrt{X_2Z_2})\theta(Z_2)+\frac{\pi^3}{2}[N_1(\sqrt{X_2|Z_2|})-iJ_1(\sqrt{X_2|Z_2|})]\theta(-Z_2)\right]\nonumber \\
&&\left[2\pi
K_0(\sqrt{B_{a9}}|\mathbf{b_1}-\mathbf{b_3'}|)\theta(B_{a9})+
\pi^2[-N_0(\sqrt{|B_{a9}|}|\mathbf{b_1}-\mathbf{b_3'}|)+iJ_0(\sqrt{|B_{a9}|}|\mathbf{b_1}-\mathbf{b_3'}|)]\theta(-B_{a9})\right]\nonumber
\\
\end{eqnarray}
with
\begin{eqnarray}
&&Z_2=C_{a9}(1-z_2)+{z_2 \over z_1(1-z_1)}[A_{a9}(1-z_1)+D_{a9}z_1]\nonumber \\
&&X_2=[\mathbf{b_1}-\mathbf{b_3'}+\mathbf{b_q}-z_1\mathbf{b_1}]^2+{z_1(1-z_1)\over
z_2}\mathbf{b_1}^2.
\end{eqnarray}

\begin{eqnarray}
&&\Omega^{a10} =\nonumber \\
&&\int_0^1 {dz_1 dz_2 \over {z_1
(1-z_1)}}\frac{\sqrt{X_2}}{\sqrt{|Z_2|}}\left[\pi^2K_1(\sqrt{X_2Z_2})\theta(Z_2)+\frac{\pi^3}{2}[N_1(\sqrt{X_2|Z_2|})-iJ_1(\sqrt{X_2|Z_2|})]\theta(-Z_2)\right]\nonumber \\
&&\left[2\pi K_0(\sqrt{A_{a10}}|\mathbf{b_2}|)\theta(A_{a10})+
\pi^2[-N_0(\sqrt{|A_{a10}|}|\mathbf{b_2}|)+iJ_0(\sqrt{|A_{a10}|}|\mathbf{b_2}|)]\theta(-A_{a10})\right]\nonumber
\\
\end{eqnarray}
with
\begin{eqnarray}
&&Z_2=B_{a10}(1-z_2)+{z_2 \over z_1(1-z_1)}[C_{a10}(1-z_1)+D_{a10}z_1]\nonumber \\
&&X_2=[\mathbf{b_3}+\mathbf{b_q}-z_1\mathbf{b_3}]^2+{z_1(1-z_1)\over
z_2}\mathbf{b_3}^2.
\end{eqnarray}

\begin{eqnarray}
&&\Omega^{a11} =\nonumber \\
&&\left[2\pi
K_0(\sqrt{A_{a11}}|\mathbf{b_2}|)\theta(A_{a11})+\pi^2\left(-N_0(\sqrt{|A_{a11}|}|\mathbf{b_2}|)+
iJ_0(\sqrt{|A_{a11}|}|\mathbf{b_2}|)
\right)\theta(-A_{a11})\right]\nonumber \\
&&[2\pi
K_0(\sqrt{B_{a11}}|\mathbf{b_1}-\mathbf{b_2}+\mathbf{b_2'}-\mathbf{b_q}|)\theta(B_{a11})+\nonumber
\\
&&\pi^2\left(-N_0(\sqrt{|B_{a11}|}|\mathbf{b_1}-\mathbf{b_2}+\mathbf{b_2'}-\mathbf{b_q}|)+iJ_0(\sqrt{|B_{a11}|}|\mathbf{b_1}-\mathbf{b_2}+\mathbf{b_2'}-\mathbf{b_q}|)
\right)\theta(-B_{a11})]\nonumber \\
&&\left[2\pi
K_0(\sqrt{C_{a11}}|\mathbf{b_2}-\mathbf{b_2'}|)\theta(C_{a11})+\pi^2\left(-N_0(\sqrt{|C_{a11}|}|\mathbf{b_2}-\mathbf{b_2'}|)+iJ_0(\sqrt{|C_{a11}|}|\mathbf{b_2}-\mathbf{b_2'}|)
\right)\theta(-C_{a11})\right]\nonumber \\
&&\left[2\pi
K_0(\sqrt{D_{a11}}|\mathbf{b_1}-\mathbf{b_2}+\mathbf{b_2'}|)\theta(D_{a11})+\pi^2\left(-N_0(\sqrt{|D_{a11}|}|\mathbf{b_1}-\mathbf{b_2}+\mathbf{b_2'}|)+iJ_0(\sqrt{|D_{a11}|}|\mathbf{b_1}-\mathbf{b_2}+\mathbf{b_2'}|)
\right)\theta(-D_{a11})\right]\nonumber \\
\end{eqnarray}

\begin{eqnarray}
&&\Omega^{a12} =\nonumber \\
&&\left[2\pi
K_0(\sqrt{A_{a12}}|\mathbf{b_2}+\mathbf{b_3'}-\mathbf{b_q}|)\theta(A_{a12})+\pi^2\left(-N_0(\sqrt{|A_{a12}|}|\mathbf{b_2}+\mathbf{b_3'}-\mathbf{b_q}|)+
iJ_0(\sqrt{|A_{a12}|}|\mathbf{b_2}+\mathbf{b_3'}-\mathbf{b_q}|)
\right)\theta(-A_{a12})\right]\nonumber \\
&&\left[2\pi
K_0(\sqrt{B_{a12}}|\mathbf{b_1}-\mathbf{b_q}|)\theta(B_{a12})+\pi^2\left(-N_0(\sqrt{|B_{a12}|}|\mathbf{b_1}-\mathbf{b_q}|)+iJ_0(\sqrt{|B_{a12}|}|\mathbf{b_1}-\mathbf{b_q}|)
\right)\theta(-B_{a12})\right]\nonumber \\
&&\left[2\pi
K_0(\sqrt{C_{a12}}|\mathbf{b_3'}-\mathbf{b_q}|)\theta(C_{a12})+\pi^2\left(-N_0(\sqrt{|C_{a12}|}|\mathbf{b_3'}-\mathbf{b_q}|)+iJ_0(\sqrt{|C_{a12}|}|\mathbf{b_3'}-\mathbf{b_q}|)
\right)\theta(-C_{a12})\right]\nonumber \\
&&\left[2\pi
K_0(\sqrt{D_{a12}}|\mathbf{b_1}|)\theta(D_{a12})+\pi^2\left(-N_0(\sqrt{|D_{a12}|}|\mathbf{b_1}|)+iJ_0(\sqrt{|D_{a12}|}|\mathbf{b_1}|)
\right)\theta(-D_{a12})\right]\nonumber \\
\end{eqnarray}

\begin{eqnarray}
&&\Omega^{a13} =\nonumber \\
&&\left[2\pi
K_0(\sqrt{A_{a13}}|\mathbf{b_2}-\mathbf{b_3}-\mathbf{b_q}|)\theta(A_{a13})+\pi^2\left(-N_0(\sqrt{|A_{a13}|}|\mathbf{b_2}-\mathbf{b_3}-\mathbf{b_q}|)+
iJ_0(\sqrt{|A_{a13}|}|\mathbf{b_2}-\mathbf{b_3}-\mathbf{b_q}|)
\right)\theta(-A_{a13})\right]\nonumber \\
&&\left[2\pi
K_0(\sqrt{B_{a13}}|\mathbf{b_3}-\mathbf{b_1'}|)\theta(B_{a13})+\pi^2\left(-N_0(\sqrt{|B_{a13}|}|\mathbf{b_3}-\mathbf{b_1'}|)+iJ_0(\sqrt{|B_{a13}|}|\mathbf{b_3}-\mathbf{b_1'}|)
\right)\theta(-B_{a13})\right]\nonumber \\
&&\left[2\pi
K_0(\sqrt{C_{a13}}|\mathbf{b_1'}+\mathbf{b_q}|)\theta(C_{a13})+\pi^2\left(-N_0(\sqrt{|C_{a13}|}|\mathbf{b_1'}+\mathbf{b_q}|)+iJ_0(\sqrt{|C_{a13}|}|\mathbf{b_1'}+\mathbf{b_q}|)
\right)\theta(-C_{a13})\right]\nonumber \\
&&\left[2\pi
K_0(\sqrt{D_{a13}}|\mathbf{b_3}|)\theta(D_{a13})+\pi^2\left(-N_0(\sqrt{|D_{a13}|}|\mathbf{b_3}|)+iJ_0(\sqrt{|D_{a13}|}|\mathbf{b_3}|)
\right)\theta(-D_{a13})\right]\nonumber \\
\end{eqnarray}

\begin{eqnarray}
&&\Omega^{a14} =\nonumber \\
&&\left[2\pi
K_0(\sqrt{A_{a14}}|\mathbf{b_2}-\mathbf{b_3}+\mathbf{b_1'}|)\theta(A_{a14})+\pi^2\left(-N_0(\sqrt{|A_{a14}|}|\mathbf{b_2}-\mathbf{b_3}+\mathbf{b_1'}|)+
iJ_0(\sqrt{|A_{a14}|}|\mathbf{b_2}-\mathbf{b_3}+\mathbf{b_1'}|)
\right)\theta(-A_{a14})\right]\nonumber \\
&&\left[2\pi
K_0(\sqrt{B_{a14}}|\mathbf{b_3}+\mathbf{b_q}|)\theta(B_{a14})+\pi^2\left(-N_0(\sqrt{|B_{a14}|}|\mathbf{b_3}+\mathbf{b_q}|)+iJ_0(\sqrt{|B_{a14}|}|\mathbf{b_3}+\mathbf{b_q}|)
\right)\theta(-B_{a14})\right]\nonumber \\
&&\left[2\pi
K_0(\sqrt{C_{a14}}|\mathbf{b_1'}+\mathbf{b_q}|)\theta(C_{a14})+\pi^2\left(-N_0(\sqrt{|C_{a14}|}|\mathbf{b_1'}+\mathbf{b_q}|)+iJ_0(\sqrt{|C_{a14}|}|\mathbf{b_1'}+\mathbf{b_q}|)
\right)\theta(-C_{a14})\right]\nonumber \\
&&\left[2\pi
K_0(\sqrt{D_{a14}}|\mathbf{b_3}|)\theta(D_{a14})+\pi^2\left(-N_0(\sqrt{|D_{a14}|}|\mathbf{b_3}|)+iJ_0(\sqrt{|D_{a14}|}|\mathbf{b_3}|)
\right)\theta(-D_{a14})\right]\nonumber \\
\end{eqnarray}

\begin{eqnarray}
&&\Omega^{a15} =\nonumber \\
&&\left[2\pi
K_0(\sqrt{A_{a15}}|\mathbf{b_2}-\mathbf{b_3'}+\mathbf{b_q}|)\theta(A_{a15})+\pi^2\left(-N_0(\sqrt{|A_{a15}|}|\mathbf{b_2}-\mathbf{b_3'}+\mathbf{b_q}|)+
iJ_0(\sqrt{|A_{a15}|}|\mathbf{b_2}-\mathbf{b_3'}+\mathbf{b_q}|)
\right)\theta(-A_{a15})\right]\nonumber \\
&&\left[2\pi
K_0(\sqrt{B_{a15}}|\mathbf{b_3}-\mathbf{b_3'}|)\theta(B_{a15})+\pi^2\left(-N_0(\sqrt{|B_{a15}|}|\mathbf{b_3}-\mathbf{b_3'}|)+iJ_0(\sqrt{|B_{a15}|}|\mathbf{b_3}-\mathbf{b_3'}|)
\right)\theta(-B_{a15})\right]\nonumber \\
&&\left[2\pi
K_0(\sqrt{C_{a15}}|\mathbf{b_3}|)\theta(C_{a15})+\pi^2\left(-N_0(\sqrt{|C_{a15}|}|\mathbf{b_3}|)+iJ_0(\sqrt{|C_{a15}|}|\mathbf{b_3}|)
\right)\theta(-C_{a15})\right]\nonumber \\
&&\left[2\pi
K_0(\sqrt{D_{a15}}|\mathbf{b_3'}-\mathbf{b_q}|)\theta(D_{a15})+\pi^2\left(-N_0(\sqrt{|D_{a15}|}|\mathbf{b_3'}-\mathbf{b_q}|)+iJ_0(\sqrt{|D_{a15}|}|\mathbf{b_3'}-\mathbf{b_q}|)
\right)\theta(-D_{a15})\right]\nonumber \\
\end{eqnarray}

\begin{eqnarray}
&&\Omega^{a16} =\nonumber \\
&&\left[2\pi
K_0(\sqrt{A_{a16}}|\mathbf{b_2}|)\theta(A_{a16})+\pi^2\left(-N_0(\sqrt{|A_{a16}|}|\mathbf{b_2}|)+
iJ_0(\sqrt{|A_{a16}|}|\mathbf{b_2}|)
\right)\theta(-A_{a16})\right]\nonumber \\
&&[2\pi
K_0(\sqrt{B_{a16}}|\mathbf{b_2}-\mathbf{b_3}-\mathbf{b_2'}+\mathbf{b_q}|)\theta(B_{a16})+\nonumber
\\
&&\pi^2\left(-N_0(\sqrt{|B_{a16}|}|\mathbf{b_2}-\mathbf{b_3}-\mathbf{b_2'}+\mathbf{b_q}|)+iJ_0(\sqrt{|B_{a16}|}|\mathbf{b_2}-\mathbf{b_3}-\mathbf{b_2'}+\mathbf{b_q}|)
\right)\theta(-B_{a16})]\nonumber \\
&&\left[2\pi
K_0(\sqrt{C_{a16}}|\mathbf{b_2}-\mathbf{b_3}-\mathbf{b_2'}|)\theta(C_{a16})+\pi^2\left(-N_0(\sqrt{|C_{a16}|}|\mathbf{b_2}-\mathbf{b_3}-\mathbf{b_2'}|)+iJ_0(\sqrt{|C_{a16}|}|\mathbf{b_2}-\mathbf{b_3}-\mathbf{b_2'}|)
\right)\theta(-C_{a16})\right]\nonumber \\
&&\left[2\pi
K_0(\sqrt{D_{a16}}|\mathbf{b_2}-\mathbf{b_2'}|)\theta(D_{a16})+\pi^2\left(-N_0(\sqrt{|D_{a16}|}|\mathbf{b_2}-\mathbf{b_2'}|)+iJ_0(\sqrt{|D_{a16}|}|\mathbf{b_2}-\mathbf{b_2'}|)
\right)\theta(-D_{a16})\right]\nonumber \\
\end{eqnarray}

\begin{eqnarray}
&&\Omega^{a17} =\nonumber \\
&&\int_0^1 {dz_1 dz_2 \over {z_1
(1-z_1)}}\frac{\sqrt{X_2}}{\sqrt{|Z_2|}}\left[\pi^2K_1(\sqrt{X_2Z_2})\theta(Z_2)+\frac{\pi^3}{2}[N_1(\sqrt{X_2|Z_2|})-iJ_1(\sqrt{X_2|Z_2|})]\theta(-Z_2)\right]\nonumber \\
&&\left[2\pi
K_0(\sqrt{B_{a17}}|\mathbf{b_3}-\mathbf{b_q}|)\theta(B_{a17})+
\pi^2[-N_0(\sqrt{|B_{a17}|}|\mathbf{b_3}-\mathbf{b_q}|)+iJ_0(\sqrt{|B_{a17}|}|\mathbf{b_3}-\mathbf{b_q}|)]\theta(-B_{a17})\right]\nonumber
\\
\end{eqnarray}
with
\begin{eqnarray}
&&Z_2=C_{a17}(1-z_2)+{z_2 \over z_1(1-z_1)}[A_{a17}(1-z_1)+D_{a17}z_1]\nonumber \\
&&X_2=[\mathbf{b_3}-z_1\mathbf{b_2}]^2+{z_1(1-z_1)\over
z_2}\mathbf{b_2}^2.
\end{eqnarray}

\begin{eqnarray}
&&\Omega^{a18} =\nonumber \\
&&\int_0^1 {dz_1 dz_2 \over {z_1
(1-z_1)}}\frac{\sqrt{X_2}}{\sqrt{|Z_2|}}\left[\pi^2K_1(\sqrt{X_2Z_2})\theta(Z_2)+\frac{\pi^3}{2}[N_1(\sqrt{X_2|Z_2|})-iJ_1(\sqrt{X_2|Z_2|})]\theta(-Z_2)\right]\nonumber \\
&&\left[2\pi K_0(\sqrt{A_{a18}}|\mathbf{b_2}|)\theta(A_{a18})+
\pi^2[-N_0(\sqrt{|A_{a18}|}|\mathbf{b_2}|)+iJ_0(\sqrt{|A_{a18}|}|\mathbf{b_2}|)]\theta(-A_{a18})\right]\nonumber
\\
\end{eqnarray}
with
\begin{eqnarray}
&&Z_2=B_{a18}(1-z_2)+{z_2 \over z_1(1-z_1)}[C_{a18}(1-z_1)+D_{a18}z_1]\nonumber \\
&&X_2=[-\mathbf{b_q}-z_1\mathbf{b_1}]^2+{z_1(1-z_1)\over
z_2}\mathbf{b_1}^2.
\end{eqnarray}

\begin{eqnarray}
&&\Omega^{a19} =\nonumber \\
&&\left[2\pi
K_0(\sqrt{A_{a19}}|\mathbf{b_2'}|)\theta(A_{a19})+\pi^2\left(-N_0(\sqrt{|A_{a19}|}|\mathbf{b_2'}|)+
iJ_0(\sqrt{|A_{a19}|}|\mathbf{b_2'}|)
\right)\theta(-A_{a19})\right]\nonumber \\
&&\left[2\pi
K_0(\sqrt{B_{a19}}|-\mathbf{b_3}+\mathbf{b_q}|)\theta(B_{a19})+\pi^2\left(-N_0(\sqrt{|B_{a19}|}|-\mathbf{b_3}+\mathbf{b_q}|)+iJ_0(\sqrt{|B_{a19}|}|-\mathbf{b_3}+\mathbf{b_q}|)
\right)\theta(-B_{a19})\right]\nonumber \\
&&\left[2\pi
K_0(\sqrt{C_{a19}}|\mathbf{b_3}|)\theta(C_{a19})+\pi^2\left(-N_0(\sqrt{|C_{a19}|}|\mathbf{b_3}|)+iJ_0(\sqrt{|C_{a19}|}|\mathbf{b_3}|)
\right)\theta(-C_{a19})\right]\nonumber \\
&&\left[2\pi
K_0(\sqrt{D_{a19}}|\mathbf{b_2}-\mathbf{b_2'}|)\theta(D_{a19})+\pi^2\left(-N_0(\sqrt{|D_{a19}|}|\mathbf{b_2}-\mathbf{b_2'}|)+iJ_0(\sqrt{|D_{a19}|}|\mathbf{b_2}-\mathbf{b_2'}|)
\right)\theta(-D_{a19})\right]\nonumber \\
\end{eqnarray}

\begin{eqnarray}
&&\Omega^{a20} =\nonumber \\
&&\left[2\pi
K_0(\sqrt{A_{a20}}|\mathbf{b_1}-\mathbf{b_2}+\mathbf{b_q}|)\theta(A_{a20})+\pi^2\left(-N_0(\sqrt{|A_{a20}|}|\mathbf{b_1}-\mathbf{b_2}+\mathbf{b_q}|)+
iJ_0(\sqrt{|A_{a20}|}|\mathbf{b_1}-\mathbf{b_2}+\mathbf{b_q}|)
\right)\theta(-A_{a20})\right]\nonumber \\
&&\left[2\pi
K_0(\sqrt{B_{a20}}|\mathbf{b_1}-\mathbf{b_1'}|)\theta(B_{a20})+\pi^2\left(-N_0(\sqrt{|B_{a20}|}|\mathbf{b_1}-\mathbf{b_1'}|)+iJ_0(\sqrt{|B_{a20}|}|\mathbf{b_1}-\mathbf{b_1'}|)
\right)\theta(-B_{a20})\right]\nonumber \\
&&\left[2\pi
K_0(\sqrt{C_{a20}}|\mathbf{b_1'}+\mathbf{b_q}|)\theta(C_{a20})+\pi^2\left(-N_0(\sqrt{|C_{a20}|}|\mathbf{b_1'}+\mathbf{b_q}|)+iJ_0(\sqrt{|C_{a20}|}|\mathbf{b_1'}+\mathbf{b_q}|)
\right)\theta(-C_{a20})\right]\nonumber \\
&&\left[2\pi
K_0(\sqrt{D_{a20}}|\mathbf{b_1}|)\theta(D_{a20})+\pi^2\left(-N_0(\sqrt{|D_{a20}|}|\mathbf{b_1}|)+iJ_0(\sqrt{|D_{a20}|}|\mathbf{b_1}|)
\right)\theta(-D_{a20})\right]\nonumber \\
\end{eqnarray}

\begin{eqnarray}
&&\Omega^{a21} =\nonumber \\
&&\left[2\pi
K_0(\sqrt{A_{a21}}|\mathbf{b_2'}|)\theta(A_{a21})+\pi^2\left(-N_0(\sqrt{|A_{a21}|}|\mathbf{b_2'}|)+
iJ_0(\sqrt{|A_{a21}|}|\mathbf{b_2'}|)
\right)\theta(-A_{a21})\right]\nonumber \\
&&\left[2\pi
K_0(\sqrt{B_{a21}}|\mathbf{b_1}+\mathbf{b_q}|)\theta(B_{a21})+\pi^2\left(-N_0(\sqrt{|B_{a21}|}|\mathbf{b_1}+\mathbf{b_q}|)+iJ_0(\sqrt{|B_{a21}|}|\mathbf{b_1}+\mathbf{b_q}|)
\right)\theta(-B_{a21})\right]\nonumber \\
&&\left[2\pi
K_0(\sqrt{C_{a21}}|\mathbf{b_1'}+\mathbf{b_q}|)\theta(C_{a21})+\pi^2\left(-N_0(\sqrt{|C_{a21}|}|\mathbf{b_1'}+\mathbf{b_q}|)+iJ_0(\sqrt{|C_{a21}|}|\mathbf{b_1'}+\mathbf{b_q}|)
\right)\theta(-C_{a21})\right]\nonumber \\
&&\left[2\pi
K_0(\sqrt{D_{a21}}|\mathbf{b_1}|)\theta(D_{a21})+\pi^2\left(-N_0(\sqrt{|D_{a21}|}|\mathbf{b_1}|)+iJ_0(\sqrt{|D_{a21}|}|\mathbf{b_1}|)
\right)\theta(-D_{a21})\right]\nonumber \\
\end{eqnarray}

\begin{eqnarray}
&&\Omega^{a22} =\nonumber \\
&&\left[2\pi
K_0(\sqrt{A_{a22}}|\mathbf{b_2'}-\mathbf{b_3'}+\mathbf{b_q}|)\theta(A_{a22})+\pi^2\left(-N_0(\sqrt{|A_{a22}|}|\mathbf{b_2'}-\mathbf{b_3'}+\mathbf{b_q}|)+
iJ_0(\sqrt{|A_{a22}|}|\mathbf{b_2'}-\mathbf{b_3'}+\mathbf{b_q}|)
\right)\theta(-A_{a22})\right]\nonumber \\
&&\left[2\pi
K_0(\sqrt{B_{a22}}|\mathbf{b_3'}|)\theta(B_{a22})+\pi^2\left(-N_0(\sqrt{|B_{a22}|}|\mathbf{b_3'}|)+iJ_0(\sqrt{|B_{a22}|}|\mathbf{b_3'}|)
\right)\theta(-B_{a22})\right]\nonumber \\
&&\left[2\pi
K_0(\sqrt{C_{a22}}|\mathbf{b_2}-\mathbf{b_2'}|)\theta(C_{a22})+\pi^2\left(-N_0(\sqrt{|C_{a22}|}|\mathbf{b_2}-\mathbf{b_2'}|)+iJ_0(\sqrt{|C_{a22}|}|\mathbf{b_2}-\mathbf{b_2'}|)
\right)\theta(-C_{a22})\right]\nonumber \\
&&[2\pi
K_0(\sqrt{D_{a22}}|\mathbf{b_2}-\mathbf{b_2'}+\mathbf{b_3'}-\mathbf{b_q}|)\theta(D_{a22})+\nonumber
\\
&&\pi^2\left(-N_0(\sqrt{|D_{a22}|}|\mathbf{b_2}-\mathbf{b_2'}+\mathbf{b_3'}-\mathbf{b_q}|)+iJ_0(\sqrt{|D_{a22}|}|\mathbf{b_2}-\mathbf{b_2'}+\mathbf{b_3'}-\mathbf{b_q}|)
\right)\theta(-D_{a22})]\nonumber \\
\end{eqnarray}

\begin{eqnarray}
&&\Omega^{a23} =\nonumber \\
&&\left[2\pi
K_0(\sqrt{A_{a23}}|\mathbf{b_2}|)\theta(A_{a23})+\pi^2\left(-N_0(\sqrt{|A_{a23}|}|\mathbf{b_2}|)+
iJ_0(\sqrt{|A_{a23}|}|\mathbf{b_2}|)
\right)\theta(-A_{a23})\right]\nonumber \\
&&\left[2\pi
K_0(\sqrt{B_{a23}}|\mathbf{b_3'}|)\theta(B_{a23})+\pi^2\left(-N_0(\sqrt{|B_{a23}|}|\mathbf{b_3'}|)+iJ_0(\sqrt{|B_{a23}|}|\mathbf{b_3'}|)
\right)\theta(-B_{a23})\right]\nonumber \\
&&\left[2\pi
K_0(\sqrt{C_{a23}}|\mathbf{b_3'}-\mathbf{b_q}|)\theta(C_{a23})+\pi^2\left(-N_0(\sqrt{|C_{a23}|}|\mathbf{b_3'}-\mathbf{b_q}|)+iJ_0(\sqrt{|C_{a23}|}|\mathbf{b_3'}-\mathbf{b_q}|)
\right)\theta(-C_{a23})\right]\nonumber \\
&&\left[2\pi
K_0(\sqrt{D_{a23}}|\mathbf{b_3}|)\theta(D_{a23})+\pi^2\left(-N_0(\sqrt{|D_{a23}|}|\mathbf{b_3}|)+iJ_0(\sqrt{|D_{a23}|}|\mathbf{b_3}|)
\right)\theta(-D_{a23})\right]\nonumber \\
\end{eqnarray}

\begin{eqnarray}
&&\Omega^{a24} =\nonumber \\
&&\left[2\pi
K_0(\sqrt{A_{a24}}|\mathbf{b_2}|)\theta(A_{a24})+\pi^2\left(-N_0(\sqrt{|A_{a24}|}|\mathbf{b_2}|)+
iJ_0(\sqrt{|A_{a24}|}|\mathbf{b_2}|)
\right)\theta(-A_{a24})\right]\nonumber \\
&&\left[2\pi
K_0(\sqrt{B_{a24}}|\mathbf{b_3'}|)\theta(B_{a24})+\pi^2\left(-N_0(\sqrt{|B_{a24}|}|\mathbf{b_3'}|)+iJ_0(\sqrt{|B_{a24}|}|\mathbf{b_3'}|)
\right)\theta(-B_{a24})\right]\nonumber \\
&&\left[2\pi
K_0(\sqrt{C_{a24}}|\mathbf{b_3'}-\mathbf{b_q}|)\theta(C_{a24})+\pi^2\left(-N_0(\sqrt{|C_{a24}|}|\mathbf{b_3'}-\mathbf{b_q}|)+iJ_0(\sqrt{|C_{a24}|}|\mathbf{b_3'}-\mathbf{b_q}|)
\right)\theta(-C_{a24})\right]\nonumber \\
&&\left[2\pi
K_0(\sqrt{D_{a24}}|\mathbf{b_1}|)\theta(D_{a24})+\pi^2\left(-N_0(\sqrt{|D_{a24}|}|\mathbf{b_1}|)+iJ_0(\sqrt{|D_{a24}|}|\mathbf{b_1}|)
\right)\theta(-D_{a24})\right]\nonumber \\
\end{eqnarray}

\begin{eqnarray}
&&\Omega^{a25} =\nonumber \\
&&\int_0^1 {dz_1 dz_2 \over {z_1
(1-z_1)}}\frac{\sqrt{X_2}}{\sqrt{|Z_2|}}\left[\pi^2K_1(\sqrt{X_2Z_2})\theta(Z_2)+\frac{\pi^3}{2}[N_1(\sqrt{X_2|Z_2|})-iJ_1(\sqrt{X_2|Z_2|})]\theta(-Z_2)\right]\nonumber \\
&&\left[2\pi K_0(\sqrt{B_{a25}}|\mathbf{b_3'}|)\theta(B_{a25})+
\pi^2[-N_0(\sqrt{|B_{a25}|}|\mathbf{b_3'}|)+iJ_0(\sqrt{|B_{a25}|}|\mathbf{b_3'}|)]\theta(-B_{a25})\right]\nonumber
\\
\end{eqnarray}
with
\begin{eqnarray}
&&Z_2=D_{a25}(1-z_2)+{z_2 \over z_1(1-z_1)}[A_{a25}(1-z_1)+C_{a25}z_1]\nonumber \\
&&X_2=[\mathbf{b_3'}-\mathbf{b_q}-z_1\mathbf{b_2'}]^2+{z_1(1-z_1)\over
z_2}\mathbf{b_2'}^2.
\end{eqnarray}

\begin{eqnarray}
&&\Omega^{a26} =\nonumber \\
&&\int_0^1 {dz_1 dz_2 \over {z_1
(1-z_1)}}\frac{\sqrt{X_2}}{\sqrt{|Z_2|}}\left[\pi^2K_1(\sqrt{X_2Z_2})\theta(Z_2)+\frac{\pi^3}{2}[N_1(\sqrt{X_2|Z_2|})-iJ_1(\sqrt{X_2|Z_2|})]\theta(-Z_2)\right]\nonumber \\
&&\left[2\pi K_0(\sqrt{A_{a26}}|\mathbf{b_2'}|)\theta(A_{a26})+
\pi^2[-N_0(\sqrt{|A_{a26}|}|\mathbf{b_2'}|)+iJ_0(\sqrt{|A_{a26}|}|\mathbf{b_2'}|)]\theta(-A_{a26})\right]\nonumber
\\
\end{eqnarray}
with
\begin{eqnarray}
&&Z_2=B_{a26}(1-z_2)+{z_2 \over z_1(1-z_1)}[D_{a26}(1-z_1)+C_{a26}z_1]\nonumber \\
&&X_2=[\mathbf{b_q}-z_1(-\mathbf{b_3'}+\mathbf{b_q})]^2+{z_1(1-z_1)\over
z_2}(\mathbf{b_3'}-\mathbf{b_q})^2.
\end{eqnarray}

\begin{eqnarray}
&&\Omega^{a27} =\nonumber \\
&&\left[2\pi
K_0(\sqrt{A_{a27}}|\mathbf{b_2}-\mathbf{b_q}|)\theta(A_{a27})+\pi^2\left(-N_0(\sqrt{|A_{a27}|}|\mathbf{b_2}-\mathbf{b_q}|)+iJ_0(\sqrt{|A_{a27}|}|\mathbf{b_2}-\mathbf{b_q}|)
\right)\theta(-A_{a27})\right]\nonumber \\
&&\left[2\pi
K_0(\sqrt{B_{a27}}|\mathbf{b_1'}|)\theta(B_{a27})+\pi^2\left(-N_0(\sqrt{|B_{a27}|}|\mathbf{b_1'}|)+iJ_0(\sqrt{|B_{a27}|}|\mathbf{b_1'}|)
\right)\theta(-B_{a27})\right]\nonumber \\
&&\left[2\pi
K_0(\sqrt{C_{a27}}|\mathbf{b_2}+\mathbf{b_1'}-\mathbf{b_2'}|)\theta(C_{a27})+\pi^2\left(-N_0(\sqrt{|C_{a27}|}|\mathbf{b_2}+\mathbf{b_1'}-\mathbf{b_2'}|)+iJ_0(\sqrt{|C_{a27}|}|\mathbf{b_2}+\mathbf{b_1'}-\mathbf{b_2'}|)
\right)\theta(-C_{a27})\right]\nonumber \\
&&[2\pi
K_0(\sqrt{D_{a27}}|-\mathbf{b_2}+\mathbf{b_2'}+\mathbf{b_q}|)\theta(D_{a27})+\nonumber
\\
&&\pi^2\left(-N_0(\sqrt{|D_{a27}|}|-\mathbf{b_2}+\mathbf{b_2'}+\mathbf{b_q}|)+iJ_0(\sqrt{|D_{a27}|}|-\mathbf{b_2}+\mathbf{b_2'}+\mathbf{b_q}|)
\right)\theta(-D_{a27})]\nonumber \\
\end{eqnarray}

\begin{eqnarray}
&&\Omega^{a28} =\nonumber \\
&&\left[2\pi
K_0(\sqrt{A_{a28}}|\mathbf{b_2'}|)\theta(A_{a28})+\pi^2\left(-N_0(\sqrt{|A_{a28}|}|\mathbf{b_2'}|)+iJ_0(\sqrt{|A_{a28}|}|\mathbf{b_2'}|)
\right)\theta(-A_{a28})\right]\nonumber \\
&&[2\pi
K_0(\sqrt{B_{a28}}|-\mathbf{b_2}-\mathbf{b_1'}+\mathbf{b_2'}+\mathbf{b_q}|)\theta(B_{a28})+\nonumber
\\
&&\pi^2\left(-N_0(\sqrt{|B_{a28}|}|-\mathbf{b_2}-\mathbf{b_1'}+\mathbf{b_2'}+\mathbf{b_q}|)+iJ_0(\sqrt{|B_{a28}|}|-\mathbf{b_2}-\mathbf{b_1'}+\mathbf{b_2'}+\mathbf{b_q}|)
\right)\theta(-B_{a28})]\nonumber \\
&&\left[2\pi
K_0(\sqrt{C_{a28}}|\mathbf{b_2}+\mathbf{b_1'}-\mathbf{b_2'}|)\theta(C_{a28})+\pi^2\left(-N_0(\sqrt{|C_{a28}|}|\mathbf{b_2}+\mathbf{b_1'}-\mathbf{b_2'}|)+iJ_0(\sqrt{|C_{a28}|}|\mathbf{b_2}+\mathbf{b_1'}-\mathbf{b_2'}|)
\right)\theta(-C_{a28})\right]\nonumber \\
&&[2\pi
K_0(\sqrt{D_{a28}}|-\mathbf{b_2}+\mathbf{b_2'}+\mathbf{b_q}|)\theta(D_{a28})+\nonumber
\\
&&\pi^2\left(-N_0(\sqrt{|D_{a28}|}|-\mathbf{b_2}+\mathbf{b_2'}+\mathbf{b_q}|)+iJ_0(\sqrt{|D_{a28}|}|-\mathbf{b_2}+\mathbf{b_2'}+\mathbf{b_q}|)
\right)\theta(-D_{a28})]\nonumber \\
\end{eqnarray}

\begin{eqnarray}
&&\Omega^{a29} =\nonumber \\
&&\left[2\pi
K_0(\sqrt{A_{a29}}|\mathbf{b_2}|)\theta(A_{a29})+\pi^2\left(-N_0(\sqrt{|A_{a29}|}|\mathbf{b_2}|)+iJ_0(\sqrt{|A_{a29}|}|\mathbf{b_2}|)
\right)\theta(-A_{a29})\right]\nonumber \\
&&\left[2\pi
K_0(\sqrt{B_{a29}}|\mathbf{b_1'}+\mathbf{b_q}|)\theta(B_{a29})+\pi^2\left(-N_0(\sqrt{|B_{a29}|}|\mathbf{b_1'}+\mathbf{b_q}|)+iJ_0(\sqrt{|B_{a29}|}|\mathbf{b_1'}+\mathbf{b_q}|)
\right)\theta(-B_{a29})\right]\nonumber \\
&&\left[2\pi
K_0(\sqrt{C_{a29}}|\mathbf{b_2}-\mathbf{b_3}+\mathbf{b_1'}|)\theta(C_{a29})+\pi^2\left(-N_0(\sqrt{|C_{a29}|}|\mathbf{b_2}-\mathbf{b_3}+\mathbf{b_1'}|)+iJ_0(\sqrt{|C_{a29}|}|\mathbf{b_2}-\mathbf{b_3}+\mathbf{b_1'}|)
\right)\theta(-C_{a29})\right]\nonumber \\
&&\left[2\pi
K_0(\sqrt{D_{a29}}|\mathbf{b_3}|)\theta(D_{a29})+\pi^2\left(-N_0(\sqrt{|D_{a29}|}|\mathbf{b_3}|)+iJ_0(\sqrt{|D_{a29}|}|\mathbf{b_3}|)
\right)\theta(-D_{a29})\right]\nonumber \\
\end{eqnarray}

\begin{eqnarray}
&&\Omega^{a30} =\nonumber \\
&&\left[2\pi
K_0(\sqrt{A_{a30}}|\mathbf{b_2}|)\theta(A_{a30})+\pi^2\left(-N_0(\sqrt{|A_{a30}|}|\mathbf{b_2}|)+iJ_0(\sqrt{|A_{a30}|}|\mathbf{b_2}|)
\right)\theta(-A_{a30})\right]\nonumber \\
&&\left[2\pi
K_0(\sqrt{B_{a30}}|\mathbf{b_1'}+\mathbf{b_q}|)\theta(B_{a30})+\pi^2\left(-N_0(\sqrt{|B_{a30}|}|\mathbf{b_1'}+\mathbf{b_q}|)+iJ_0(\sqrt{|B_{a30}|}|\mathbf{b_1'}+\mathbf{b_q}|)
\right)\theta(-B_{a30})\right]\nonumber \\
&&[2\pi
K_0(\sqrt{C_{a30}}|-\mathbf{b_1}+\mathbf{b_2}+\mathbf{b_1'}|)\theta(C_{a30})+\nonumber
\\
&&\pi^2\left(-N_0(\sqrt{|C_{a30}|}|-\mathbf{b_1}+\mathbf{b_2}+\mathbf{b_1'}|)+iJ_0(\sqrt{|C_{a30}|}|-\mathbf{b_1}+\mathbf{b_2}+\mathbf{b_1'}|)
\right)\theta(-C_{a30})]\nonumber \\
&&\left[2\pi
K_0(\sqrt{D_{a30}}|\mathbf{b_1}|)\theta(D_{a30})+\pi^2\left(-N_0(\sqrt{|D_{a30}|}|\mathbf{b_1}|)+iJ_0(\sqrt{|D_{a30}|}|\mathbf{b_1}|)
\right)\theta(-D_{a30})\right]\nonumber \\
\end{eqnarray}

\begin{eqnarray}
&&\Omega^{a31} =\nonumber \\
&&[2\pi
K_0(\sqrt{A_{a31}}|\mathbf{b_2}+\mathbf{b_1'}-\mathbf{b_3'}+\mathbf{b_q}|)\theta(A_{a31})+\nonumber
\\
&&\pi^2\left(-N_0(\sqrt{|A_{a31}|}|\mathbf{b_2}+\mathbf{b_1'}-\mathbf{b_3'}+\mathbf{b_q}|)+iJ_0(\sqrt{|A_{a31}|}|\mathbf{b_2}+\mathbf{b_1'}-\mathbf{b_3'}+\mathbf{b_q}|)
\right)\theta(-A_{a31})]\nonumber \\
&&\left[2\pi
K_0(\sqrt{B_{a31}}|\mathbf{b_3'}|)\theta(B_{a31})+\pi^2\left(-N_0(\sqrt{|B_{a31}|}|\mathbf{b_3'}|)+iJ_0(\sqrt{|B_{a31}|}|\mathbf{b_3'}|)
\right)\theta(-B_{a31})\right]\nonumber \\
&&\left[2\pi
K_0(\sqrt{C_{a31}}|\mathbf{b_2}+\mathbf{b_1'}|)\theta(C_{a31})+\pi^2\left(-N_0(\sqrt{|C_{a31}|}|\mathbf{b_2}+\mathbf{b_1'}|)+iJ_0(\sqrt{|C_{a31}|}|\mathbf{b_2}+\mathbf{b_1'}|)
\right)\theta(-C_{a31})\right]\nonumber \\
&&[2\pi
K_0(\sqrt{D_{a31}}|-\mathbf{b_1'}+\mathbf{b_3'}-\mathbf{b_q}|)\theta(D_{a31})+\nonumber
\\
&&\pi^2\left(-N_0(\sqrt{|D_{a31}|}|-\mathbf{b_1'}+\mathbf{b_3'}-\mathbf{b_q}|)+iJ_0(\sqrt{|D_{a31}|}|-\mathbf{b_1'}+\mathbf{b_3'}-\mathbf{b_q}|)
\right)\theta(-D_{a31})]\nonumber \\
\end{eqnarray}

\begin{eqnarray}
&&\Omega^{a32} =\nonumber \\
&&[2\pi
K_0(\sqrt{A_{a32}}|\mathbf{b_2}-\mathbf{b_1'}+\mathbf{b_3'}-\mathbf{b_q}|)\theta(A_{a32})+\nonumber
\\
&&\pi^2\left(-N_0(\sqrt{|A_{a32}|}|\mathbf{b_2}-\mathbf{b_1'}+\mathbf{b_3'}-\mathbf{b_q}|)+iJ_0(\sqrt{|A_{a32}|}|\mathbf{b_2}-\mathbf{b_1'}+\mathbf{b_3'}-\mathbf{b_q}|)
\right)\theta(-A_{a32})]\nonumber \\
&&\left[2\pi
K_0(\sqrt{B_{a32}}|\mathbf{b_3'}|)\theta(B_{a32})+\pi^2\left(-N_0(\sqrt{|B_{a32}|}|\mathbf{b_3'}|)+iJ_0(\sqrt{|B_{a32}|}|\mathbf{b_3'}|)
\right)\theta(-B_{a32})\right]\nonumber \\
&&\left[2\pi
K_0(\sqrt{C_{a32}}|\mathbf{b_1'}-\mathbf{b_3'}+\mathbf{b_q}|)\theta(C_{a32})+\pi^2\left(-N_0(\sqrt{|C_{a32}|}|\mathbf{b_1'}-\mathbf{b_3'}+\mathbf{b_q}|)+iJ_0(\sqrt{|C_{a32}|}|\mathbf{b_1'}-\mathbf{b_3'}+\mathbf{b_q}|)
\right)\theta(-C_{a32})\right]\nonumber \\
&&\left[2\pi
K_0(\sqrt{D_{a32}}|\mathbf{b_2}+\mathbf{b_3'}-\mathbf{b_q}|)\theta(D_{a32})+\pi^2\left(-N_0(\sqrt{|D_{a32}|}|\mathbf{b_2}+\mathbf{b_3'}-\mathbf{b_q}|)+iJ_0(\sqrt{|D_{a32}|}|\mathbf{b_2}+\mathbf{b_3'}-\mathbf{b_q}|)
\right)\theta(-D_{a32})\right]\nonumber \\
\end{eqnarray}

\begin{eqnarray}
&&\Omega^{a33} =\nonumber \\
&&[2\pi
K_0(\sqrt{A_{a33}}|-\mathbf{b_2}+\mathbf{b_3}+\mathbf{b_q}|)\theta(A_{a33})+\nonumber
\\
&&\pi^2\left(-N_0(\sqrt{|A_{a33}|}|-\mathbf{b_2}+\mathbf{b_3}+\mathbf{b_q}|)+iJ_0(\sqrt{|A_{a33}|}|-\mathbf{b_2}+\mathbf{b_3}+\mathbf{b_q}|)
\right)\theta(-A_{a33})]\nonumber \\
&&\left[2\pi
K_0(\sqrt{B_{a33}}|\mathbf{b_3}-\mathbf{b_1'}|)\theta(B_{a33})+\pi^2\left(-N_0(\sqrt{|B_{a33}|}|\mathbf{b_3}-\mathbf{b_1'}|)+iJ_0(\sqrt{|B_{a33}|}|\mathbf{b_3}-\mathbf{b_1'}|)
\right)\theta(-B_{a33})\right]\nonumber \\
&&\left[2\pi
K_0(\sqrt{C_{a33}}|\mathbf{b_2}-\mathbf{b_3}+\mathbf{b_1'}|)\theta(C_{a33})+\pi^2\left(-N_0(\sqrt{|C_{a33}|}|\mathbf{b_2}-\mathbf{b_3}+\mathbf{b_1'}|)+iJ_0(\sqrt{|C_{a33}|}|\mathbf{b_2}-\mathbf{b_3}+\mathbf{b_1'}|)
\right)\theta(-C_{a33})\right]\nonumber \\
&&\left[2\pi
K_0(\sqrt{D_{a33}}|\mathbf{b_3}|)\theta(D_{a33})+\pi^2\left(-N_0(\sqrt{|D_{a33}|}|\mathbf{b_3}|)+iJ_0(\sqrt{|D_{a33}|}|\mathbf{b_3}|)
\right)\theta(-D_{a33})\right]\nonumber \\
\end{eqnarray}

\begin{eqnarray}
&&\Omega^{a34} =\nonumber \\
&&[2\pi
K_0(\sqrt{A_{a34}}|\mathbf{b_1}-\mathbf{b_2}+\mathbf{b_q}|)\theta(A_{a34})+\nonumber
\\
&&\pi^2\left(-N_0(\sqrt{|A_{a34}|}|\mathbf{b_1}-\mathbf{b_2}+\mathbf{b_q}|)+iJ_0(\sqrt{|A_{a34}|}|\mathbf{b_1}-\mathbf{b_2}+\mathbf{b_q}|)
\right)\theta(-A_{a34})]\nonumber \\
&&\left[2\pi
K_0(\sqrt{B_{a34}}|\mathbf{b_1}-\mathbf{b_1'}|)\theta(B_{a34})+\pi^2\left(-N_0(\sqrt{|B_{a34}|}|\mathbf{b_1}-\mathbf{b_1'}|)+iJ_0(\sqrt{|B_{a34}|}|\mathbf{b_1}-\mathbf{b_1'}|)
\right)\theta(-B_{a34})\right]\nonumber \\
&&[2\pi
K_0(\sqrt{C_{a34}}|-\mathbf{b_1}+\mathbf{b_2}+\mathbf{b_1'}|)\theta(C_{a34})+\nonumber
\\
&&\pi^2\left(-N_0(\sqrt{|C_{a34}|}|-\mathbf{b_1}+\mathbf{b_2}+\mathbf{b_1'}|)+iJ_0(\sqrt{|C_{a34}|}|-\mathbf{b_1}+\mathbf{b_2}+\mathbf{b_1'}|)
\right)\theta(-C_{a34})]\nonumber \\
&&\left[2\pi
K_0(\sqrt{D_{a34}}|\mathbf{b_1}|)\theta(D_{a34})+\pi^2\left(-N_0(\sqrt{|D_{a34}|}|\mathbf{b_1}|)+iJ_0(\sqrt{|D_{a34}|}|\mathbf{b_1}|)
\right)\theta(-D_{a34})\right]\nonumber \\
\end{eqnarray}

\begin{eqnarray}
&&\Omega^{a35} =\nonumber \\
&&[2\pi
K_0(\sqrt{A_{a35}}|-\mathbf{b_2}-\mathbf{b_1'}+\mathbf{b_3'}|)\theta(A_{a35})+\nonumber
\\
&&\pi^2\left(-N_0(\sqrt{|A_{a35}|}|-\mathbf{b_2}-\mathbf{b_1'}+\mathbf{b_3'}|)+iJ_0(\sqrt{|A_{a35}|}|-\mathbf{b_2}-\mathbf{b_1'}+\mathbf{b_3'}|)
\right)\theta(-A_{a35})]\nonumber \\
&&\left[2\pi
K_0(\sqrt{B_{a35}}|-\mathbf{b_3'}+\mathbf{b_q}|)\theta(B_{a35})+\pi^2\left(-N_0(\sqrt{|B_{a35}|}|-\mathbf{b_3'}+\mathbf{b_q}|)+iJ_0(\sqrt{|B_{a35}|}|-\mathbf{b_3'}+\mathbf{b_q}|)
\right)\theta(-B_{a35})\right]\nonumber \\
&&\left[2\pi
K_0(\sqrt{C_{a35}}|\mathbf{b_2}+\mathbf{b_1'}|)\theta(C_{a35})+\pi^2\left(-N_0(\sqrt{|C_{a35}|}|\mathbf{b_2}+\mathbf{b_1'}|)+iJ_0(\sqrt{|C_{a35}|}|\mathbf{b_2}+\mathbf{b_1'}|)
\right)\theta(-C_{a35})\right]\nonumber \\
&&[2\pi
K_0(\sqrt{D_{a35}}|-\mathbf{b_1'}+\mathbf{b_3'}-\mathbf{b_q}|)\theta(D_{a35})+\nonumber
\\
&&\pi^2\left(-N_0(\sqrt{|D_{a35}|}|-\mathbf{b_1'}+\mathbf{b_3'}-\mathbf{b_q}|)+iJ_0(\sqrt{|D_{a35}|}|-\mathbf{b_1'}+\mathbf{b_3'}-\mathbf{b_q}|)
\right)\theta(-D_{a35})]\nonumber \\
\end{eqnarray}

\begin{eqnarray}
&&\Omega^{a36} =\nonumber \\
&&\left[2\pi
K_0(\sqrt{A_{a36}}|\mathbf{b_2}-\mathbf{b_q}|)\theta(A_{a36})+\pi^2\left(-N_0(\sqrt{|A_{a36}|}|\mathbf{b_2}-\mathbf{b_q}|)+iJ_0(\sqrt{|A_{a36}|}|\mathbf{b_2}-\mathbf{b_q}|)
\right)\theta(-A_{a36})\right]\nonumber \\
&&\left[2\pi
K_0(\sqrt{B_{a36}}|\mathbf{b_2'}|)\theta(B_{a36})+\pi^2\left(-N_0(\sqrt{|B_{a36}|}|\mathbf{b_2'}|)+iJ_0(\sqrt{|B_{a36}|}|\mathbf{b_2'}|)
\right)\theta(-B_{a36})\right]\nonumber \\
&&[2\pi
K_0(\sqrt{C_{a36}}|\mathbf{b_2}-\mathbf{b_2'}+\mathbf{b_3'}-\mathbf{b_q}|)\theta(C_{a36})+\nonumber
\\
&&\pi^2\left(-N_0(\sqrt{|C_{a36}|}|\mathbf{b_2}-\mathbf{b_2'}+\mathbf{b_3'}-\mathbf{b_q}|)+
iJ_0(\sqrt{|C_{a36}|}|\mathbf{b_2}-\mathbf{b_2'}+\mathbf{b_3'}-\mathbf{b_q}|)
\right)\theta(-C_{a36})]\nonumber \\
&&\left[2\pi
K_0(\sqrt{D_{a36}}|\mathbf{b_3'}-\mathbf{b_q}|)\theta(D_{a36})+\pi^2\left(-N_0(\sqrt{|D_{a36}|}|\mathbf{b_3'}-\mathbf{b_q}|)+iJ_0(\sqrt{|D_{a36}|}|\mathbf{b_3'}-\mathbf{b_q}|)
\right)\theta(-D_{a36})\right]\nonumber \\
\end{eqnarray}

\begin{eqnarray}
&&\Omega^{b1} =\nonumber \\
&&\int_0^1 {dz_1 dz_2 dz_3 \over z_1
(1-z_1)z_2(1-z_2)}\frac{\sqrt{X_3}}{\sqrt{|Z_3|}}\left[\pi^3K_1(\sqrt{X_3Z_3})\theta(Z_3)+\frac{\pi^4}{2}[N_1(\sqrt{X_3|Z_3|})-iJ_1(\sqrt{X_3|Z_3|})]\theta(-Z_3)\right]\nonumber \\
\end{eqnarray}
with
\begin{eqnarray}
&&Z_3=B_{b1}(1-z_3)+{z_3 \over z_2(1-z_2)}\left[C_{b1}(1-z_2)+{z_2\over z_1(1-z_1)}[A_{b1}(1-z_1)+D_{b1}z_1]\right]\nonumber \\
&&X_3=[-\mathbf{b_1'}+z_2(\mathbf{b_1'}+\mathbf{b_q})-(\mathbf{b_2}-\mathbf{b_q})z_2(1-z_1)]^2+{z_2(1-z_2)\over
z_3}[-\mathbf{b_1'}-\mathbf{b_q}-(\mathbf{b_2}-\mathbf{b_q})z_1]^2.
\end{eqnarray}

\begin{eqnarray}
&&\Omega^{b2} =\nonumber \\
&&\int_0^1
dz\frac{|\mathbf{b_1'}+\mathbf{b_q}|}{\sqrt{|Z_1|}}\left[\pi
K_1(\sqrt{Z_1}|\mathbf{b_1'}+\mathbf{b_q}|)\theta(Z_1)+
\frac{\pi^2}{2}[N_1(\sqrt{|Z_1|}|\mathbf{b_1'}+\mathbf{b_q}|)-iJ_1(\sqrt{|Z_1|}|\mathbf{b_1'}+\mathbf{b_q}|)]\theta(-Z_1)\right]\nonumber \\
&&\left[2\pi
K_0(\sqrt{A_{b2}}|\mathbf{b_2}+\mathbf{b_1'}|)\theta(A_{b2})+\pi^2\left(-N_0(\sqrt{|A_{b2}|}|\mathbf{b_2}+\mathbf{b_1'}|)+
iJ_0(\sqrt{|A_{b2}|}|\mathbf{b_2}+\mathbf{b_1'}|)
\right)\theta(-A_{b2})\right]\nonumber \\
&&\left[2\pi
K_0(\sqrt{B_{b2}}|\mathbf{b_q}|)\theta(B_{b2})+\pi^2\left(-N_0(\sqrt{|B_{b2}|}|\mathbf{b_q}|)+
iJ_0(\sqrt{|B_{b2}|}|\mathbf{b_q}|)
\right)\theta(-B_{b2})\right]\nonumber \\
\end{eqnarray}
with
\begin{eqnarray}
&&Z_1=C_{b2}z+D_{b2}(1-z).
\end{eqnarray}

\begin{eqnarray}
&&\Omega^{b3} =\nonumber \\
&&\left[2\pi
K_0(\sqrt{A_{b3}}|\mathbf{b_2}-\mathbf{b_3}-\mathbf{b_q}|)\theta(A_{b3})+\pi^2\left(-N_0(\sqrt{|A_{b3}|}|\mathbf{b_2}-\mathbf{b_3}-\mathbf{b_q}|)+
iJ_0(\sqrt{|A_{b3}|}|\mathbf{b_2}-\mathbf{b_3}-\mathbf{b_q}|)
\right)\theta(-A_{b3})\right]\nonumber \\
&&\left[2\pi
K_0(\sqrt{B_{b3}}|\mathbf{b_3}-\mathbf{b_1'}|)\theta(B_{b3})+\pi^2\left(-N_0(\sqrt{|B_{b3}|}|\mathbf{b_3}-\mathbf{b_1'}|)+iJ_0(\sqrt{|B_{b3}|}|\mathbf{b_3}-\mathbf{b_1'}|)
\right)\theta(-B_{b3})\right]\nonumber \\
&&\left[2\pi
K_0(\sqrt{C_{b3}}|\mathbf{b_1'}+\mathbf{b_q}|)\theta(C_{b3})+\pi^2\left(-N_0(\sqrt{|C_{b3}|}|\mathbf{b_1'}+\mathbf{b_q}|)+iJ_0(\sqrt{|C_{b3}|}|\mathbf{b_1'}+\mathbf{b_q}|)
\right)\theta(-C_{b3})\right]\nonumber \\
&&\left[2\pi
K_0(\sqrt{D_{b3}}|\mathbf{b_3}|)\theta(D_{b3})+\pi^2\left(-N_0(\sqrt{|D_{b3}|}|\mathbf{b_3}|)+iJ_0(\sqrt{|D_{b3}|}|\mathbf{b_3}|)
\right)\theta(-D_{b3})\right]\nonumber \\
\end{eqnarray}

\begin{eqnarray}
&&\Omega^{b4} =\nonumber \\
&&\left[2\pi
K_0(\sqrt{A_{b4}}|\mathbf{b_2}-\mathbf{b_3}+\mathbf{b_1'}|)\theta(A_{b4})+\pi^2\left(-N_0(\sqrt{|A_{b4}|}|\mathbf{b_2}-\mathbf{b_3}+\mathbf{b_1'}|)+
iJ_0(\sqrt{|A_{b4}|}|\mathbf{b_2}-\mathbf{b_3}+\mathbf{b_1'}|)
\right)\theta(-A_{b4})\right]\nonumber \\
&&\left[2\pi
K_0(\sqrt{B_{b4}}|\mathbf{b_3}+\mathbf{b_q}|)\theta(B_{b4})+\pi^2\left(-N_0(\sqrt{|B_{b4}|}|\mathbf{b_3}+\mathbf{b_q}|)+iJ_0(\sqrt{|B_{b4}|}|\mathbf{b_3}+\mathbf{b_q}|)
\right)\theta(-B_{b4})\right]\nonumber \\
&&\left[2\pi
K_0(\sqrt{C_{b4}}|\mathbf{b_1'}+\mathbf{b_q}|)\theta(C_{b4})+\pi^2\left(-N_0(\sqrt{|C_{b4}|}|\mathbf{b_1'}+\mathbf{b_q}|)+iJ_0(\sqrt{|C_{b4}|}|\mathbf{b_1'}+\mathbf{b_q}|)
\right)\theta(-C_{b4})\right]\nonumber \\
&&\left[2\pi
K_0(\sqrt{D_{b4}}|\mathbf{b_3}|)\theta(D_{b4})+\pi^2\left(-N_0(\sqrt{|D_{b4}|}|\mathbf{b_3}|)+iJ_0(\sqrt{|D_{b4}|}|\mathbf{b_3}|)
\right)\theta(-D_{b4})\right]\nonumber \\
\end{eqnarray}

\begin{eqnarray}
&&\Omega^{b5} =\nonumber \\
&&\left[2\pi
K_0(\sqrt{A_{b5}}|\mathbf{b_1}-\mathbf{b_2}+\mathbf{b_q}|)\theta(A_{b5})+\pi^2\left(-N_0(\sqrt{|A_{b5}|}|\mathbf{b_1}-\mathbf{b_2}+\mathbf{b_q}|)+
iJ_0(\sqrt{|A_{b5}|}|\mathbf{b_1}-\mathbf{b_2}+\mathbf{b_q}|)
\right)\theta(-A_{b5})\right]\nonumber \\
&&\left[2\pi
K_0(\sqrt{B_{b5}}|\mathbf{b_1}-\mathbf{b_1'}|)\theta(B_{b5})+\pi^2\left(-N_0(\sqrt{|B_{b5}|}|\mathbf{b_1}-\mathbf{b_1'}|)+iJ_0(\sqrt{|B_{b5}|}|\mathbf{b_1}-\mathbf{b_1'}|)
\right)\theta(-B_{b5})\right]\nonumber \\
&&\left[2\pi
K_0(\sqrt{C_{b5}}|\mathbf{b_1'}+\mathbf{b_q}|)\theta(C_{b5})+\pi^2\left(-N_0(\sqrt{|C_{b5}|}|\mathbf{b_1'}+\mathbf{b_q}|)+iJ_0(\sqrt{|C_{b5}|}|\mathbf{b_1'}+\mathbf{b_q}|)
\right)\theta(-C_{b5})\right]\nonumber \\
&&\left[2\pi
K_0(\sqrt{D_{b5}}|\mathbf{b_1}|)\theta(D_{b5})+\pi^2\left(-N_0(\sqrt{|D_{b5}|}|\mathbf{b_1}|)+iJ_0(\sqrt{|D_{b5}|}|\mathbf{b_1}|)
\right)\theta(-D_{b5})\right]\nonumber \\
\end{eqnarray}

\begin{eqnarray}
&&\Omega^{b6} =\nonumber \\
&&\left[2\pi
K_0(\sqrt{A_{b6}}|\mathbf{b_2'}|)\theta(A_{b6})+\pi^2\left(-N_0(\sqrt{|A_{b6}|}|\mathbf{b_2'}|)+
iJ_0(\sqrt{|A_{b6}|}|\mathbf{b_2'}|)
\right)\theta(-A_{b6})\right]\nonumber \\
&&\left[2\pi
K_0(\sqrt{B_{b6}}|\mathbf{b_1}+\mathbf{b_q}|)\theta(B_{b6})+\pi^2\left(-N_0(\sqrt{|B_{b6}|}|\mathbf{b_1}+\mathbf{b_q}|)+iJ_0(\sqrt{|B_{b6}|}|\mathbf{b_1}+\mathbf{b_q}|)
\right)\theta(-B_{b6})\right]\nonumber \\
&&\left[2\pi
K_0(\sqrt{C_{b6}}|\mathbf{b_1'}+\mathbf{b_q}|)\theta(C_{b6})+\pi^2\left(-N_0(\sqrt{|C_{b6}|}|\mathbf{b_1'}+\mathbf{b_q}|)+iJ_0(\sqrt{|C_{b6}|}|\mathbf{b_1'}+\mathbf{b_q}|)
\right)\theta(-C_{b6})\right]\nonumber \\
&&\left[2\pi
K_0(\sqrt{D_{b6}}|\mathbf{b_1}|)\theta(D_{b6})+\pi^2\left(-N_0(\sqrt{|D_{b6}|}|\mathbf{b_1}|)+iJ_0(\sqrt{|D_{b6}|}|\mathbf{b_1}|)
\right)\theta(-D_{b6})\right]\nonumber \\
\end{eqnarray}

\begin{eqnarray}
&&\Omega^{b7} =\nonumber \\
&&\left[2\pi
K_0(\sqrt{A_{b7}}|\mathbf{b_2}-\mathbf{b_q}|)\theta(A_{b7})+\pi^2\left(-N_0(\sqrt{|A_{b7}|}|\mathbf{b_2}-\mathbf{b_q}|)+iJ_0(\sqrt{|A_{b7}|}|\mathbf{b_2}-\mathbf{b_q}|)
\right)\theta(-A_{b7})\right]\nonumber \\
&&\left[2\pi
K_0(\sqrt{B_{b7}}|\mathbf{b_1'}|)\theta(B_{b7})+\pi^2\left(-N_0(\sqrt{|B_{b7}|}|\mathbf{b_1'}|)+iJ_0(\sqrt{|B_{b7}|}|\mathbf{b_1'}|)
\right)\theta(-B_{b7})\right]\nonumber \\
&&\left[2\pi
K_0(\sqrt{C_{b7}}|\mathbf{b_2}+\mathbf{b_1'}-\mathbf{b_2'}|)\theta(C_{b7})+\pi^2\left(-N_0(\sqrt{|C_{b7}|}|\mathbf{b_2}+\mathbf{b_1'}-\mathbf{b_2'}|)+iJ_0(\sqrt{|C_{b7}|}|\mathbf{b_2}+\mathbf{b_1'}-\mathbf{b_2'}|)
\right)\theta(-C_{b7})\right]\nonumber \\
&&[2\pi
K_0(\sqrt{D_{b7}}|-\mathbf{b_2}+\mathbf{b_2'}+\mathbf{b_q}|)\theta(D_{b7})+\nonumber
\\
&&\pi^2\left(-N_0(\sqrt{|D_{b7}|}|-\mathbf{b_2}+\mathbf{b_2'}+\mathbf{b_q}|)+iJ_0(\sqrt{|D_{b7}|}|-\mathbf{b_2}+\mathbf{b_2'}+\mathbf{b_q}|)
\right)\theta(-D_{b7})]\nonumber \\
\end{eqnarray}

\begin{eqnarray}
&&\Omega^{b8} =\nonumber \\
&&\left[2\pi
K_0(\sqrt{A_{b8}}|\mathbf{b_2'}|)\theta(A_{b8})+\pi^2\left(-N_0(\sqrt{|A_{b8}|}|\mathbf{b_2'}|)+iJ_0(\sqrt{|A_{b8}|}|\mathbf{b_2'}|)
\right)\theta(-A_{b8})\right]\nonumber \\
&&[2\pi
K_0(\sqrt{B_{b8}}|-\mathbf{b_2}-\mathbf{b_1'}+\mathbf{b_2'}+\mathbf{b_q}|)\theta(B_{b8})+\nonumber
\\
&&\pi^2\left(-N_0(\sqrt{|B_{b8}|}|-\mathbf{b_2}-\mathbf{b_1'}+\mathbf{b_2'}+\mathbf{b_q}|)+iJ_0(\sqrt{|B_{b8}|}|-\mathbf{b_2}-\mathbf{b_1'}+\mathbf{b_2'}+\mathbf{b_q}|)
\right)\theta(-B_{b8})]\nonumber \\
&&\left[2\pi
K_0(\sqrt{C_{b8}}|\mathbf{b_2}+\mathbf{b_1'}-\mathbf{b_2'}|)\theta(C_{b8})+\pi^2\left(-N_0(\sqrt{|C_{b8}|}|\mathbf{b_2}+\mathbf{b_1'}-\mathbf{b_2'}|)+iJ_0(\sqrt{|C_{b8}|}|\mathbf{b_2}+\mathbf{b_1'}-\mathbf{b_2'}|)
\right)\theta(-C_{b8})\right]\nonumber \\
&&[2\pi
K_0(\sqrt{D_{b8}}|-\mathbf{b_2}+\mathbf{b_2'}+\mathbf{b_q}|)\theta(D_{b8})+\nonumber
\\
&&\pi^2\left(-N_0(\sqrt{|D_{b8}|}|-\mathbf{b_2}+\mathbf{b_2'}+\mathbf{b_q}|)+iJ_0(\sqrt{|D_{b8}|}|-\mathbf{b_2}+\mathbf{b_2'}+\mathbf{b_q}|)
\right)\theta(-D_{b8})]\nonumber \\
\end{eqnarray}

\begin{eqnarray}
&&\Omega^{c1} =\nonumber \\
&&{1\over 2}\left[2\pi K_0(\sqrt{A_{c1}}{1\over
2}|\mathbf{b_2}-\mathbf{b_1'}+\mathbf{b_2'}|)\theta(A_{c1})+\pi^2\left(-N_0(\sqrt{|A_{c1}|}{1\over
2}|\mathbf{b_2}-\mathbf{b_1'}+\mathbf{b_2'}|)+iJ_0(\sqrt{|A_{c1}|}{1\over
2}|\mathbf{b_2}-\mathbf{b_1'}+\mathbf{b_2'}|)
\right)\theta(-A_{c1})\right]\nonumber \\
&&\left[2\pi K_0(\sqrt{B_{c1}}{1\over
2}|\mathbf{b_2}-\mathbf{b_1'}-\mathbf{b_2'}|)\theta(B_{c1})+\pi^2\left(-N_0(\sqrt{|B_{c1}|}{1\over
2}|\mathbf{b_2}-\mathbf{b_1'}-\mathbf{b_2'}|)+iJ_0(\sqrt{|B_{c1}|}{1\over
2}|\mathbf{b_2}-\mathbf{b_1'}-\mathbf{b_2'}|)
\right)\theta(-B_{c1})\right]\nonumber \\
&&[2\pi K_0(\sqrt{C_{c1}}{1\over
2}|-\mathbf{b_2}+\mathbf{b_1'}+\mathbf{b_2'}+2\mathbf{b_q}|)\theta(C_{c1})+\nonumber
\\
&&\pi^2\left(-N_0(\sqrt{|C_{c1}|}{1\over
2}|-\mathbf{b_2}+\mathbf{b_1'}+\mathbf{b_2'}+2\mathbf{b_q}|)+iJ_0(\sqrt{|C_{c1}|}{1\over
2}|-\mathbf{b_2}+\mathbf{b_1'}+\mathbf{b_2'}+2\mathbf{b_q}|)
\right)\theta(-C_{c1})]\nonumber \\
&&\left[2\pi
K_0(\sqrt{D_{c1}}|\mathbf{b_2}-\mathbf{b_2'}-\mathbf{b_q}|)\theta(D_{c1})+\pi^2\left(-N_0(\sqrt{|D_{c1}|}|\mathbf{b_2}-\mathbf{b_2'}-\mathbf{b_q}|)+iJ_0(\sqrt{|D_{c1}|}|\mathbf{b_2}-\mathbf{b_2'}-\mathbf{b_q}|)
\right)\theta(-D_{c1})\right]\nonumber \\
\end{eqnarray}

\begin{eqnarray}
&&\Omega^{c2} =\nonumber \\
&&{1\over 2}[2\pi K_0(\sqrt{A_{c2}}{1\over
2}|2\mathbf{b_2}-\mathbf{b_1'}-\mathbf{b_q}|)\theta(A_{c2})+\nonumber
\\
&&\pi^2\left(-N_0(\sqrt{|A_{c2}|}{1\over
2}|2\mathbf{b_2}-\mathbf{b_1'}-\mathbf{b_q}|)+iJ_0(\sqrt{|A_{c2}|}{1\over
2}|2\mathbf{b_2}-\mathbf{b_1'}-\mathbf{b_q}|)
\right)\theta(-A_{c2})]\nonumber \\
&&[2\pi K_0(\sqrt{B_{c2}}{1\over
2}|2\mathbf{b_3}-\mathbf{b_1'}+\mathbf{b_q}|)\theta(B_{c2})+\nonumber
\\
&&\pi^2\left(-N_0(\sqrt{|B_{c2}|}{1\over
2}|2\mathbf{b_3}-\mathbf{b_1'}+\mathbf{b_q}|)+iJ_0(\sqrt{|B_{c2}|}{1\over
2}|2\mathbf{b_3}-\mathbf{b_1'}+\mathbf{b_q}|)
\right)\theta(-B_{c2})]\nonumber \\
&&\left[2\pi K_0(\sqrt{C_{c2}}{1\over
2}|\mathbf{b_1'}+2\mathbf{b_q}|)\theta(C_{c2})+\pi^2\left(-N_0(\sqrt{|C_{c2}|}{1\over
2}|\mathbf{b_1'}+2\mathbf{b_q}|)+iJ_0(\sqrt{|C_{c2}|}{1\over
2}|\mathbf{b_1'}+2\mathbf{b_q}|)
\right)\theta(-C_{c2})\right]\nonumber \\
&&\left[2\pi
K_0(\sqrt{D_{c2}}|\mathbf{b_3}|)\theta(D_{c2})+\pi^2\left(-N_0(\sqrt{|D_{c2}|}|\mathbf{b_3}|)+iJ_0(\sqrt{|D_{c2}|}|\mathbf{b_3}|)
\right)\theta(-D_{c2})\right]\nonumber \\
\end{eqnarray}

\noindent{\bf Appendix E: Expressions for $H_F^{ij}$}\\

$H^{ij}_F$  corresponding to the form factors defined in
eq.(\ref{formfactor}) with a ''tilde'' for $\overline{B^0}\to
\Lambda_c^+ \overline{p}$. The expressions for diagrams
(a1)$\sim$(a36) in Fig. \ref{diagrams1}, (b1)$\sim$(b8) in Fig.
\ref{diagrams2} and (c1), (c2) in Fig. \ref{diagrams3} are listed
in the following.

For the hard amplitudes of Fig.1(a1):
\begin{eqnarray}
&&\widetilde{A^V}=\widetilde{A^A}=\widetilde{B^V}=\widetilde{B^A}=16m_B^5r^2(-1+r^2)(1-y) \nonumber\\
&&\widetilde{A^T}=-\widetilde{B^T}=32m_B^5r^2(-1+r^2)(1-y)
\end{eqnarray}

For the hard amplitudes of Fig.1(a2):
\begin{eqnarray}
&&\widetilde{A^V}=-\widetilde{A^A}=-\widetilde{B^V}=\widetilde{B^A}=16
m_B^5 r(-1 + r^2)^2x_3'(-1 + x_2)
\end{eqnarray}

For the hard amplitudes of Fig.1(a3):
\begin{eqnarray}
&&\widetilde{A^V}=\widetilde{A^A}=\widetilde{B^V}=\widetilde{B^A}=0\nonumber
\\
&&\widetilde{A^T}=\widetilde{B^T}=0
\end{eqnarray}

For the hard amplitudes of Fig.1(a4):
\begin{eqnarray}
&&\widetilde{A^V}=\widetilde{A^A}=\widetilde{B^V}=\widetilde{B^A}=0\nonumber
\\
&&\widetilde{A^T}=\widetilde{B^T}=0
\end{eqnarray}

For the hard amplitudes of Fig.1(a5):
\begin{eqnarray}
&&\widetilde{A^V}=-\widetilde{A^A}=-\widetilde{B^V}=\widetilde{B^A}=-16m_B^5r(r^2-1)^2x_2x_3'
\end{eqnarray}

For the hard amplitudes of Fig.1(a6):
\begin{eqnarray}
&&\widetilde{A^V}=\widetilde{B^A}=-16m_B^5r(-1 + r^2)(r(-1 + x_2 +
y) +(-1 + x_2)(x_2' - r^2x_2' + r(-1 + x_2 + y)))\nonumber
\\
&&\widetilde{A^A}=\widetilde{B^V}=-16m_B^5r(-1 + r^2)(r(-1 + x_2 +
y) +(-1 + x_2)(-x_2' + r^2x_2' + r(-1 + x_2 + y))) \nonumber
\\
&&\widetilde{A^T}=-\widetilde{B^T}=-32m_B^5r^2(-1 + r^2)x_2(-1 +
x_2 +y)
\end{eqnarray}

For the hard amplitudes of Fig.1(a7):
\begin{eqnarray}
&&\widetilde{A^V}=-\widetilde{A^A}=-\widetilde{B^V}=\widetilde{B^A}=-16m_B^5r(-1
+ r^2)x_2
\end{eqnarray}

For the hard amplitudes of Fig.1(a8):
\begin{eqnarray}
\widetilde{A^V}&=&\widetilde{B^A}=-16m_B^4(1 - r^2)(m_c r +
m_B(r^2(-1 + x_3') - x_3') + (x_2 - y)(-(m_c(-1 + r)r) +\nonumber
\\
&&m_B(r^2(-1 + x_3') - x_3' + r^3(-1 + x_1 + y))))\nonumber
\\
\widetilde{A^A}&=&\widetilde{B^V}=16m_B^4(1 - r^2)(-m_c r +
m_B(-r^2(-1 + x_3') + x_3') + (x_2 - y)(-(m_c r(1 + r)) +\nonumber
\\
&&m_B(-(r^2(-1 + x_3')) + x_3' + r^3(-1 + x_1 + y))))
\end{eqnarray}

For the hard amplitudes of Fig.1(a9):
\begin{eqnarray}
\widetilde{A^V}&=&-\widetilde{A^A}=-\widetilde{B^V}=\widetilde{B^A}=16m_B^3(-1
+ r^2)(m_c^2r + m_cm_B(r^2(-1 + x_3') - x_3') + \nonumber
\\
&&m_cm_B(1 + (-1 + r^2)x_1' + r^2(-x_3 + y)) - m_B^2r(-1 + x_1 +
y)(1 + (-1 + r^2)x_1' + r^2(-x_3 + y)))
\end{eqnarray}

For the hard amplitudes of Fig.1(a10):
\begin{eqnarray}
\widetilde{A^V}&=&-\widetilde{A^A}=-\widetilde{B^V}=\widetilde{B^A}=16m_cm_B^4r^2(-1+r^2)(1-y)
\end{eqnarray}

For the hard amplitudes of Fig.1(a11):
\begin{eqnarray}
\widetilde{A^V}&=&-\widetilde{A^A}=-\widetilde{B^V}=\widetilde{B^A}=-16m_B^4(-1
+ r^2)^2x_2'(-m_c + m_Br(-1 + x_1 + y))
\end{eqnarray}

For the hard amplitudes of Fig.1(a12):
\begin{eqnarray}
\widetilde{A^V}&=&\widetilde{A^A}=\widetilde{B^V}=\widetilde{B^A}=-16m_B^4(-1
+ r^2)x_2(m_c r + m_B(r^2(-1 + x_3') - x_3'))
\end{eqnarray}

For the hard amplitudes of Fig.1(a13):
\begin{eqnarray}
\widetilde{A^V}&=&-\widetilde{A^A}=-\widetilde{B^V}=\widetilde{B^A}=-16m_B^4(1
- r^2)(m_Br^3(x_3 - y)(-1 + x_2 + y) + m_c(x_2' + r^2(-1 + x_2 -
x_2' + y)))\nonumber \\
\end{eqnarray}

For the hard amplitudes of Fig.1(a14):
\begin{eqnarray}
\widetilde{A^V}&=&-\widetilde{A^A}=-\widetilde{B^V}=\widetilde{B^A}=-16m_B^5r(-1
+ r^2)(1 + (-1 + r^2)x_1')x_2
\end{eqnarray}

For the hard amplitudes of Fig.1(a15):
\begin{eqnarray}
\widetilde{A^V}&=&\widetilde{B^A}=16m_B^5(-1 + r^2)((1 - r^2)x_3'
-(x_2 - y)(-x_3' + r^2x_3' + r^3(-1 + x_3 + y)))\nonumber
\\
\widetilde{A^A}&=&\widetilde{B^V}=-16m_B^5(-1 + r^2)((-1 +
r^2)x_3' -(x_2 - y)(x_3' - r^2x_3' + r^3(-1 + x_3 + y)))
\end{eqnarray}

For the hard amplitudes of Fig.1(a16):
\begin{eqnarray}
\widetilde{A^V}&=&-\widetilde{A^A}=-\widetilde{B^V}=\widetilde{B^A}=16m_cm_B^4r^2(-1
+ r^2)(-1 + x_3 + y)
\end{eqnarray}

For the hard amplitudes of Fig.1(a17):
\begin{eqnarray}
\widetilde{A^V}&=&-\widetilde{A^A}=-\widetilde{B^V}=\widetilde{B^A}=16m_B^5r^3(-1
+ r^2)(-x_1 + y)(-1 + x_3 + y)
\end{eqnarray}

For the hard amplitudes of Fig.1(a18):
\begin{eqnarray}
\widetilde{A^V}&=&-\widetilde{A^A}=-\widetilde{B^V}=\widetilde{B^A}=-16m_B^5r(-1
+ r^2)^2x_2'(x_1-y)
\end{eqnarray}

For the hard amplitudes of Fig.1(a19):
\begin{eqnarray}
\widetilde{A^V}&=&\widetilde{A^A}=\widetilde{B^V}=\widetilde{B^A}=16m_B^5(-1
+ r^2)^2x_2x_3'
\end{eqnarray}

For the hard amplitudes of Fig.1(a20):
\begin{eqnarray}
\widetilde{A^V}&=&-\widetilde{A^A}=-\widetilde{B^V}=\widetilde{B^A}=-16m_B^5r(1
- r^2)(-x_1 + y)(x_2' + r^2(-1 + x_2 - x_2' + y))
\end{eqnarray}

For the hard amplitudes of Fig.1(a21):
\begin{eqnarray}
&&\widetilde{A^V}=\widetilde{A^A}=\widetilde{B^V}=\widetilde{B^A}=0\nonumber
\\
&&\widetilde{A^T}=\widetilde{B^T}=0
\end{eqnarray}

For the hard amplitudes of Fig.1(a22):
\begin{eqnarray}
&&\widetilde{A^V}=\widetilde{A^A}=\widetilde{B^V}=\widetilde{B^A}=-16m_B^5r^2(-1
+ r^2)(1-y)\nonumber
\\
&&\widetilde{A^T}=-\widetilde{B^T}=-32m_B^5r^2(-1 + r^2)(1-y)
\end{eqnarray}

For the hard amplitudes of Fig.1(a23):
\begin{eqnarray}
&&\widetilde{A^V}=\widetilde{A^A}=\widetilde{B^V}=\widetilde{B^A}=16m_B^4(-1
+ r^2)(m_c r + m_B(-x_2' + r^2(-2 + x_2' + x_3)))(1 - y)\nonumber
\\
&&\widetilde{A^T}=-\widetilde{B^T}=-32m_B^4r(-1 + r^2)(2m_c +
m_Br(-1 + x_3))(1 - y)
\end{eqnarray}

For the hard amplitudes of Fig.1(a24):
\begin{eqnarray}
&&\widetilde{A^V}=\widetilde{A^A}=\widetilde{B^V}=\widetilde{B^A}=-16m_B^5(-1
+ r^2)^2x_2'(1 - y)\nonumber
\\
&&\widetilde{A^T}=-\widetilde{B^T}=64m_B^5r^2(-1 +
r^2)(-1+x_1)(1-y)
\end{eqnarray}

For the hard amplitudes of Fig.1(a25):
\begin{eqnarray}
&&\widetilde{A^V}=\widetilde{A^A}=\widetilde{B^V}=\widetilde{B^A}=16m_B^5r^2(-1
+ r^2)(1 - y)(-1 + x_2 + y)\nonumber
\\
&&\widetilde{A^T}=-\widetilde{B^T}=32m_B^5r^2(-1 + r^2)(1 - y)(-1
+ x_2 + y)
\end{eqnarray}

For the hard amplitudes of Fig.1(a26):
\begin{eqnarray}
&&\widetilde{A^V}=-\widetilde{A^A}=-\widetilde{B^V}=\widetilde{B^A}=16m_B^5r(-1
+ r^2)^2x_2
\end{eqnarray}

For the hard amplitudes of Fig.1(a27):
\begin{eqnarray}
&&\widetilde{A^V}=\widetilde{A^A}=\widetilde{B^V}=\widetilde{B^A}=16m_B^5r^2(-1
+ r^2)(-1+x_2+y)^2\nonumber
\\
&&\widetilde{A^T}=-\widetilde{B^T}=32m_B^5r^2(-1 +
r^2)(-1+x_2+y)^2
\end{eqnarray}

For the hard amplitudes of Fig.1(a28):
\begin{eqnarray}
&&\widetilde{A^V}=-\widetilde{A^A}=-\widetilde{B^V}=\widetilde{B^A}=-16m_B^5r(-1
+ r^2)^2x_2
\end{eqnarray}

For the hard amplitudes of Fig.1(a29):
\begin{eqnarray}
&&\widetilde{A^V}=-\widetilde{A^A}=-\widetilde{B^V}=\widetilde{B^A}=16m_B^4r(-1
+ r^2)(1 - y)(-m_c r + m_B(1 + (-1 + r^2)x_1' + r^2(-x_3 + y)))
\end{eqnarray}

For the hard amplitudes of Fig.1(a30):
\begin{eqnarray}
&&\widetilde{A^V}=-\widetilde{A^A}=-\widetilde{B^V}=\widetilde{B^A}=-16m_B^5r^3(-1
+ r^2)(1 - y)(-x_1 + y)
\end{eqnarray}

For the hard amplitudes of Fig.1(a31):
\begin{eqnarray}
&&\widetilde{A^V}=-\widetilde{A^A}=-\widetilde{B^V}=\widetilde{B^A}=16m_B^5r(-1
+ r^2)(1 - y)\nonumber \\
&&\widetilde{A^T}=-\widetilde{B^T}=32m_B^5r^2(-1 + r^2)(1 - y)x_2
\end{eqnarray}

For the hard amplitudes of Fig.1(a32):
\begin{eqnarray}
&&\widetilde{A^V}=-\widetilde{A^A}=-\widetilde{B^V}=\widetilde{B^A}=16m_B^5r(-1 + r^2)^2(1 - y)\nonumber \\
&&\widetilde{A^T}=-\widetilde{B^T}=-32m_B^5r^2(-1 + r^2)(1 - y)(-1
+ x_2 + y)
\end{eqnarray}

For the hard amplitudes of Fig.1(a33):
\begin{eqnarray}
&&\widetilde{A^V}=-\widetilde{A^A}=-\widetilde{B^V}=\widetilde{B^A}=-16m_B^4m_c(-1+r^2)^2x_3'
\end{eqnarray}

For the hard amplitudes of Fig.1(a34):
\begin{eqnarray}
&&\widetilde{A^V}=-\widetilde{A^A}=-\widetilde{B^V}=\widetilde{B^A}=-16m_B^5r(-1
+ r^2)^2x_3'(-x_1 + y)
\end{eqnarray}

For the hard amplitudes of Fig.1(a35):
\begin{eqnarray}
&&\widetilde{A^V}=\widetilde{B^A}=-16m_B^5r(1 - r^2)(r(-1 + x_2 +
y) +(-1 +
x_2)(-x_3' + r^2x_3' + r(-1 + x_2 + y)))\nonumber \\
&&\widetilde{A^A}=\widetilde{B^V}=-16m_B^5r(1 - r^2)(r(-1 + x2 +
y) +(-1 + x_2)(x_3' - r^2x_3' + r(-1 + x_2 + y)))
\end{eqnarray}

For the hard amplitudes of Fig.1(a36):
\begin{eqnarray}
&&\widetilde{A^V}=\widetilde{B^A}=\widetilde{A^A}=\widetilde{B^V}=-16m_B^5r^2(-1
+ r^2)(-1 + x_2 + y)^2
\end{eqnarray}

For the hard amplitudes of Fig.2(b1):
\begin{eqnarray}
&&\widetilde{A^V}=\widetilde{A^A}=\widetilde{B^V}=\widetilde{B^A}=0\nonumber
\\
&&\widetilde{A^T}=\widetilde{B^T}=0
\end{eqnarray}

For the hard amplitudes of Fig.2(b2):
\begin{eqnarray}
&&\widetilde{A^V}=-\widetilde{A^A}=-\widetilde{B^V}=\widetilde{B^A}=-16m_B^5r(-1+r^2)x_2^2
\end{eqnarray}

For the hard amplitudes of Fig.2(b3):
\begin{eqnarray}
&&\widetilde{A^V}=\widetilde{A^A}=\widetilde{B^V}=\widetilde{B^A}=-16m_B^4(-1
+ r^2)x_2(-m_cr + m_B((-1 + r^2)x_1' + y))
\end{eqnarray}

For the hard amplitudes of Fig.2(b4):
\begin{eqnarray}
&&\widetilde{A^T}=-\widetilde{B^T}=32m_B^4r(-1 + r^2)x_2(m_c +
m_Br(-1 + x_3))
\end{eqnarray}

For the hard amplitudes of Fig.2(b5):
\begin{eqnarray}
&&\widetilde{A^V}=\widetilde{A^A}=\widetilde{B^V}=\widetilde{B^A}=16m_B^5(-1
+ r^2)x_2(r^2(-1 + x_1') - x_1' + y)
\end{eqnarray}

For the hard amplitudes of Fig.2(b6):
\begin{eqnarray}
&&\widetilde{A^T}=-\widetilde{B^T}=-32m_B^5r^2(-1 +
r^2)x_2(-1+x_1)
\end{eqnarray}

For the hard amplitudes of Fig.2(b7):
\begin{eqnarray}
&&\widetilde{A^V}=\widetilde{A^A}=\widetilde{B^V}=\widetilde{B^A}=0\nonumber
\\
&&\widetilde{A^T}=\widetilde{B^T}=0
\end{eqnarray}

For the hard amplitudes of Fig.2(b8):
\begin{eqnarray}
&&\widetilde{A^V}=-\widetilde{A^A}=-\widetilde{B^V}=\widetilde{B^A}=16m_B^5r(-1
+ r^2)x_2^2
\end{eqnarray}

For the hard amplitudes of Fig.3(c1):
\begin{eqnarray}
&&\widetilde{A^V}=\widetilde{B^A}=8m_B^5r(1 - r^2)(1 + x_2 - y -
r(-2 + x_2 + 2y) - (-1 + x_2)(x_2' - x_3' +r^2(-x_2' + x_3') +
r(-2 + x_2 +
2y)))\nonumber \\
&&\widetilde{A^A}=\widetilde{B^V}=-8m_B^5r(1 - r^2)(1 + x_2 - y +
r(-2 + x_2 + 2y) + (-1 + x_2)(-x_2' + r^2(x_2' - x_3') + x_3' +
r(-2 + x_2 +
2y)))\nonumber \\
&&\widetilde{A^T}=-\widetilde{B^T}=16m_B^5r^2(-1 + r^2)x_2(-2 +
x_2 + 2y)
\end{eqnarray}

For the hard amplitudes of Fig.3(c2):
\begin{eqnarray}
&&\widetilde{A^V}=-\widetilde{A^A}=-\widetilde{B^V}=\widetilde{B^A}=-8m_B^4(-1
+ r^2)(m_Br((1 + (-1 + r^2)x_1' + r^2(x_3 - y))(1 + x_2 - y)
+\nonumber \\
&& 3r^2(x_3 - y)(-1 + y)) +m_c(x_2' - x_3' + r^2(-2 + x_2 - x_2' +
x_3' + 2y)))
\end{eqnarray}

\end{document}